\documentclass[%
 reprint,
 superscriptaddress, 
 amsmath,amssymb,
 aps,
 pra,
 floatfix,
]{revtex4-2}
\usepackage{adjustbox}
\usepackage[table,dvipsnames]{xcolor}
\usepackage{array}
\usepackage{graphicx}%
\usepackage{dcolumn}%
\usepackage{bm}%
\usepackage{enumitem}
\usepackage{multirow}
\usepackage{hyperref}
\hypersetup{hidelinks,hypertexnames=false}
\usepackage{amsthm}
\usepackage{amsmath}
\usepackage{tabularx}
\usepackage[linesnumbered,ruled,vlined]{algorithm2e}
\usepackage{stackengine}

\DeclareMathOperator{\Hom}{Hom}
\newtheorem{definition}{Definition}

\newtheorem{lemma}{Lemma}
\usepackage{amsmath, amssymb}
\usepackage{hyperref}
\usepackage{amsmath,amsthm}
\usepackage{float}
\usepackage{epstopdf}
\usepackage{longtable}
\usepackage{mathtools}
\usepackage{colortbl}
\usepackage{booktabs}
\usepackage{mathtools,amssymb}
\usepackage{amsthm}
\usepackage[caption=false]{subfig}
\usepackage{tikz}
\usepackage{amsmath,amssymb}
\usetikzlibrary{positioning,arrows.meta,decorations.pathreplacing,calligraphy,calc}
\usetikzlibrary{cd}
\DeclareMathOperator{\im}{im}
\usepackage{tikz}
\usetikzlibrary{cd}
\usepackage{graphicx}%
\usepackage{dcolumn}%
\usepackage{bm}%
\newtheorem{statement}[lemma]{Statement}
\usepackage{enumitem}
\usepackage{multirow}
\usepackage{hyperref}
\usepackage{float}
\usepackage{csquotes}%
\usepackage[T1]{fontenc}
\makeatletter\AtBeginDocument{\g@addto@macro{\UrlBreaks}{\do\-}}\makeatother
\MakeOuterQuote{"}

\usepackage[linesnumbered, ruled, vlined]{algorithm2e}
\SetKw{KwDownTo}{downto}
\SetKwInput{KwInput}{Input}
\SetKwInput{KwOutput}{Output}
\usepackage{graphicx} 
\usepackage{mathtools}
\usepackage{amsmath}
\usepackage{amsfonts}
\usepackage{amssymb}
\usepackage{amsthm}
\usepackage{makecell}
\usepackage{physics}
\newcommand\editcolor[1]{{\color{blue}#1}}

\DeclareMathOperator{\id}{id}

\newtheorem{proposition}{Proposition}[section]

\newtheorem{remark}{Remark}
\theoremstyle{definition}
\newtheorem{example}{Example}
\theoremstyle{definition}
\def\id{{\mathchoice {\rm 1\mskip-4mu l} {\rm 1\mskip-4mu l} {\rm 1\mskip-4.5mu l} {\rm 1\mskip-5mu l}}}
\usepackage{tikz}
\usetikzlibrary{arrows.meta, positioning}
\usepackage{braket}

\usepackage{listings}
\usepackage{color}
\definecolor{mygreen}{rgb}{0,0.6,0}
\definecolor{myred}{rgb}{0.9,0.2,0.3}
\definecolor{mymauve}{rgb}{0.28,0,0.92}
\definecolor{mygray}{rgb}{0.5,0.5,0.5}

\lstdefinelanguage{JuliaREPL}{
  keywords={julia},
  keywordstyle=\color{mygreen}\bfseries,
  keywords=[2]{Stabilizer, GeneralizedStabilizer, naive_encoding_circuit, PauliOperator, Destabilizer, project, CliffordOperator, apply},
  keywordstyle=[2]\color{mymauve}\bfseries,
  identifierstyle=\color{black},
  sensitive=true,
  mathescape=true,
  comment=[l]{\#},
  morecomment=[s]{/*}{*/},
  commentstyle=\color{mygray}\ttfamily,
  stringstyle=\color{myred}\ttfamily,
  morestring=[b]',
  morestring=[b]"
}

\usepackage{csquotes}%

\usepackage[textsize=tiny]{todonotes}

\definecolor{customcolorblue}{HTML}{4573ae}

\newcommand{\MMTableStyle}{%
  \color{black}%
  \fontsize{9}{11}\selectfont
  \renewcommand{\arraystretch}{1.2}%
  \setlength{\tabcolsep}{2pt}%
  \setlength{\extrarowheight}{2pt}%
  \setlength{\LTleft}{\fill}%
  \setlength{\LTright}{\fill}%
}

\DeclareRobustCommand{\revisionblue}[1]{{\color{blue}#1}}

\colorlet{designblue}{blue}
\colorlet{blue}{black}
\begin{document}

\title{Multivariate Multicycle Codes for Complete Single-Shot Decoding}

\author{Feroz Ahmed Mian}
\email{fmian@umass.edu}
\affiliation{Manning College of Information and Computer Sciences, University of Massachusetts Amherst, 140 Governors Drive
Amherst, Massachusetts 01003, USA}

\author{Owen Gwilliam}
\email{ogwilliam@umass.edu}
\affiliation{Department of Mathematics and Statistics, University of Massachusetts Amherst, Lederle Graduate Research Tower, 1654, 710 N Pleasant St, Amherst, Massachusetts 01003, USA}

\author{Stefan Krastanov}
\email{skrastanov@umass.edu}
\affiliation{Manning College of Information and Computer Sciences, University of Massachusetts Amherst, 140 Governors Drive
Amherst, Massachusetts 01003, USA}
\affiliation{Department of Physics, University of Massachusetts Amherst, 1126 Lederle Graduate Research Tower,
710 North Pleasant Street
Amherst, Massachusetts 01003, USA}

\date{\today}

\begin{abstract}
We introduce multivariate multicycle (MM) codes, \revisionblue{a Koszul-complex formulation of abelian multicycle (AMC) codes, which are multidimensional scalar lifted-product codes over a finite abelian group algebra,} that unifies bivariate bicycle, multivariate bicycle, abelian two-block group algebra, generalized bicycle, trivariate tricycle, and toric codes. MM codes are Calderbank-Shor-Steane (CSS) codes defined from length-\revisionblue{$(t+1)$} chain complexes\revisionblue{, and for $t\geq4$ they possess metachecks for both types of checks, which can be used for single-shot decoding}. We offer a framework that facilitates the construction of long-length chain complexes through the use of \revisionblue{the} Koszul complex. In particular, obtaining explicit boundary maps (parity check and metacheck matrices) is particularly straightforward in our approach. \revisionblue{We also determine when an MM code encodes logical qubits, count the independent check relations not generated by its metachecks, and show that middle-degree codes with even $t$ have equal $X$ and $Z$ distances.} This simple but very general parameterization of codes permitted us to efficiently perform a numerical search, where we identify several MM code candidates \revisionblue{with} high rates and high code distances. Examples of new codes with parameters $[[n,k,d]]$ include $[[96,12,8]]$, $[[144,12,12]]$, \revisionblue{$[[486,24,12]]$, $[[630,70,9]]$, and $[[648,60,9]]$. For selected codes, we report confinement profiles and study single-shot and few-shot decoding under phenomenological and circuit-level noise. Our codes also include explicit collapsed 5D through 9D instances, and our 5D instances have} check weights \revisionblue{lower than} those of recent small instances of quantum Tanner codes.
\end{abstract}

\maketitle

\section{Introduction}

The development of quantum low-density parity-check (qLDPC) codes~\cite{Tillich_2014,Breuckmann_2021, gottesman2013fault, Panteleev_2021, breuckmann2021balanced,dinur2021locallytestablecodesconstant, panteleev2022asymptoticallygoodquantumlocally, Leverrier_2022, leverrier2022efficientdecodingconstantfraction, dinur2022goodquantumldpccodes, gu2022efficientdecoderlineardistance, leverrier2022quantumtannercodes, leverrier2025smallquantumtannercodes, radebold2025explicit, evra2020decodablequantumldpccodes, liang2025generalized, guemard2025lifts, guémard2025moderatelengthliftedquantumtanner, Hastings_2021, gulshen2025quantum, 10756140, liang2025planar, wang2026check} represents a pivotal advancement in quantum error correction (QEC), offering a route to high-threshold, high-rate logical memories and scalable low-overhead fault-tolerant quantum processors. One of their many advantages is the possibility of single-shot or few-shot decoding~\cite{Bombin_2013, bombin2015single, Campbell_2019, Hong_2025}, which can suppress errors using only one or a few rounds of noisy syndrome extraction. The frontier of the field has now shifted from memory to computation, where the central challenge is the construction of codes that retain their single-shot property during fault-tolerant logical operations. Current approaches, including limited transversal gates~\cite{eastin2009restrictions,Breuckmann_2024}, code automorphisms~\cite{sayginel2025faulttolerantlogicalcliffordgates, eberhardt2024logicaloperatorsfoldtransversalgates}, and code surgery~\cite{Ide_2025,cowtan2024css,cowtan2025parallel, Cohen_2022, zheng2025highratesurgeryconstantoverheadlogical, cross2025improvedqldpcsurgerylogical, williamson2024lowoverheadfaulttolerantquantumcomputation}, must interface with the decoding problem, making the code's performance under a single-shot decoder~\cite{breuckmann2016local, brown2016fault, Herold_2017, 9585444,Vasmer_2021, Quintavalle_2021, Higgott_2023, gu2022efficientdecoderlineardistance, lin2024single, Guo_2026} \revisionblue{a key metric} for a scalable architecture~\cite{tan2025single, yoder2025tourgrossmodularquantum}.

A significant line of research has focused on constructing such codes using algebraic techniques, leading to celebrated families of codes such as bivariate bicycle (BB)~\cite{Bravyi_2024, eberhardt2024logicaloperatorsfoldtransversalgates,postema2025existencecharacterisationbivariatebicycle, symons2025sequences, liang2025selfdualbivariatebicyclecodes}, abelian and non-abelian two-block group algebra (2BGA)~\cite{lin2023quantumtwoblockgroupalgebra}, and the recently introduced trivariate tricycle (TT) \cite{jacob2025singleshotdecodingfaulttolerantgates, menon2025magictricyclesefficientmagic}. A key motivation for exploring higher-dimensional algebraic constructions (such as three-dimensional (3D) gauge color ~\cite{Brown_2016}, 4D surface, and 4D toric~\cite{Dennis_2002, breuckmann2016local, Berthusen_2024}, double homological product~\cite{Campbell_2019}, homological product~\cite{xu2024fastparallelizablelogicalcomputation}, 3D surface and 3D toric~\cite{Quintavalle_2021}, 3D subsystem toric~\cite{Kubica_2022}, abelian multicycle~\cite{lin2025abelianmulticyclecodessingleshot}, intertwined toric~\cite{Stahl_2024}, 4D lifted homological product (4D-LHP)~\cite{11249819}, 4D hyperbolic~\cite{9585444}) is the inherent presence of local redundancies among parity checks. These linear redundancies, often formalized as ``checks on checks'' or metachecks, are important for single-shot error correction~\cite{PhysRevX.5.031043, Campbell_2019}.

\revisionblue{Single-shot error correction with metachecks for both types of checks} has been recently demonstrated in practice through the implementation of the $[[33,1,4]]$ 4D surface code on a trapped-ion quantum computer~\cite{Berthusen_2024}.

In this work, we establish that \revisionblue{several} of the algebraic constructions mentioned above~\cite{Bravyi_2024, voss2025multivariate, PhysRevA.88.012311, lin2023quantumtwoblockgroupalgebra, jacob2025singleshotdecodingfaulttolerantgates, rabeti2025listdecodingnewbicycle, aydin2025cyclic} can be unified \revisionblue{within a framework of} Koszul complexes over quotient rings with cyclic relations. We call the codes obtained in this way Multivariate Multicycle (MM) codes. Our formalism employs tools from commutative algebra and homological algebra~\cite{eisenbud2013commutative, weibel1994introduction}. \revisionblue{Over a finite abelian group algebra, the Koszul complex coincides with the multi-block complex of the abelian multicycle (AMC) construction of Lin et al.~\cite{lin2025abelianmulticyclecodessingleshot}, giving the same codes up to basis ordering (Proposition~\ref{prop:mm_amc}).} We \revisionblue{use} the standard isomorphism between tensor products of \revisionblue{two-term} chain complexes and Koszul complexes to provide a systematic framework for the construction of these codes. \revisionblue{Each factor is defined by a $1\times1$ matrix containing one defining polynomial. Thus, the construction is an iterated lifted product of chain complexes over the group algebra in the sense of Panteleev and Kalachev~\cite{panteleev2022almostlinear} (Sec.~\ref{sec:balanced_product_of_complexes}).}

Compared to the \revisionblue{recursive procedures} for finding the boundary maps (i.e.\ parity checks and metachecks) in previous work (e.g.~\cite{lin2025abelianmulticyclecodessingleshot, jacob2025singleshotdecodingfaulttolerantgates, menon2025magictricyclesefficientmagic} or Fig.~2 of~\cite{11249819}), the Koszul complex provides simple closed-form expressions for the boundary maps. \revisionblue{Standard properties of these complexes also determine when the codes encode logical qubits and count the independent check relations not generated by their metachecks (Proposition~\ref{prop:mm_homology}). For middle-degree codes with even $t$, we show that the $X$ and $Z$ distances are equal (Proposition~\ref{prop:xz_duality}).} This simplicity enabled us to efficiently perform a numerical search for good \revisionblue{codes}, as seen in our numerical experiments section.

For example, TT codes \revisionblue{have adjacent Koszul metachecks only for their $Z$-type checks and have been studied for} partial single-shot decoding~\cite{jacob2025singleshotdecodingfaulttolerantgates}. \revisionblue{For MM codes with $t\geq4$, the complex supplies metachecks for both check types, with both maps nonzero whenever at least one defining polynomial is nonzero (Remark~\ref{prop:mm_metachecks}). Their existence alone does not establish single-shot decoding: the decoder must also control the residual data errors after processing a noisy syndrome (Sec.~\ref{sec:single_shot_error_correction}). We investigate single-shot decoding under phenomenological noise and few-shot decoding under phenomenological and circuit-level noise for selected finite instances.}

Koszul complex theory~\cite{green1984koszul, aprodu2010koszul, august2015differential} has recently been applied to several other problems in fault-tolerant quantum computation. We mention other applications here for general context, noting they represent uses distinct from our approach: In context of BB codes, Ref.~\cite{Chen_2025} employs a Koszul duality property yielding an $R$-module isomorphism between the homology groups $H(a,b;g,f)$ and $H(g,f;a,b)$, where $(a,b)=(x^\ell-1,y^m-1)$ encode periodic boundary conditions and $(g,f)$ the stabilizer checks, relating equivalence classes of logical operators to equivalence classes of anyon configurations. In proof theory~\cite{negri2008structural}, \revisionblue{Koszul complexes are used} to model cut-elimination in multiplicative proof nets as error-correction processes~\cite{murfet2024linearlogicquantumerror}. Furthermore, Ref.~\cite{breuckmann2025logical} establishes that tile codes correspond to higher global sections of the shifted Koszul complex \(K^\bullet(\mathbb{P}^1\times\mathbb{P}^1,f,g)\otimes\mathcal{S}\) of vector bundles on \(\mathbb{P}^1\times\mathbb{P}^1\), where \(f,g\in\mathbb{F}_2[x^{\pm1},y^{\pm1}]\) encode stabilizer generators and \(\mathcal{S}\) is a line bundle controlling code dimensions and boundaries. Their framework yields logical operators isomorphic to \(\mathbb{F}_2[x^{\pm1},y^{\pm1}]/(f,g)\) and enables derived automorphisms---operations implementable via fault-tolerant lattice surgery~\cite{Horsman_2012, Litinski_2019} that induce logical CNOT circuits.

This paper is structured as follows: In Section~\ref{sec:ha_preliminaries}, we provide the mathematical preliminaries, first from the perspective of homological algebra and single-shot error correction\revisionblue{, where we also define the notions of metacheck structure, reduced weight, irreducible errors, and confinement used later,} and then from the perspective of commutative algebra in Section~\ref{sec:ca_preliminaries}. Section~\ref{sec:previous_constructions} surveys previous constructions of abelian 2BGA codes, including BB codes, multivariate bicycle (MB) codes, generalized bicycle (GB) codes, and TT codes, establishing the context for our work. Section~\ref{sec:koszul_complex_formalism_for_quantum_codes} introduces our Koszul complex formalism for quantum codes. After defining these complexes using the language of exterior algebras and tensor products, it reinterprets the code constructions from \revisionblue{the} previous section as specific manifestations of this general framework. In Section~\ref{sec:multivariate_multicycle_codes}, using the method established in the previous sections, we then present our main contribution: the Multivariate Multicycle (MM) codes. \revisionblue{We prove that their metachecks exist for $t\geq4$, establish their relation to AMC and lifted-product codes, and show how they unify} various families of previous QECCs. Section~\ref{sec:numerical_simulations} provides numerical simulations using the belief propagation with ordered statistics decoding (BP+OSD)~\cite{Roffe_LDPC_Python_tools_2022} and Tesseract~\cite{beni2025tesseractsearchbaseddecoderquantum} decoders over MM codes found from \revisionblue{a} search over MM polynomials for $t=4$. We compare these results to other metacheck-capable codes\revisionblue{, and report finite-weight syndrome profiles together with single-shot and overlapping-window decoding experiments}. In Section~\ref{sec:discussion}, we discuss our results in relation to other codes of contemporary interest \revisionblue{and outline} directions for future research. The Appendix further includes additional instances of MM codes, \revisionblue{the search and distance-verification procedures, and the simulation settings.}

\section{Preliminaries from homological algebra\label{sec:ha_preliminaries}}

This section presents the preliminaries for our study of classical and quantum codes through the lens of homological algebra. For standard introductions to homological algebra, the reader may refer to~\cite{eisenbud2013commutative, weibel1994introduction}. Readers familiar with other works that employ chain complexes for the creation of codes can skip this section\revisionblue{, except for Sec.~\ref{sec:single_shot_error_correction}, where we fix the single-shot terminology (metacheck structure, reduced weight, irreducible errors, confinement, and soundness) used in the rest of the paper}.

\subsection{Chain Complexes and Homology\label{sec:chain_complexes_and_homology}}
We give a formal definition of chain complexes and homology. The definition below is quite dry and concise, but the next subsection restates these definitions in more familiar terms related to error correcting codes. The main intuition is that chain complexes, when used for creating stabilizer codes, are a natural way to build sequences of vector spaces with linear constraints at each step of the sequence. These constraints will be the parity checks or metachecks of our codes.

Let $R$ be a commutative ring. In the category $R\text{-}Mod$ of $R$-modules, a \emph{chain complex} $(\mathcal{C}_\bullet, \partial_\bullet)$ is a sequence of $R$-modules and $R$-linear maps
\begin{equation}
\cdots \leftarrow C_{i-1} \xleftarrow{\partial_i} C_i \xleftarrow{\partial_{i+1}} C_{i+1} \leftarrow \cdots
\end{equation}
satisfying $\partial_i \circ \partial_{i+1} = 0$ for all $i$ (see Definition 1.1.1 of~\cite{weibel1994introduction}). This perspective allows us to define the fundamental homological invariants as modules:
\begin{equation}
Z_i(\mathcal{C}) = \ker \partial_i, \quad 
B_i(\mathcal{C}) = \im \partial_{i+1}, \quad H_i(\mathcal{C}) = Z_i(\mathcal{C}) / B_i(\mathcal{C})
\end{equation}
called \emph{\(i\)-cycles}, \emph{\(i\)-boundaries}, and \emph{\(i\)-th homology}, respectively.

The contravariant functor $\Hom_R(-, R): R\text{-}Mod \to R\text{-}Mod$ yields the \emph{cochain complex} $(\mathcal{C}^\bullet, \delta^\bullet)$ with $C^i = \Hom_R(C_i, R)$ and coboundary maps $\delta^i = \partial_{i+1}^*$. The cohomology modules are:
\begin{equation}
Z^i(\mathcal{C}) = \ker \delta^i, \quad B^i(\mathcal{C}) = \im \delta^{i-1}, \quad H^i(\mathcal{C}) = Z^i(\mathcal{C}) / B^i(\mathcal{C})
\end{equation}

When $R = \mathbb{F}_2$, we work in the category of $\mathbb{F}_2$-vector spaces. If each $C_i$ is equipped with a basis, we may identify $C^i \cong C_i$ via the canonical pairing\revisionblue{, so that $\delta^i$ is the transpose of $\partial_{i+1}$. Over a field, the universal coefficient theorem gives $H^i(\mathcal{C})\cong\Hom_{\mathbb{F}_2}(H_i(\mathcal{C}),\mathbb{F}_2)$, so $H^i(\mathcal{C})$ and $H_i(\mathcal{C})$ have the same dimension~\cite{weibel1994introduction, breuckmann2021balanced}.}

\subsection{Codes from Chain Complexes\label{sec:codes_from_chain_complexes}}

\subsubsection{Classical Codes\label{sec:classical_codes}}

A classical $[n,k,d]$ linear code $\mathcal{C} \subset \mathbb{F}_2^n$ corresponds to a 2-term chain complex with $R=\mathbb{F}_2$:
\begin{equation}
\mathcal{C}_{\text{classical}}: \quad \revisionblue{\mathbb{F}_2^{m}} \xleftarrow{P} \mathbb{F}_2^n
\end{equation}
where $P$ is \revisionblue{an $m\times n$} parity-check matrix\revisionblue{, whose rows need not be independent, so that $m\geq n-k$}. The codespace is $\mathcal{C} = Z_1(\mathcal{C}_{\text{classical}}) = \ker P$ with dimension:
\begin{equation}
k = \dim H_1(\mathcal{C}_{\text{classical}})
\end{equation}
and code distance given by
\begin{equation}
d = \min\{ |c| : c \in Z_1(\mathcal{C}_{\text{classical}}) \setminus B_1(\mathcal{C}_{\text{classical}}) \}
\end{equation}

\subsubsection{Quantum CSS Codes}
A CSS $[[n,k,d]]$ code~\cite{calderbank1996good, steane1996simple} is specified by two binary matrices $P_X$ and $P_Z$ of sizes $m_X \times n$ and $m_Z \times n$ respectively, satisfying the orthogonality condition  
\begin{equation}
P_X P_Z^{\top} = 0.
\end{equation}
These matrices correspond to a 3-term chain complex of $\mathbb{F}_2$-vector spaces  
\begin{equation}
\label{eqn: CSS as chain complex}
\mathcal{C}_{\mathrm{CSS}}  : \mathbb{F}_2^{m_X}
\xleftarrow{\;\partial_1 = P_X\;}
\mathbb{F}_2^{n}
\xleftarrow{\;\partial_2 = P_Z^{\top}\;} 
\mathbb{F}_2^{m_Z} 
\end{equation}
where the condition $\partial_1 \partial_2 = 0$ corresponds to the orthogonality of $P_X$ and $P_Z$.  
The \revisionblue{standard basis vectors of} $\mathbb{F}_2^n$ in degree 1 \revisionblue{label the $n$ physical qubits, and a binary vector specifies the support of an $X$- or $Z$-type Pauli operator}.
The homology group in degree 1 is the quotient vector space
\begin{equation}
H_1(\mathcal{C}_{\mathrm{CSS}}) = \frac{\ker \partial_1}{\operatorname{im} \partial_2}
   = \frac{\ker P_X}{\operatorname{im} P_Z^{\top}}.
\end{equation}
Its nonzero elements represent homology classes of nontrivial $Z$-type logical operators---cycles that cannot be expressed as boundaries.

Dually, the cohomology group in degree 1 is the quotient vector space
\begin{equation}
H^1(\mathcal{C}_{\mathrm{CSS}}) = \frac{\ker \partial_2^{\!\top}}{\operatorname{im} \partial_1^{\!\top}}
   = \frac{\ker P_Z}{\operatorname{im} P_X^{\top}}.
\end{equation}
Its nonzero elements correspond to cohomology classes of nontrivial $X$-type logical operators. 
The code parameters are  
\begin{equation}
k = \dim H_1(\mathcal{C}_{\mathrm{CSS}}) = \dim H^1(\mathcal{C}_{\mathrm{CSS}})
\end{equation}
and the $Z$ and $X$ distances are defined as the minimum Hamming weights among representatives of nontrivial (co)homology classes:  
\begin{align}
d_Z = \min\bigl\{ |z| : z \in \ker P_X \setminus \operatorname{im} P_Z^{\top} \bigr\}, \\
d_X = \min\bigl\{ |x| : x \in \ker P_Z \setminus \operatorname{im} P_X^{\top} \bigr\}.
\end{align}
The overall code distance is $d = \min(d_X, d_Z)$.

A family of CSS quantum codes is \emph{low-density parity-check (LDPC)} if there exists a constant $w$ such that each row and column of $P_X$ and $P_Z$ has Hamming weight at most $w$.

\subsection{Tensor Product of Complexes\label{sec:balanced_product_of_complexes}}

The tensor product of complexes constructs higher-length complexes from a few lower-length complexes. As such, it is a convenient tool to create codes with extra structure (e.g.\ quantum codes or even codes with metachecks) out of simpler building blocks (e.g.\ classical codes).
 
The tensor product of complexes is built as follows~\cite{panteleev2022almostlinear, eberhardt2024logicaloperatorsfoldtransversalgates}. Consider two-term complexes in $R\text{-}Mod$:
\begin{align}
\mathcal{A}_\bullet: &\quad R \xleftarrow{a} R \\
\mathcal{B}_\bullet: &\quad R \xleftarrow{b} R
\end{align}

Their tensor product $\mathcal{A}_\bullet \otimes_R \mathcal{B}_\bullet$ is the following double complex:
\begin{equation}
\begin{array}{ccc}
R \otimes_R R & \xleftarrow{a \otimes \id} & R \otimes_R R \\
\downarrow{\scriptstyle \id \otimes b} & & \downarrow{\scriptstyle \id \otimes b} \\
R \otimes_R R & \xleftarrow{a \otimes \id} & R \otimes_R R
\end{array}
\end{equation}

Here, the differentials should follow the Koszul sign rule, but when \(R\) has characteristic \(2\) (which is the case for complexes used to build stabilizer codes), the sign distinctions vanish, simplifying the expressions.

Taking the total complex $\operatorname{Tot}(\mathcal{A}_\bullet \otimes_R \mathcal{B}_\bullet)$ yields a complex of higher length:
\begin{equation}
R \otimes_R R \xleftarrow{(a \otimes \id, \id \otimes b)} 
(R \otimes_R R) \oplus (R \otimes_R R) \xleftarrow{\begin{pmatrix} \id \otimes b \\ a \otimes \id \end{pmatrix}} R \otimes_R R.
\end{equation}
This ``total complex'' is frequently also called ``the tensor product complex''. 
A quantum code is associated to $\operatorname{Tot}(\mathcal{A}_\bullet \otimes_R \mathcal{B}_\bullet)$, where $\mathcal{A}_\bullet$ and $\mathcal{B}_\bullet$ are chain complexes representing classical linear codes.

\revisionblue{In quantum coding, tensor products of this kind were formalized by Panteleev and Kalachev, who introduced lifted products over finite-dimensional algebras, such as $R=\mathbb{F}_2[G]$, and described the lifted product of two chain complexes as their tensor product over $R$~\cite{panteleev2022almostlinear}; Appendix~B of that work gives the algebraic characterization of these codes. For a commutative algebra $R$, the scalar case $\mathrm{LP}(a,b)$ is exactly the tensor product of the two-term complexes $R\xleftarrow{a}R$ and $R\xleftarrow{b}R$ shown above, and it includes the generalized bicycle codes.

The coefficient rings used in this paper have the form $R=\mathbb{F}_2[G]$ with $G=\mathbb{Z}_{\ell_1}\times\cdots\times\mathbb{Z}_{\ell_D}$, which in the product-group basis is isomorphic to $\mathbb{F}_2[x_1,\dots,x_D]/\langle x_1^{\ell_1}-1,\dots,x_D^{\ell_D}-1\rangle$. Tensor products of more than two scalar complexes over such rings have also been studied: Eberhardt and Steffan describe the $t$-fold product for general $t$, work out the maps for $t=3$, and identify the adjacent differentials as relations among checks~\cite[Appendix~B]{eberhardt2024logicaloperatorsfoldtransversalgates}; Li, Preskill and Xu construct three-dimensional lifted products with general constituent matrices over finite abelian group algebras and note that the construction extends to more factors~\cite{li2025transversaldimensionjump}; and Ref.~\cite{wu2025generalconstruction} gives a related construction from several classical codes.

In this language, the complex that defines an MM code is the $t$-fold tensor product over $\mathbb{F}_2[G]$ of scalar two-term complexes, that is, an iterated lifted product (or balanced product~\cite{breuckmann2021balanced}) of $1\times1$ matrices, with qubits in degree $q=\lfloor t/2\rfloor$. We make this identification precise in Proposition~\ref{prop:mm_amc} and the discussion following it.}

\subsection{Single-Shot Error Correction\label{sec:single_shot_error_correction}}

Quantum codes constructed via sufficiently long chain complexes naturally have metachecks for both the X and Z parity checks. Let the code be defined on the middle space $C_q$ of the chain complex in consideration with $P_X = \partial_q$ and $P_Z = \partial_{q+1}^\top$
where $P_X$ and $P_Z$ are the X and Z-type parity check matrices, respectively. The natural candidates for metacheck matrices are the adjacent boundary maps
\begin{align}
M_X = \partial_{q-1} \quad\text{and}\quad M_Z = \partial_{q+2}^{\top}
\end{align}
so that $M_X P_X=0$ and $M_Z P_Z=0$ \revisionblue{by the chain-complex condition $\partial_{q-1}\partial_q = 0$ and $\partial_{q+1}\partial_{q+2}=0$}.
When the chain complex is sufficiently long \revisionblue{(at least five terms, in degrees $0,\ldots,t$ with $t\geq4$ and qubits in degree $q=\lfloor t/2\rfloor$), both $M_X$ and $M_Z$ are available; for $t=3$ only $M_Z$ is available, and for $t=2$ neither is. For MM codes, both maps are nonzero as soon as one of the defining polynomials is (Remark~\ref{prop:mm_metachecks}).} These metachecks encode the redundancies among stabilizers that facilitate single-shot error correction~\cite{Campbell_2019, Quintavalle_2021}.

\begingroup\color{blue}
\begin{definition}[Partial and complete metacheck structure]
\label{def:metacheck_structure}
A CSS check presentation equipped with the adjacent chain-complex metacheck maps $M_X$ and $M_Z$ defined above has a \emph{complete} (two-sided) metacheck structure if both $M_X$ and $M_Z$ are nonzero, and a \emph{partial} (one-sided) metacheck structure if exactly one of them is nonzero.
\end{definition}
In these terms, TT codes have partial metacheck structure: their Koszul complex supplies $M_Z$ but no $M_X$ in our convention, and Ref.~\cite{jacob2025singleshotdecodingfaulttolerantgates} accordingly describes them as supporting partial single-shot decoding. MM codes with $t\geq4$ and a nonzero defining polynomial have complete metacheck structure (Remark~\ref{prop:mm_metachecks}). The notion concerns the adjacent maps of the chain complex, which need not account for every linear relation among the checks. Checks can be redundant without such maps: the BB $[[144,12,12]]$ code, whose Koszul complex has only three terms, has six independent relations among its $X$-type checks. Conversely, when $k>0$ the metachecks of an MM code capture only part of the redundancy of its checks, and Proposition~\ref{prop:mm_homology} counts the relations they miss.

The role of the metachecks in a single round of noisy measurements is easy to describe. Write the observed syndromes as $\sigma_X=P_Xe_Z+m_X$ and $\sigma_Z=P_Ze_X+m_Z$, where $e_X,e_Z$ are data errors and $m_X,m_Z$ are measurement errors. The chain-complex identities give $M_X\sigma_X=M_Xm_X$ and $M_Z\sigma_Z=M_Zm_Z$, so each metasyndrome depends only on the corresponding measurement error. It reveals every measurement error outside the kernel of the metacheck matrix, while measurement errors in the kernel leave it unchanged.
\endgroup

The single-shot distances are defined by

\begin{align}
d_{ss}^X = \min \{ |s| : M_X s = 0,\ s \notin \revisionblue{\im P_X} \}, \\
d_{ss}^Z = \min \{ |s| : M_Z s = 0,\ s \notin \revisionblue{\im P_Z} \}.
\end{align}

These quantify the code's robustness against measurement errors during a single measurement round~\cite{Campbell_2019}\revisionblue{: they are the minimum weights of invalid syndromes that satisfy all the corresponding metachecks. Decoding performance depends in addition on the noise model, the decoder, and how well the code controls the residual data errors.}

\begingroup\color{blue}
\textit{Reduced weight, irreducible errors, and confinement.} To make the last point precise, and because these notions are used throughout Sec.~\ref{sec:numerical_simulations}, we now recall how the size of an error is compared with the size of its syndrome. In the following, $H$ denotes one of the two check matrices of a CSS code: $H=P_X$ when we consider $Z$-type errors and $H=P_Z$ when we consider $X$-type errors. An error is a binary vector $e$ on the qubits, its syndrome is $He$, and $|\cdot|$ denotes Hamming weight.

\begin{definition}[Reduced weight~\cite{Campbell_2019, Quintavalle_2021}]
\label{def:reduced_weight}
The \emph{reduced weight} of an error $e$ is the smallest weight of an error with the same syndrome,
\begin{equation}
|e|_{\mathrm{red}} = \min\{\,|e'| : He' = He\,\} = \min\{\,|e+c| : c \in \ker H\,\}.
\end{equation}
An error is \emph{reduced} if $|e| = |e|_{\mathrm{red}}$.
\end{definition}
The kernel of $H$ contains the stabilizers of the opposite type ($\im P_Z^{\top}$ for $Z$-type errors and $\im P_X^{\top}$ for $X$-type errors), so stabilizer-equivalent errors have the same reduced weight. Because the minimum also runs over logical operators, $|e|_{\mathrm{red}}$ can be smaller than the minimum weight over the stabilizer-equivalence class of $e$.

\begin{definition}[Irreducible error]
\label{def:irreducible}
A nonzero error $e$ is \emph{reducible} with respect to $H$ if it can be written as $e = e_1 + e_2$, where $e_1, e_2$ are nonzero, have disjoint supports, and no row of $H$ has a nonzero entry on both supports. Otherwise $e$ is \emph{irreducible}. Equivalently, $e$ is irreducible exactly when its support induces a connected subgraph of the graph on the qubits in which two qubits are adjacent whenever some check in $H$ acts on both.
\end{definition}
This is the error analogue of the irreducible codewords used in the connected-cluster distance algorithm~\cite{Dumer_2017, Pryadko-2025-distm4ri}. If $e = e_1 + e_2$ is reducible, the syndromes $He_1$ and $He_2$ have disjoint supports, so $|He| = |He_1| + |He_2|$. Every error therefore splits into irreducible components whose syndrome weights add up to the syndrome weight of the error, and it is natural to study the syndrome weights of irreducible errors.

\begin{definition}[Confinement~\cite{Quintavalle_2021}]
\label{def:confinement}
Let $\Delta$ be an integer and $f$ an increasing function with $f(0)=0$. The code has \emph{$(\Delta,f)$-confinement} with respect to $H$ if
\begin{equation}
f(|He|) \geq |e|_{\mathrm{red}} \quad \text{for every error $e$ with } |e|_{\mathrm{red}} \leq \Delta.
\end{equation}
It has \emph{linear confinement} with constant $c>0$ if $f(x) = x/c$, that is, if $|He| \geq c\,|e|_{\mathrm{red}}$ whenever $|e|_{\mathrm{red}}\leq \Delta$. An infinite family of codes has \emph{good confinement} if each member has $(\Delta,f)$-confinement with $\Delta \in \Omega(n^{b})$ for some $b>0$ and with $f$ independent of the block length $n$.
\end{definition}
Confinement with respect to $P_X$ (for $Z$-type errors) and with respect to $P_Z$ (for $X$-type errors) are the boundary and coboundary confinement of Ref.~\cite[Definitions~3.44--3.45]{sriram2025diffusioncodesselfcorrectionsmallerset}. \revisionblue{Since every error can be replaced by a reduced error with the same syndrome, and by the additivity noted after Definition~\ref{def:irreducible}, it suffices to check a linear bound on irreducible errors of Hamming weight at most $\Delta$.}

\begin{definition}[Soundness~\cite{Campbell_2019}]
\label{def:soundness}
Let $\Delta$ be an integer and $f$ a function with $f(0)=0$. The code is \emph{$(\Delta,f)$-sound} with respect to $H$ if $f(|He|) \geq |e|_{\mathrm{red}}$ for every error $e$ with $|He| \leq \Delta$.
\end{definition}
Confinement constrains the syndromes of small errors, whereas soundness constrains the errors behind small syndromes. A sound LDPC code is also confined~\cite{Quintavalle_2021}, but the converse need not hold.

In Sec.~\ref{sec:numerical_simulations} we probe Definition~\ref{def:confinement} on finite instances through syndrome profiles (Table~\ref{tab:confinement}): for each error weight $w = 1, 2, \ldots$, the smallest nonzero syndrome weight found by a connected-cluster search among irreducible errors of Hamming weight $w$. Such profiles are a finite-size diagnostic rather than a confinement bound: they use Hamming rather than reduced weight, come from a time-limited search, and describe one code at a time. Establishing confinement or soundness bounds that hold uniformly over MM code families is left for future work (Sec.~\ref{sec:discussion}).
\endgroup

\begin{remark}
Notably, single-shot decoding does not necessarily require higher-dimensional complexes. Quantum expander codes exhibit single-shot decoding under stochastic noise due to their expansion property of the underlying expander graphs~\cite{Leverrier_2015, leverrier2022quantumtannercodes, Gu_2024, tan2025singleshotuniversalityquantumldpc}. Planar \editcolor{GB} codes, for example, achieve single-shot and highly effective two-shot decoding when stabilizer redundancy gives a maximum syndrome distance \(d_{S}= 3\)~\cite{lin2025singleshottwoshotdecodinggeneralized}.

For a unifying perspective on these cases, we refer the reader to the notion of confinement introduced in Ref.~\cite{Quintavalle_2021} \revisionblue{(Definition~\ref{def:confinement})}, which captures the principle that \revisionblue{small errors must produce correspondingly large syndromes}. It is closely related to the notion of soundness introduced earlier by Campbell~\cite{Campbell_2019} \revisionblue{(Definition~\ref{def:soundness}), which for LDPC codes is the stronger of the two conditions.} While quantum expander codes have confinement but not soundness~\cite{Quintavalle_2021}, strong expansion is a sufficient condition for confinement~\cite{sriram2025diffusioncodesselfcorrectionsmallerset}.
\end{remark}

\section{Preliminaries from commutative algebra\label{sec:ca_preliminaries}}

While the previous section describes the utility of chain complexes for the construction of codes in the abstract, this section enumerates a number of algebraic structures that are particularly useful for the explicit construction of specific instances of chain complexes.

This section presents preliminaries from commutative algebra, including polynomial rings, quotient rings, and group rings, which underlie our MM code framework and are used extensively in later sections. The tensor algebra serves as a convenient framework for constructing exterior algebras and wedge products, which are then used to define the Koszul complex over a ring $R$. Readers already familiar with these algebraic tools may skip this section. For a standard introduction to commutative algebra, the reader may refer to~\cite{dummit2004abstract, eisenbud2013commutative}.

\subsection{Group Rings}

Many contemporary QECCs can be described using chain complexes of modules over an algebra $\mathbb{F}_2[G]$ associated to a group~$G$. 
Here, we formally define such group algebras (or group rings).
Let $R$ be a commutative ring with identity, and let $G = \{g_1, g_2, \dots, g_n\}$ be a finite group. The group ring $RG$ (also denoted $R[G]$) is the set of all formal sums
\begin{align}
\sum_{i=1}^n a_i g_i, \qquad a_i \in R
\end{align}
where the $a_i$ are coefficients from $R$ and the $g_i$ are elements of $G$. 
Addition and multiplication are defined by
\begin{align}
\sum_{i=1}^n a_i g_i + \sum_{i=1}^n b_i g_i
&= \sum_{i=1}^n (a_i+b_i) g_i, \\
\left(\sum_{i=1}^n a_i g_i\right)
\left(\sum_{j=1}^n b_j g_j\right)
&= \sum_{k=1}^n
\left(\sum_{g_i g_j = g_k} a_i b_j \right) g_k .
\end{align}
With these operations, $RG$ forms a group ring~\cite{dummit2004abstract}. This ring is commutative if $G$ is an abelian group.

When the coefficient ring $R$ is a field $F$, the group ring $RG$ is often called
the $F$-algebra (a.k.a.~group algebra over $F$), denoted $F[G]$. In this case, $F[G]$ is a vector space
over $F$ with basis $G$, so that its dimension as an $F$-vector space is $|G|$, and
the elements of $F$ commute with all elements of $F[G]$. These two properties make
$F[G]$ an $F$-algebra~\cite{dummit2004abstract}.

\subsection{Polynomial Rings}
Before introducing quotient rings, we recall the definition of the polynomial ring
$R[x]$ in one indeterminate $x$ over a commutative ring $R$.
An element, known as a polynomial, is a formal sum
\begin{align}
f(x) = \sum_{i=0}^n a_i x^i
\end{align}
where $a_i \in R$ and where $n$ can be any natural number~\cite{dummit2004abstract}.

Addition is defined componentwise:
\begin{align}
\sum_{i=0}^n a_i x^i + \sum_{i=0}^n b_i x^i
= \sum_{i=0}^n (a_i + b_i)x^i
\end{align}
where coefficients may be zero to accommodate polynomials of different degrees.
Multiplication is determined by the rule
\begin{align}
(a x^i)(b x^j) = ab\, x^{i+j}
\end{align}
and extended to all polynomials by distributivity, so that
\begin{align}
\left( \sum_{i=0}^n a_i x^i \right)
\left( \sum_{j=0}^m b_j x^j \right)
= \sum_{k=0}^{n+m} \left( \sum_{i+j=k} a_i b_j \right) x^k .
\end{align}
If $a_n \neq 0$, the integer $n$ is called the degree of $f(x)$ and $a_n$ is called its
leading coefficient. 
The zero polynomial is defined to have leading coefficient~$0$. 
A polynomial is monic if its leading coefficient is $1$. Identifying each $r \in R$ with the constant polynomial $r$, the ring $R[x]$ contains $R$ as a subring and hence has an identity~\cite{dummit2004abstract}.

\begin{remark}
Let $\mathbb{K}$ be a field and let $V$ be an $n$-dimensional $\mathbb{K}$-vector space with basis ${x_1,\dots,x_n}$. Then the symmetric algebra on $V$ is isomorphic to the polynomial ring in $n$ variables over $\mathbb{K}$:
\begin{align}
 \operatorname{Sym}(V) \cong \mathbb{K}[x_1,\dots,x_n].   
\end{align}
\end{remark}

\subsection{Quotient Rings}
In quantum code construction, quotient rings are frequently used to encode periodic boundary conditions and ``cyclic'' algebraic relations. Here we define the quotient ring formally.

Let $R$ be a commutative ring with identity and let $I\subset R$. The subset $I$ is an ideal when $I$ is an additive subgroup of $R$ and is closed under multiplication by arbitrary elements of $R$, i.e., $ra \in I$ for all $r\in R$ and $a\in I$. Consider the set of additive cosets
\begin{align} 
r+I=\{r+a \,:\, a\in I\}
\end{align} 
with $r \in R$. 
With addition and multiplication of cosets defined by
\begin{align}
\label{eq:quotient_ring_addmul}
(r+I)+(s+I)&=(r+s)+I \\ (r+I)(s+I)&=rs+I
\end{align}
the collection of cosets becomes the quotient ring~$R/I$. 

\subsection{Tensor Algebra}
Before defining the exterior algebra, 
which plays a central role in this paper,
we first introduce the tensor algebra, which provides a convenient method for constructing the exterior algebra as a quotient algebra. 
We first review graded rings, graded ideals, and graded homomorphisms, which provide the language for formulating tensor and exterior algebras.

A ring $S$ is called graded if
\begin{align}
S &= \bigoplus_{k=0}^\infty S_k, \quad 
S_i S_j \subseteq S_{i+j}, \quad \forall i,j \ge 0
\end{align}
where $S_k$ is called the homogeneous component of degree $k$ and its elements are homogeneous of degree $k$.

An ideal $I \subseteq S$ is a graded ideal if
\begin{align}
I &= \bigoplus_{k=0}^\infty (I \cap S_k).
\end{align}
A ring homomorphism $\varphi: S \to T$ is a graded homomorphism if
\begin{align}
\varphi(S_k) &\subseteq T_k, \quad \forall k \ge 0.
\end{align}
Now we define a graded algebra formally built from a given module.

Let $R$ be a commutative ring with identity and let $M$ be an $R$-module. For each integer $k \ge 1$, define
\begin{align}
\label{eq:tensors}
T^k(M) &= M \otimes_R \cdots \otimes_R M \quad {(k \text{ factors})}
\end{align}
and set $T^0(M) = R$~\cite{dummit2004abstract}. The tensor algebra of $M$ is a graded, non-commutative algebra with
\begin{align}
\label{eq:tensor_algebra}
T(M) &= \bigoplus_{k=0}^\infty T^k(M)
\end{align}
where multiplication is simply the tensor product itself:
\begin{align}
(m_1 \otimes \cdots &\otimes m_k) \cdot (m'_1 \otimes \cdots \otimes m'_\ell) \\
&= m_1 \otimes \cdots \otimes m_k \otimes m'_1 \otimes \cdots \otimes m'_\ell.
\end{align}
Elements of $T^k(M)$ are called $k$-tensors. 
Identifying $M$ with $T^1(M)$, the module $M$ is an $R$-submodule of $T(M)$. 
For further discussion on the multiplicative structure and the universal property of $T(M)$, see~\cite[Theorem 31]{dummit2004abstract}.

\subsection{Exterior Algebra}
Having defined the tensor algebra $T(M)$, we next construct the exterior algebra as a quotient of $T(M)$ that enforces skew-commutativity (a.k.a. anti-commutativity).

Let $M$ be an $R$-module over a commutative ring $R$.
The exterior algebra of $M$ is the quotient of the tensor algebra $T(M)$ by the two-sided ideal $J(M)$ generated by all elements of the form
\begin{align}
m \otimes m, \quad \text{with }m \in M.
\end{align}
We denote the exterior algebra by $\bigwedge M$, and write the image of $m_1 \otimes \cdots \otimes m_k$ in $\bigwedge M$ as $m_1 \wedge \cdots \wedge m_k$.

The ideal $J(M)$ enforces skew-commutativity. 
For any $x, y \in M$, the identity
\begin{align}
(x + y) \otimes (x + y) = x \otimes x + x \otimes y + y \otimes x + y \otimes y
\end{align}
holds in $T(M)$ and hence induces the identity 
\begin{align}
0 = (x + y) \wedge (x + y) = x \wedge x + x \wedge y + y \wedge x + y \wedge y
\end{align}
in~$\bigwedge M$.
Since $x \wedge x = 0 = y \wedge y$, we obtain that
\begin{align}
0 &\equiv x \wedge y + y \wedge x \\
&\implies x \wedge y = -\, y \wedge x \quad \forall\, x, y \in M
\end{align}
in the quotient ring~$\bigwedge M$.
Thus, $\bigwedge M$ is a skew-commutative algebra obtained from $T(M)$ by imposing the relations $m \otimes m = 0$ for all $m \in M$.

Of note is the fact that the exterior algebra is graded \cite[Proposition 33]{dummit2004abstract}, with $k$th homogeneous component
\editcolor{\begin{align}
\label{eq:exterior_algebra_homogeneous_component}
\bigwedge^k(M) = T^k(M) / J^k(M)
\end{align}}
where $J^k$ is the $k$th graded component of the ideal $J(M)$ and we define $\wedge^0(M) = R$ and $\wedge^1(M) = M$, treating $M$ as an $R$-submodule of $\bigwedge(M)$.
The wedge product (a.k.a.\ exterior product, as in ``exterior algebra'') that we just defined above is an alternative to the tensor product when building higher-length chain complexes. In particular, in a later section we will introduce the Koszul complex, building it either in terms of tensor products or in terms of wedge products. We will see that the wedge product notation is useful for simplifying the process of building longer-length chain complexes.

More formally, the multiplication in $\bigwedge(M)$, called the ``wedge product'', is alternating: $m_1 \wedge \cdots \wedge m_k = 0$ if $m_i = m_j$ for some $i \neq j$, and simple wedge products satisfy anticommutativity $m \wedge m' = - m' \wedge m$ for $m,m' \in M$ \cite[Theorem 36]{dummit2004abstract}.

\begin{remark}
\revisionblue{In characteristic~2 the exterior algebra is commutative, since $m \wedge m' = m' \wedge m$, but it is not the symmetric (polynomial) algebra, because every generator still squares to zero. 
Hence 
\[
\textstyle\bigwedge(M) \cong \operatorname{Sym}(M)/(e_1^2,\dots,e_t^2)
\]
for a free module $M$ with basis $e_1,\dots,e_t$.
}
\end{remark}

\section{Previous Explicit Constructions\label{sec:previous_constructions}}

In this section we survey previously discovered families of qLDPC codes, anticipating rederiving them as special cases of our Multivariate Multicycle (MM) codes in an upcoming section. The reader may skip this section if they are familiar with the recent discoveries of various qLDPC codes.

\subsection{Abelian Two-Block Group Algebra Codes\label{sec:abelian_2bga}}

The construction of 2BGA codes~\cite{lin2023quantumtwoblockgroupalgebra} considers two length-1 complexes $\mathcal{A}_1 \xrightarrow{a} \mathcal{A}_0$ and $\mathcal{B}_1 \xrightarrow{b} \mathcal{B}_0$. Their tensor product complex has the form:
\begin{equation}
\mathcal{A}_0 \otimes_G \mathcal{B}_0 \xleftarrow{\;\,\partial_1\;\,} (\mathcal{A}_0 \otimes_G \mathcal{B}_1) \oplus (\mathcal{A}_1 \otimes_G \mathcal{B}_0) \xleftarrow{\;\,\partial_2\;\,} \mathcal{A}_1 \otimes_G \mathcal{B}_1
\end{equation}

Using $\mathbb{F}_2[G] \otimes_G \mathbb{F}_2[G] \simeq \mathbb{F}_2[G]$, this yields the total complex which has the following boundary maps:
\begin{align}
\partial_1 = \begin{bmatrix} a & b \end{bmatrix}, \quad 
\partial_2 = \begin{bmatrix} b \\ a \end{bmatrix}
\end{align}

For $G = \mathbb{Z}_\ell \times \mathbb{Z}_m$, this construction gives BB codes with parity-check matrices $P_X = \partial_1$ and $P_Z = \partial_2^\top$.

\subsection{Multivariate Bicycle Codes\label{sec:mb_code}}

MB~\cite{voss2025multivariate} codes generalize the BB construction by writing the two defining polynomials in more than two variables. \revisionblue{The group is unchanged: MB codes, like BB codes, are defined over $\mathbb{F}_2[\mathbb{Z}_\ell \times \mathbb{Z}_m]$, and the additional variables are monomials in the two BB generators. The word ``trivariate'' therefore counts variables, not cyclic factors of the group.}

For the trivariate bicycle (TB) case, which is the one explicitly explored in Table 2 of~\cite{voss2025multivariate}, the matrices are:
\begin{equation}
x = S_l \otimes I_m,\quad y = I_l \otimes S_m,\quad z = S_l \otimes S_m
\end{equation}
where $S_l$ and $S_m$ are cyclic shift matrices. \revisionblue{Thus $z=xy$ lies in the group $\mathbb{Z}_l\times\mathbb{Z}_m$ generated by $x$ and $y$, and all three matrices belong to the regular representation of $\mathbb{F}_2[\mathbb{Z}_l\times\mathbb{Z}_m]$.}
The stabilizer generator matrices maintain the standard bicycle form:
\begin{align}
P_X = [A|B], \quad P_Z = [B^\top|A^\top]
\end{align}
where \editcolor{$A(x,y,z)$} and $B(x,y,z)$ are polynomials in $x$, $y$, and $z$\revisionblue{, with $z=xy$}.

\subsection{Generalized Bicycle Codes\label{sec:gb_code}}

GB codes~\cite{PhysRevA.88.012311} constitute a special case of MM codes where one cyclic dimension is trivial. The standard GB construction employs a single cyclic group $\mathbb{Z}_\ell$ to generate commuting circulant matrices $A$ and $B$, forming stabilizer matrices:
\begin{equation}
P_X = [A|B], \quad P_Z = [B^\top|A^\top]
\end{equation}
yielding quantum codes with parameters $[[2\ell, k, d]]$.

The GB code construction uses the quotient ring $\mathbb{F}_2[x,y]/\langle x^\ell-1, y-1\rangle \cong\mathbb{F}_2[x]/\langle x^\ell-1\rangle$ due to $y=1$. The generating polynomials employ only the non-trivial variable $x$.

\subsection{Trivariate Tricycle Codes\label{sec:tt_code}}

The work of~\cite{jacob2025singleshotdecodingfaulttolerantgates, menon2025magictricyclesefficientmagic} extends the BB codes to three length-1 complexes $\mathcal{A}_\bullet$, $\mathcal{B}_\bullet$, $\mathcal{C}_\bullet$. Their tensor product (after some algebraic manipulations) has the form~\cite{jacob2025singleshotdecodingfaulttolerantgates}:

\editcolor{\begin{equation}
\begin{aligned}
\label{eq:equation_35}
&\bigoplus_{i+j+k=2} \mathcal{A}_i \otimes_G \mathcal{B}_j \otimes_G \mathcal{C}_k \xleftarrow{\partial_3} \mathcal{A}_1 \otimes_G \mathcal{B}_1 \otimes_G \mathcal{C}_1 \\
&\qquad \mathcal{A}_0 \otimes_G \mathcal{B}_0 \otimes_G \mathcal{C}_0 \xleftarrow{\partial_1} \bigoplus_{i+j+k=1} \mathcal{A}_i \otimes_G \mathcal{B}_j \otimes_G \mathcal{C}_k \xleftarrow{\partial_2}
\end{aligned}
\end{equation}}

The boundary maps of this tensor product complex are given by~\cite{jacob2025singleshotdecodingfaulttolerantgates}:

\begin{align}
\label{eq:tt_code}
\quad \partial_3 = \begin{bmatrix} a \\ b \\ c \end{bmatrix}, 
\quad \partial_2 = \begin{bmatrix} 0 & c & b \\ c & 0 & a \\ b & a & 0 \end{bmatrix}, 
\quad \partial_1 = \begin{bmatrix} a & b & c \end{bmatrix}
\end{align}
For TT codes, the parity-check matrices are $P_X = \partial_1$, $P_Z = \partial_2^\top$, and $M_Z = \partial_3^\top$ provides Z-metachecks. 
\begin{figure*}
\centering
\begin{tikzpicture}[
  font=\normalsize,
  box/.style={
    rectangle, rounded corners=6pt,
    draw,
    very thick,
    align=center,
    text width=6.4cm,
    inner sep=10pt,
    minimum height=1.1cm
  },
  amc/.style={box, fill=red!3, draw=red!60!black},
  mm/.style={box, fill=teal!3, draw=teal!60!black},
  css/.style={box, fill=violet!4, draw=violet!60!black, text width=13.5cm},
  arr/.style={-{Stealth[length=6pt]}, thick, draw=gray!60},
  header/.style={font=\bfseries\large}
]
\node[header] at (-4.2,0.7) {AMC \revisionblue{description}};
\node[header] at ( 4.2,0.7) {\revisionblue{Koszul (MM) description}};
\node[amc] (a1) at (-4.2,-0.8)
  {Abelian \textcolor{red}{group algebra}};
\node[amc] (a2) at (-4.2,-2.6)
  {\textcolor{red}{Regular representation} of group algebra elements};
\node[amc] (a3) at (-4.2,-4.4)
  {Multi-block complex (MBC) via \textcolor{red}{iterative tensor products of chain complexes}};
\node[mm] (m1) at (4.2,-0.8)
  {\textcolor{designblue}{Polynomial ring} R (free commutative algebra)};
\node[mm] (m2) at (4.2,-2.6)
  {\textcolor{designblue}{Quotient} R by an \textcolor{designblue}{ideal} I \\ (Circulant matrix representation of polynomials in R/I)};
\node[mm] (m3) at (4.2,-4.4)
  {\textcolor{designblue}{Koszul complex} over R/I};
\node[css] (css) at (0,-6.4)
  {CSS quantum code};
\draw[arr] (a1) -- (a2);
\draw[arr] (a2) -- (a3);
\draw[arr] (a3) -- (a3.south |- css.north);

\draw[arr] (m1) -- (m2);
\draw[arr] (m2) -- (m3);
\draw[arr] (m3) -- (m3.south |- css.north);
\draw[gray!30, line width=0.6pt] (0,-0.4) -- (0,-5.9);
\end{tikzpicture}
\caption{\revisionblue{\textbf{Two descriptions of the same CSS codes.}} Left (AMC codes~\cite{lin2025abelianmulticyclecodessingleshot}): To make a candidate for a good qLDPC CSS code, start by picking an abelian group algebra, then pick a regular representation that will be used to convert elements of the algebra to binary matrices, then pick a few elements of that algebra, and finally, use the MBC rule~\cite{lin2025abelianmulticyclecodessingleshot} to map the binary matrix representation of these elements to an arbitrary length chain complex. Right (\revisionblue{Koszul description, MM codes}): To make a candidate code, start by picking a polynomial ring $R$ (i.e.\ picking a fixed number of symbolic variables $x,\ y,\ z,\ \dots$), then pick an ideal \revisionblue{$I$ of cyclic relations} (equivalent to specifying lattice size for the physical qubits), then pick $t$ polynomials from the quotient of the polynomial ring and the ideal (potentially at random), and finally use \revisionblue{the Koszul complex} to make a chain complex \revisionblue{with $t+1$ terms}. \revisionblue{By the fundamental theorem of finitely generated abelian groups~\cite[Sec.~5.2]{dummit2004abstract}, every finite abelian group is isomorphic to a direct product of cyclic groups, $G\cong\mathbb{Z}_{\ell_1}\times\cdots\times\mathbb{Z}_{\ell_D}$. The orders $\ell_i$ are unique only once a normal form, such as the invariant factors or the elementary divisors, is fixed, but any such decomposition can be used here, and it identifies the group algebra with the quotient ring $\mathbb{F}_2[G]\cong\mathbb{F}_2[x_1,\ldots,x_D]/\langle x_1^{\ell_1}-1,\ldots,x_D^{\ell_D}-1\rangle$. The two routes therefore yield the same complexes up to basis ordering (Proposition~\ref{prop:mm_amc}). What the Koszul description adds is a closed-form expression for every boundary map, which makes the search over polynomials convenient (Tables~\ref{tab:table2}--\ref{tab:table11}). With qubits in the middle degree, complexes with at least five terms have metachecks for both types of checks.}}

\label{fig:MMcodesconstruction}
\end{figure*}

\subsection{Challenges in creating higher-length complexes}

The TT codes provide only partial single-shot decoding (they have metachecks only for the Z parity checks). 
A longer complex lets one provide metachecks for both the X and Z parity checks. The chief difficulty now is to figure out an ``easy'' technique to construct sufficiently long complexes. 
The technique of tensor products employed in the discovery of the TT codes can, in principle, be used to create such even longer complexes~\cite{lin2025abelianmulticyclecodessingleshot}, but explicitly describing the boundary maps (i.e.\ the parity check and metacheck matrices) of such complexes can grow complicated with those methods. Another alternative is the more general approach of AMC codes~\cite{lin2025abelianmulticyclecodessingleshot}, using a so-called ``multiblock complex'' (MBC) rule, also generating long enough complexes, at the cost of following a \revisionblue{recursive} procedure. However, using the Koszul complex over a quotient ring, as we advocate in this paper, offers a simpler, efficient way to describe these matrices that has already been in use for a long time in other fields of mathematics. \revisionblue{It produces the same complexes as the MBC rule (Proposition~\ref{prop:mm_amc}), but with every boundary map written down directly in terms of the defining polynomials, so no new set of block matrices has to be derived for each length.} In particular, it substantially expedites generating random samples of such codes, as we demonstrate through the many codes we have \revisionblue{found}, listed in Tables~\revisionblue{\ref{tab:table2}--\ref{tab:table11}}. For a comparison, see Fig.~\ref{fig:MMcodesconstruction}. Moreover, our construction \revisionblue{recovers} many families of known codes (see Fig.~\ref{fig:conceptual_diagram_of_subfamilies}) and \revisionblue{provides} explicit instances of collapsed high-dimensional quantum codes \revisionblue{(defined in Sec.~\ref{sec:main_construction})} with low check weights.

\section{Koszul Complex Formalism for Quantum Codes\label{sec:koszul_complex_formalism_for_quantum_codes}}

We establish that the abstract chain complexes (and, equivalently, their boundary maps which provide parity checks and potentially metachecks)
defining the recently introduced 2-block and 3-block group algebra codes (BB~\cite{Bravyi_2024}, TT codes~\cite{jacob2025singleshotdecodingfaulttolerantgates, menon2025magictricyclesefficientmagic} and more),
are manifestations of a fundamental object in commutative algebra and algebraic geometry: the Koszul complex. This connection places these previous QECC constructions within a well-studied mathematical framework, allowing us to systematically derive their properties, generalize them to higher dimensions (thus providing both X and Z metachecks), and perform efficient numerical searches for new codes.

\subsection{Koszul Complexes\label{sec:koszul_complexes}}

Koszul complexes can be defined in a few different ways. A definition (shown below) in terms of tensor products makes it easy to restate previous code constructions in terms of Koszul complexes. Another equivalent definition, in terms of exterior algebra, makes the derivation of boundary maps more explicit. Here we present both definitions and use the exterior algebra approach to restate the codes from Sec.~\ref{sec:previous_constructions} in new, simpler, more general terms, making them also much easier to construct through modern computer algebra systems like OSCAR~\cite{OSCAR,OSCAR-book}.

Let $R$ be a commutative Noetherian ring with identity, and let
\editcolor{$\underline{x} = (x_1,\dots,x_t)$} be a finite sequence of elements of $R$.
The associated Koszul complex is a standard $R$-complex admitting several equivalent constructions~\cite{atiyah2018introduction, bruns1998cohen, eisenbud2013commutative, sather2009homological, stacks-project}. We recall the tensor product and exterior algebra
descriptions, which will be used interchangeably.
\begin{definition}[Tensor product construction~\cite{sather2009homological}]
For \revisionblue{$x_1\in R$}, let
\begin{equation}
K_\bullet(x_1; R)=\bigl(0 \longleftarrow R \xleftarrow{x_1} R \longleftarrow 0\bigr)
\end{equation}
denote the $R$-complex concentrated in degrees $1$ and $0$.
For \revisionblue{$\underline{x}=(x_1,\dots,x_t)$}, define
\begin{equation}
\revisionblue{K_\bullet(\underline{x}; R)
\coloneqq
K_\bullet(x_1; R)\otimes_R \cdots \otimes_R K_\bullet(x_t; R)}
\end{equation}
where the tensor product is taken in the category of $R$-complexes.
This $R$-complex is called the \emph{Koszul complex} on \revisionblue{$\underline{x}$}.
\end{definition}
An equivalent description is obtained using wedge product from exterior algebra. The reader may consult~\cite{abraham2012manifolds, bruns1998cohen, eisenbud2013commutative, darling1994differential} for more details.
\begin{definition}[Exterior algebra construction~\cite{sather2009homological}]
Let $\underline{x}$ be a sequence of elements in $R$. Define the $R$-modules:
\editcolor{\begin{align}
K_0(\underline{x}; R) &= R \quad\text{ with basis }\{1\}    \\
K_1(\underline{x}; R) &= R^t \quad\text{with basis }\{e_1,\dots,e_t\}
\end{align}}
and for $1 < i \leq t$, consider $K_i(\underline{x}) = R^{\binom{t}{i}}$ with basis
\begin{align}
{\{e_{j_1} \wedge \cdots \wedge e_{j_i} \mid 1 \leq j_1 < \cdots < j_i \leq t\}}  
\end{align}
The Koszul complex $K_\bullet(\underline{x}; R)$ is the $R$-complex
\editcolor{\begin{align}
0 \leftarrow K_0(\underline{x};R) \xleftarrow{\partial_1} K_1(\underline{x}; R) \xleftarrow{\partial_2} \cdots \xleftarrow{\partial_t} K_t(\underline{x}; R) \leftarrow 0
\end{align}}
where the differentials $\partial_i: K_i(\underline{x}) \to K_{i-1}(\underline{x})$ are defined on basis elements by
\begin{align}
\label{eq:explicit_formula}
\partial_i(e_{j_1} \wedge \cdots \wedge e_{j_i}) = \sum_{s=1}^i (-1)^{s+1} x_{j_s} \; e_{j_1} \wedge \cdots \wedge \widehat{e}_{j_s} \wedge \cdots \wedge e_{j_i}
\end{align}
where $\widehat{e}_{j_s}$ indicates that the factor $e_{j_s}$ is omitted from the wedge product.
Notice that each $K_i(\underline{x}; R)$ is a free $R$-module, so the homomorphism $\partial_i$ is completely determined by its values on the basis. In particular, $\partial_1: K_1(\underline{x};R) \to K_0(\underline{x}; R)$ is given by $\partial_1(e_j) = x_j$ for each $j = 1,\dots,t$.
\end{definition}

If $R=\mathbb{F}_2[G]$ with $G$ a finite abelian group, then all sign factors vanish.

\subsection{Examples of Boundary Maps and Code Constructions\label{sec:boundary_maps_and_code_constructions}}

The exterior algebra description provides explicit formulas for the boundary maps, which correspond directly to the parity check and metacheck matrices in CSS code constructions. \editcolor{We form the standard $(t+1)$-term Koszul complex over a polynomial quotient ring}, which, as we saw above, has a simple explicit form for its boundary maps. To then generate a specific instance of a code from this class of complexes, the variables appearing in boundary maps are replaced by circulant matrix representations of polynomials from multivariate quotient rings. Thus deciding on a $t$ and on these polynomials completely defines a code instance (and if $t$ is large enough, we also have metachecks, see Eq.~\eqref{eq:explicit_formula}).

We restate all of the previously established code constructions (see Sec.~\ref{sec:previous_constructions}) in our Koszul complex formalism:

\begin{example}[BB codes~\cite{Bravyi_2024} (\(t=2\))]
Consider $R = \mathbb{F}_2[x,y]$, and let $S = R/\langle x^\ell-1, y^m-1 \rangle 
\cong \mathbb{F}_2[\mathbb{Z}_\ell \times \mathbb{Z}_m]$. Let $\pi: R \rightarrow S$ be the natural projection mapping. Note that $\ker(\pi) = \langle x^\ell-1, y^m-1 \rangle$, since for any ring $R$ and ideal $I \subseteq R$, the kernel of $r \mapsto r + I$ from $R$ to $R/I$ is $I$ \cite[Theorem 15.4]{gallian2021contemporary}. Given bivariate polynomials $f(x,y)$, $g(x,y) \in R$, we define their images in $S$ as $F = \pi(f(x,y))$ and $G = \pi(g(x,y))$.
 
Form the Koszul complex $K_\bullet([F,G]; S)$:
\begin{equation}\label{eq:koszul-complex-sequence-t2}
0 \longleftarrow K_0 \xleftarrow{\partial_1} K_1 \xleftarrow{\partial_2} K_2 \longleftarrow 0
\end{equation}
which has the standard $S$-modules and bases (for $t=2$):
\begin{alignat}{3}
&K_0 \cong S &\text{ with basis }&\{1\} \\
&K_1 \cong S^2 &\text{ with basis }&\{e_1, e_2\} \\
&K_2 \cong S &\text{ with basis }&\{e_1 \wedge e_2\}
\end{alignat}
and which has the standard boundary maps:
\begin{align}\label{eq:boundary-maps-t2}
\revisionblue{\partial_1} = \begin{bmatrix} F & G \end{bmatrix}, \quad 
\revisionblue{\partial_2} = \begin{bmatrix} G \\ F \end{bmatrix}.
\end{align}
To construct the BB code,
\revisionblue{we note that $F$ and $G$ already lie in $S$, so the relations $x^\ell = 1$ and $y^m = 1$ hold throughout and this is a chain complex of $S$-modules. To obtain binary matrices we use the regular representation of $S$: the monomial $x^ay^b$ acts on the group basis by the permutation matrix $S_\ell^a\otimes S_m^b$, where $S_\ell$ and $S_m$ are cyclic shift matrices, so each polynomial becomes a sum of Kronecker products of shift matrices (a two-level circulant matrix). For a single variable this is} the standard isomorphism between the ring of $n \times n$ circulant matrices over \revisionblue{$\mathbb{F}_2$} and the group algebra $\mathbb{F}_2[\mathbb{Z}_n]$ of the cyclic group~\cite{hurley2006group}\revisionblue{,~\cite[Theorem~2.1]{albuquerque2011hopfalgebraicpropertiescirculant}}. As a result, we obtain a CSS code with \revisionblue{$P_X = \partial_1$ and $P_Z = \partial_2^\top$}. 
\end{example}

\begin{example}[TT codes~\cite{jacob2025singleshotdecodingfaulttolerantgates, menon2025magictricyclesefficientmagic} (\(t=3\))]
Consider $R = \mathbb{F}_2[x,y,z]$, and let 
$S = R/\langle x^\ell-1, y^m-1, z^p-1 \rangle 
\cong \mathbb{F}_2[\mathbb{Z}_\ell \times \mathbb{Z}_m \times \mathbb{Z}_p]$.
Let $\pi: R \rightarrow S$ be the natural projection mapping. Given trivariate polynomials $f(x,y,z)$, $g(x,y,z)$, $h(x,y,z) \in R$, we define their images in $S$ as $F = \pi(f(x,y,z))$, $G = \pi(g(x,y,z))$ and $H = \pi(h(x,y,z))$.

Form the Koszul complex $K_\bullet([F,G,H]; S)$:
\begin{equation}
0 \longleftarrow K_0 \xleftarrow{\partial_1} K_1 \xleftarrow{\partial_2} K_2 
\xleftarrow{\partial_3} K_3 \longleftarrow 0
\end{equation}
which has the standard $S$-modules and bases (for $t=3$):
\begin{alignat}{3}
&K_0 \cong S &\text{ with basis }&\{1\} \\
&K_1 \cong S^3 &\text{ with basis }&\{e_1, e_2, e_3\} \\
&K_2 \cong S^3 &\text{ with basis }&\{e_1 \wedge e_2,\ e_1 \wedge e_3,\ e_2 \wedge e_3\} \\
&K_3 \cong S &\text{ with basis }&\{e_1 \wedge e_2 \wedge e_3\}
\end{alignat}
and which has the standard boundary maps:
\begin{align}
\revisionblue{\partial_1} &= \begin{bmatrix} F & G & H \end{bmatrix}, &
\partial_2 &= \begin{bmatrix} 
G & H & 0 \\ 
F & 0 & H \\ 
0 & F & G 
\end{bmatrix}, &
\revisionblue{\partial_3} &= \begin{bmatrix} H \\ G \\ F \end{bmatrix}
\end{align}
To obtain the TT code, we replace each polynomial $F(x,y,z)$, $G(x,y,z)$, $H(x,y,z)$ \revisionblue{in $S$, where the cyclic relations $x^\ell = 1$, $y^m = 1$ and $z^p=1$ already hold, with its regular representation, a sum of Kronecker products $S_\ell^a\otimes S_m^b\otimes S_p^c$ of cyclic shift matrices (a three-level circulant matrix).}
This produces a chain complex over $S$ that encodes a CSS code with \revisionblue{$P_X = \partial_1$, $P_Z = \partial_2^\top$, 
and $Z$-metacheck $M_Z = \partial_3^\top$}.
\end{example}
\begin{example}[\revisionblue{Multivariate multicycle (MM) codes} at \(t=4\)]
Consider the polynomial ring $R = \mathbb{F}_2[w,x,y,z]$ and let
$S = R/\langle w^\ell-1, x^m-1, y^p-1, z^r-1 \rangle$, 
which is isomorphic to $\mathbb{F}_2[\mathbb{Z}_\ell \times \mathbb{Z}_m \times \mathbb{Z}_p \times \mathbb{Z}_r]$. Let $\pi: R \rightarrow S$ be the natural projection mapping. Given multivariate polynomials $f(w,x,y,z)$, $g(w,x,y,z)$, $h(w,x,y,z)$, $i(w,x,y,z) \in R$, we define their images in $S$ as $F = \pi(f(w,x,y,z))$, $G = \pi(g(w,x,y,z))$, $H = \pi(h(w,x,y,z))$ and $I = \pi(i(w,x,y,z))$.
Form the Koszul complex $K_\bullet([F,G,H,I]; S)$:
\begin{equation}
0 \longleftarrow K_0 \xleftarrow{\partial_1} K_1 \xleftarrow{\partial_2} K_2 
\xleftarrow{\partial_3} K_3 \xleftarrow{\partial_4} K_4 \longleftarrow 0
\end{equation}
which has the standard $S$-modules and bases (for $t=4$):
\begin{alignat}{3}
&K_0 \cong S &\text{ with basis }&\{1\} \\
&K_1 \cong S^4 &\text{ with basis }&\{e_1, e_2, e_3, e_4\} \\
&K_2 \cong S^6 &\text{ with basis }&\{e_1 \wedge e_2,\ e_1 \wedge e_3,\ e_1 \wedge e_4, \nonumber \\
&&&\ e_2 \wedge e_3,\ e_2 \wedge e_4,\ e_3 \wedge e_4\} \\
&K_3 \cong S^4 &\text{ with basis }&\{e_1 \wedge e_2 \wedge e_3,\ e_1 \wedge e_2 \wedge e_4, \nonumber \\
&&&\ e_1 \wedge e_3 \wedge e_4,\ e_2 \wedge e_3 \wedge e_4\} \\
&K_4 \cong S &\text{ with basis }&\{e_1 \wedge e_2 \wedge e_3 \wedge e_4\}
\end{alignat}
and which has the standard boundary maps:
\begin{align}
\label{eq:4d_koszul_complex}
\revisionblue{\partial_1} &= \begin{bmatrix} F & G & H & I \end{bmatrix}, &
\revisionblue{\partial_2} &= \begin{bmatrix} 
G & H & I & 0 & 0 & 0 \\ 
F & 0 & 0 & H & I & 0 \\ 
0 & F & 0 & G & 0 & I \\ 
0 & 0 & F & 0 & G & H 
\end{bmatrix} \\
\revisionblue{\partial_3} &= \begin{bmatrix} 
H & I & 0 & 0 \\ 
G & 0 & I & 0 \\ 
0 & G & H & 0 \\ 
F & 0 & 0 & I \\ 
0 & F & 0 & H \\ 
0 & 0 & F & G 
\end{bmatrix}, &
\revisionblue{\partial_4} &= \begin{bmatrix} I \\ H \\ G \\ F \end{bmatrix}
\end{align}
To obtain the MM code, following the process we've seen twice now, 
we replace each polynomial $F(w,x,y,z)$, $G(w,x,y,z)$, $H(w,x,y,z)$, $I(w,x,y,z)$ with \revisionblue{its regular representation, a four-level circulant matrix built from Kronecker products of cyclic shift matrices.}
This produces a chain complex over $S$ that encodes a CSS code with \revisionblue{$P_X = \partial_2$, $P_Z = \partial_3^\top$, 
and metachecks $M_X = \partial_1$, $M_Z = \partial_4^\top$}.
\end{example}

\section{Multivariate Multicycle Codes\label{sec:multivariate_multicycle_codes}}

Finally, we show how the mechanistic procedure shown in Sec.~\ref{sec:koszul_complex_formalism_for_quantum_codes} can be used not only for representing many classes of known codes in simpler and more unified terms, but also for the creation of new codes, possessing valuable properties (like \revisionblue{metachecks for both the X and Z parity checks when $t \geq 4$}, and possibly non-Clifford fault-tolerant logical gates). In this section we simply elaborate upon the $t=4$ example above.

\subsection{Metacheck CSS Codes (mCSS Codes)}
We introduce the notion of a \emph{metacheck CSS code (mCSS code)} as a 5-term chain complex that extends the standard CSS code construction.

Recall that a CSS code involves two parity-check matrices $P_Z$ and $P_X$ satisfying $P_X P_Z^{\top} = 0$,
and hence we could rephrase it as a three-term chain complex, as in Eq.~\eqref{eqn: CSS as chain complex}.
Metachecks are additional maps $M_X$ and $M_Z$ that satisfy $M_X P_X = 0$ and $M_Z P_Z = 0$.
We can encode a CSS code with metachecks as a five-term chain complex.

\begin{definition}[mCSS code]
An \emph{mCSS code} is a 5-term chain complex of finite-dimensional vector spaces over $\mathbb{F}_2$:
\begin{align}
C_{-1} &\xleftarrow{\partial_0} C_0 \xleftarrow{\partial_1} C_1 \xleftarrow{\partial_2} C_2 \xleftarrow{\partial_3} C_3
\end{align}
whose inner 3-term complex
\begin{align}
C_0 \xleftarrow{\partial_1} C_1 \xleftarrow{\partial_2} C_2
\end{align}
defines a conventional CSS code with parity-check matrices $P_X = \partial_1$ and $P_Z = \partial_2^\top$, 
and whose remaining differentials determine metacheck matrices $M_X = \partial_0$ and \revisionblue{$M_Z =\partial_3^\top$}.
\end{definition} 

Any sufficiently long chain complex of vector spaces over $\mathbb{F}_2$ contains a 5-term \revisionblue{segment},
which thus provides an mCSS code.
\revisionblue{For $t\geq4$ over a nonzero coefficient ring, the Koszul complex has at least five module terms, so it contains a five-term segment of nonzero modules.}

It can be convenient to ask for the following property.

\begin{definition}[Balanced mCSS Code]
An mCSS code
\begin{align}
C_{-1} &\xleftarrow{M_X} C_0 \xleftarrow{P_X} C_1 \xleftarrow{\revisionblue{P_Z^{\top}}} C_2 \xleftarrow{\revisionblue{M_Z^{\top}}} C_3
\end{align}
is {\em balanced} if $\dim C_{-1} = \dim C_3$ and $\dim C_0 = \dim C_2$.
\end{definition}

Balanced codes are convenient because the $X$‑ and $Z$‑stabilizer generator matrices have identical matrix dimensions up to transposition.

A simple method to obtain a balanced mCSS code is to extract the middle 5-term segment of a Koszul complex as follows.

\begin{proposition}
Let $S$ be a commutative algebra over $\mathbb{F}_2$ such that $s:=\dim_{\mathbb{F}_2}(S)$ is finite.
Let $t = 2q$ \revisionblue{with $q \geq 2$ (equivalently $t \geq 4$)} and $\underline{a} = (a_1, \ldots, a_t)$ be a sequence of elements of $S$.
The middle 5-term segment
\begin{align}
K_{q-2} \leftarrow K_{q-1} \leftarrow K_{q} 
\leftarrow K_{q+1} \leftarrow K_{q+2}  
\end{align}
of the Koszul complex $K_\bullet(\underline{a};S)$ is a balanced mCSS code.
\end{proposition}

More generally, unbalanced mCSS codes can be constructed by selecting any 5-term segment from such a Koszul complex or, more broadly, from any chain complex.
\begin{statement}
\label{statementone}
A useful fact is that for the Koszul complex $K(x_1, \dots, x_t)$, the rank of the free module in homological degree $k$ is
\begin{equation}
\rank K_k(x_1, \dots, x_t) = \binom{t}{k}
\end{equation}
\end{statement}
\subsection{The Multivariate Multicycle (MM) Construction\label{sec:main_construction}}

We now systematically generalize the BB and TT codes by applying the proposition to a special class of commutative algebras~$S$.

\begin{definition}[Multivariate multicycle code]\label{def:mm}
Let $t \geq 2$ be the number of polynomials and \revisionblue{$D \geq 1$} be the number of variables with exponents $\ell_1, \dots, \ell_D \in \mathbb{Z}_{\geq 1}$ defining the ideal $I = \langle x_1^{\ell_1} - 1, \dots, x_D^{\ell_D} - 1 \rangle$.
The polynomial ring $R = \mathbb{F}_2[x_1, \dots, x_D]$ has a quotient ring 
\editcolor{$S = R/I \cong \mathbb{F}_2[\mathbb{Z}_{\ell_1} \times \cdots \times \mathbb{Z}_{\ell_D}]$}\revisionblue{, the group algebra of a finite abelian group}.
Let $\pi: R \rightarrow S$ be the natural projection mapping.
Take the polynomials $f_1, \dots, f_t \in R$, and define their images in $S$ as:
\begin{equation}
F_i = \pi(f_i(x_1,\dots,x_D)) \quad (i = 1, \dots, t).
\end{equation}

Form the Koszul complex \revisionblue{$K_\bullet([F_1, \ldots, F_t]; S)$} over the quotient ring $S$ as follows

\begin{align}
0 \longleftarrow K_0 \xleftarrow{\partial_1} K_1 \xleftarrow{\partial_2} K_2 
\xleftarrow{\partial_3} \cdots \xleftarrow{\partial_t} K_t \longleftarrow 0
\end{align}
where \revisionblue{$K_j = S^{\binom{t}{j}}$}. Let $q = \lfloor t/2 \rfloor$.
The MM code is the CSS code obtained by placing physical qubits on $K_q$ with parity-check matrices
\begin{align}
P_X &= \revisionblue{\partial_q, \quad \partial_q: K_q \longrightarrow K_{q-1}} \\
P_Z &= \revisionblue{\partial_{q+1}^{\!\top}, \quad \partial_{q+1}: K_{q+1} \longrightarrow K_q}
\end{align}

To construct the MM code, replace each polynomial $F_i$ in the boundary maps with \revisionblue{its regular representation: the monomial $x_1^{a_1}\cdots x_D^{a_D}$ becomes the permutation matrix $S_{\ell_1}^{a_1}\otimes\cdots\otimes S_{\ell_D}^{a_D}$, where $S_\ell$ is the $\ell\times\ell$ cyclic shift matrix. For $D=1$ this is the circulant representation of Ref.~\cite[Theorem~2.1]{albuquerque2011hopfalgebraicpropertiescirculant}; for $D\geq2$ it is a multilevel circulant matrix.}
\end{definition}
\revisionblue{In the column convention used throughout, each matrix maps the space on its right to the space on its left, and the spaces and maps of an MM code with $t \geq 4$ are}
\begin{widetext}
{\color{blue}
\begin{equation}
\label{eq:mm_diagram}
\underbrace{K_{q-2}}_{\text{$X$-metachecks}}
\xleftarrow{\;M_X = \partial_{q-1}\;}
\underbrace{K_{q-1}}_{\text{$X$-syndromes}}
\xleftarrow{\;P_X = \partial_{q}\;}
\underbrace{K_{q}}_{\text{qubits}}
\xleftarrow{\;P_Z^{\top} = \partial_{q+1}\;}
\underbrace{K_{q+1}}_{\text{$Z$-syndromes}}
\xleftarrow{\;M_Z^{\top} = \partial_{q+2}\;}
\underbrace{K_{q+2}}_{\text{$Z$-metachecks}} .
\end{equation}
}
\end{widetext}
\revisionblue{A $Z$-type error $e_Z \in K_q$ has $X$-type syndrome $P_X e_Z \in K_{q-1}$, an $X$-type error $e_X \in K_q$ has $Z$-type syndrome $P_Z e_X \in K_{q+1}$, and the metachecks act on these syndrome spaces.}

The number of physical qubits is \editcolor{$n = \dim_{\mathbb{F}_2} K_q = \binom{t}{q} \prod_{i=1}^{D} \ell_i$}. If $t \geq 4$, the code has an X-metacheck matrix
\begin{align}
M_X = \revisionblue{\partial_{q-1}: K_{q-1} \longrightarrow K_{q-2}}
\end{align}
and if $t \geq 3$, it has a Z-metacheck matrix
\begin{align}
M_Z = \revisionblue{\partial_{q+2}^{\!\top}, \quad \partial_{q+2}: K_{q+2} \longrightarrow K_{q+1}}
\end{align}
These satisfy
\begin{align}
M_X P_X = 0, \quad M_Z P_Z = 0
\end{align}
providing \revisionblue{metachecks for both types of checks, that is, complete metacheck structure in the sense of Definition~\ref{def:metacheck_structure}. The identities follow from the chain-complex condition alone; what is specific to MM codes is that both maps are nonzero, as the following remark shows. How much of the redundancy among the checks these maps capture is the subject of Proposition~\ref{prop:mm_homology} in Sec.~\ref{sec:mm_homology}.}

\begingroup\color{blue}\begin{remark}[Metachecks of MM codes]
\label{prop:mm_metachecks}
Let $t \geq 4$, $q = \lfloor t/2 \rfloor$, and let $F_1,\dots,F_t \in S$ be arbitrary. Then the MM code of Definition~\ref{def:mm} has an $X$-metacheck matrix $M_X = \partial_{q-1}$ and a $Z$-metacheck matrix $M_Z = \partial_{q+2}^{\top}$ satisfying $M_X P_X = 0$ and $M_Z P_Z = 0$. Both are nonzero whenever at least one $F_i$ is nonzero. For $t = 3$, the Koszul complex supplies only $M_Z$, and for $t = 2$ it supplies neither adjacent metacheck map, although the checks may still satisfy other relations.
\end{remark}
\begin{proof}
For $t \geq 4$ we have $q \geq 2$ and $q + 2 \leq t$, so $K_{q-2}$ and $K_{q+2}$ are nonzero and the maps $\partial_{q-1}$ and $\partial_{q+2}$ are defined. Then $M_X P_X = \partial_{q-1}\partial_q = 0$ and $M_Z P_Z = \partial_{q+2}^{\top}\partial_{q+1}^{\top} = (\partial_{q+1}\partial_{q+2})^{\top} = 0$ by the chain-complex condition, which holds for every Koszul complex. Each Koszul differential has an entry equal to $F_i$ for every $i$, so it is nonzero if some $F_i \neq 0$; this property is preserved by the circulant representation. For $t = 3$ we have $q = 1$, so $K_{q-2}$ is zero and only $M_Z$ exists; for $t = 2$ both $K_{q-2}$ and $K_{q+2}$ vanish.
\end{proof}\endgroup

For general parameters $t$ and $D$ where $t$ is the number of polynomials and \revisionblue{$D$ is the number of cyclic factors in the presentation of $S$}, 
let \editcolor{$s = \prod_{i=1}^{D} \ell_i = \dim_{\mathbb{F}_2}(S)$} denote the dimension of the quotient ring $S$ 
as a finite-dimensional $\mathbb{F}_2$-vector space, admitting the standard monomial basis 
$\{x_1^{a_1} \cdots x_D^{a_D} : 0 \leq a_i < \ell_i\}$.
Recall from Statement~\ref{statementone} that each space \revisionblue{$K_j$ has dimension $\dim_{\mathbb{F}_2}(K_j) = \binom{t}{j} s$}. The parity-check matrices are as follows:
\begin{align}
P_X &= \revisionblue{\partial_q} \text{\ \ of size\ \ } \binom{t}{q-1} s \times \binom{t}{q} s, \\
P_Z &= \revisionblue{\partial_{q+1}^{\!\top}} \text{\ \ of size\ \ } \binom{t}{q+1} s \times \binom{t}{q} s.
\end{align}
When $t \geq 4$, the corresponding metacheck matrices are
\begin{align}
M_X &= \revisionblue{\partial_{q-1}} \text{\ \ of size\ \ } \binom{t}{q-2} s \times \binom{t}{q-1} s, \\
M_Z &= \revisionblue{\partial_{q+2}^{\!\top}} \text{\ \ of size\ \ } \binom{t}{q+2} s \times \binom{t}{q+1} s.
\end{align}
The CSS orthogonality condition \(P_X P_Z^{\!\top} = 0\) and metacheck conditions \(M_X P_X = 0\), \(M_Z P_Z = 0\) follow from the fact that the Koszul complex over the multivariate polynomial ring satisfies \(\partial_i \partial_{i+1} = 0\) for all \(i\).
\begin{remark}
    The case \editcolor{$D=1$} (a single cyclic group) is equivalent to $D=2$ with $\ell_2=1$ via the isomorphism
$\mathbb{F}_2[x]/\langle x^{\ell_1}-1 \rangle \cong \mathbb{F}_2[x,y]/\langle x^{\ell_1}-1, y-1 \rangle$. \revisionblue{Thus Definition~\ref{def:mm} with $D\geq1$ includes the univariate rings, and they may equally be written with more variables of order one, as in Tables~\ref{tab:table6}--\ref{tab:table11}.}
\end{remark}
\revisionblue{\textit{Collapsed codes.} \revisionblue{The codes in Tables~\ref{tab:table6}--\ref{tab:table11} are labelled $t$-dimensional (4D to 9D) because of their chain complex, not their geometry. Two numbers enter the construction: the number $t$ of defining polynomials, which fixes the length of the complex (degrees $0,\ldots,t$) and hence its metacheck structure, and the number of invariant factors of the group $G$ that indexes the qubits, that is, the smallest number of cyclic groups whose direct product is isomorphic to $G$~\cite[Sec.~5.2]{dummit2004abstract}. For the 4D toric code with four equal side lengths (Remark~\ref{mm:4dtoric}), both numbers equal four. In Tables~\ref{tab:table6}--\ref{tab:table11} the second number is one: the group $G\cong\mathbb{Z}_\ell$ is cyclic and $S\cong\mathbb{F}_2[\mathbb{Z}_\ell]$. These codes therefore have the chain complex and metachecks of a $t$-dimensional code, with qubits indexed by a single cyclic group.}}

\revisionblue{For $D=1$, the stabilizer group of an MM code is a quasi-cyclic code of index $2\binom{t}{q}$; Appendix~\ref{app:quasicyclic} describes this connection and gives an example.}\begin{example}[MM code (\(t=4, D=4\))]
A newly discovered $[[648, 60, (9,9)]]$ MM code: Let $\ell_1=3,\ell_2=3,\ell_3=3,\ell_4=4$ and 
$R = \mathbb{F}_2[w,x,y,z]$ and $S= R/\langle w^3-1,x^3-1,y^3-1,z^4-1\rangle$. Let $\pi: R \rightarrow S$ be the natural projection mapping. Given multivariate polynomials $f(w,x,y,z),g(w,x,y,z), h(w,x,y,z), i(w,x,y,z) \in R$, we define their images in $S$ as $F = \pi(f(w,x,y,z))$, $G = \pi(g(w,x,y,z))$, $H = \pi(h(w,x,y,z))$ and $I = \pi(i(w,x,y,z))$. Consider the polynomials
\begin{align}
&F(w,x,y,z) = (1 + x)(1 + yz) \\
&G(w,x,y,z) = (1 + y)(1 + zw) \\
&H(w,x,y,z) = (1 + z)(1 + wx) \\
&I(w,x,y,z) = (1 + w)(1 + xy)
\end{align}
The associated Koszul complex $K_\bullet([F,G,H,I]; S)$ has dimensions:
\begin{align}
\dim_{\mathbb{F}_2} C_0 &= \tbinom{4}{0}\prod_{i=1}^{4} \ell_i = 108 \\
\dim_{\mathbb{F}_2} C_1 &= \tbinom{4}{1}\prod_{i=1}^{4} \ell_i = 432 \\
\dim_{\mathbb{F}_2} C_2 &= \tbinom{4}{2}\prod_{i=1}^{4} \ell_i = 648 \\
\dim_{\mathbb{F}_2} C_3 &= \tbinom{4}{3}\prod_{i=1}^{4} \ell_i = 432 \\
\dim_{\mathbb{F}_2} C_4 &= \tbinom{4}{4}\prod_{i=1}^{4} \ell_i = 108
\end{align}
The differentials of this complex have the following matrix dimensions:
\revisionblue{$\partial_1 = M_X$ of size $108 \times 432$, $\partial_2 = P_X$ of size $432 \times 648$, $\partial_3 = P_Z^\top$ of size $648 \times 432$, and $\partial_4 = M_Z^\top$ of size $432 \times 108$.}

With $q = \lfloor 4/2 \rfloor = 2$, physical qubits are placed in $C_2$, giving:  \editcolor{$n = \dim C_2 = 648$ and $k = 60$ logical qubits, and distance $d = 9$}. \revisionblue{Verified logical operators of weight $9$ exist in both sectors, and an exhaustive connected-cluster search with \texttt{dist-m4ri} finds none of weight at most $8$, so $d_X = d_Z = 9$ exactly (Appendix~\ref{sec:appendix_search}).} The parity-check matrices are:
\begin{align}
\quad P_X &= \revisionblue{\partial_2, \quad \partial_2: K_2 \to K_1} \\
\quad P_Z &= \revisionblue{\partial_3^{\!\top}, \quad \partial_3: K_3 \to K_2} \quad
\end{align}
Metachecks (both exist for $t=4$\revisionblue{, by Remark~\ref{prop:mm_metachecks}}):
\begin{align}
\quad M_X &= \revisionblue{\partial_1: K_1 \to K_0}
\\
\quad M_Z &= \revisionblue{\partial_4^{\!\top}, \quad \partial_4: K_4 \to K_3}
\end{align}
These satisfy $P_X P_Z^{\!\top}=0$, $M_X P_X=0$, and $M_Z P_Z =0$.
\end{example}
\begin{figure}[tbp]
    \centering
    \begin{tikzpicture}[
        x=1cm,y=1cm,
        family/.style={font=\normalfont\small\bfseries,text=white,align=center}
    ]
        \definecolor{mmA}{HTML}{5E0F20}
        \definecolor{mmB}{HTML}{8A1D36}
        \definecolor{mmC}{HTML}{A93A52}
        \definecolor{mmD}{HTML}{C9677B}
        \path[fill=mmA] (0,0) ellipse (4.15 and 2.75);
        \node[family,font=\normalfont\large\bfseries] at (0,2.25) {MM / AMC};
        \path[fill=mmB] (-1.15,-0.3) ellipse (2.6 and 2.05);
        \node[family] at (-1.15,1.25) {$t=2$: abelian 2BGA};
        \path[fill=mmC] (-1.95,-0.25) ellipse (1.25 and 0.95);
        \node[family] at (-1.95,0.35) {MB};
        \path[fill=mmD] (-1.95,-0.45) ellipse (0.7 and 0.45);
        \node[family] at (-1.95,-0.45) {BB};
        \node[family] at (0.1,0.45) {GB};
        \node[family] at (0.1,-0.35) {C2 / CxR};
        \node[family] at (-1.1,-1.65) {Haah's cubic code};
        \path[fill=mmB] (2.75,0.55) ellipse (1.0 and 0.62);
        \node[family] at (2.75,0.55) {$t=3$: TT};
        \path[fill=mmB] (2.75,-1.0) ellipse (1.0 and 0.62);
        \node[family] at (2.75,-1.0) {$t\geq4$};
    \end{tikzpicture}
   \caption{\revisionblue{\textbf{Subfamilies of multivariate multicycle (MM) codes.} Nested regions denote inclusion. The MM construction is a Koszul-complex reformulation of abelian multicycle (AMC) codes~\cite{lin2025abelianmulticyclecodessingleshot} over a finite abelian group algebra, so the two share the outer region (Proposition~\ref{prop:mm_amc}). Inside it, the codes are grouped by the number $t$ of defining polynomials. For $t=2$ they are the abelian two-block group-algebra (2BGA) codes~\cite{lin2023quantumtwoblockgroupalgebra}, which include multivariate bicycle (MB) codes~\cite{voss2025multivariate} and, among these, bivariate bicycle (BB) codes~\cite{Bravyi_2024}, as well as generalized bicycle (GB) codes, cyclic hypergraph product codes (C2 and CxR)~\cite{aydin2025cyclic}, and Haah's cubic codes on an $L\times L\times L$ lattice~\cite{panteleev2022asymptoticallygoodquantumlocally}. For $t=3$ they are the trivariate tricycle (TT) codes~\cite{jacob2025singleshotdecodingfaulttolerantgates, menon2025magictricyclesefficientmagic}, and for $t\geq4$ they have metachecks for both types of checks.}}
\label{fig:conceptual_diagram_of_subfamilies}
\end{figure}
Here is another example to provide more clarity about the code construction process:
\begin{example}
\begingroup\color{blue}With $S = \mathbb{F}_2[x,y]/\langle x^{21}-1,y^{18}-1\rangle$ and
$F=x^3+y^{10}+y^{17}$, $G=x^{19}+x^3+y^5$, the column-convention Koszul maps are
\begin{align}
\partial_1&=\begin{pmatrix}F&G\end{pmatrix}:S^2\to S,\\
\partial_2&=\begin{pmatrix}G\\F\end{pmatrix}:S\to S^2.
\end{align}
Let $A$ and $B$ be the $378\times378$ binary regular-representation matrices of $F$ and $G$. Then
\begin{align}
P_X&=\widetilde{\partial}_1=\begin{pmatrix}A&B\end{pmatrix},\\
P_Z&=\widetilde{\partial}_2^\top=\begin{pmatrix}B^\top&A^\top\end{pmatrix}.
\end{align}
Both check matrices have size $378\times756$, and $P_XP_Z^\top=AB+BA=0$ because $S$ is commutative and has characteristic two. This gives the $[[756,16,\leq34]]$ code from Table 3 of~\cite{Bravyi_2024}.\endgroup
\end{example}
\subsection{Subfamilies of MM codes}
\label{sec:subfamilies_of_mm_code}
Now that we have introduced a \revisionblue{five-term} complex, we continue the restating of known codes in terms of \revisionblue{the MM construction}. This is a natural extension of Sec.~\ref{sec:boundary_maps_and_code_constructions}, but we wanted to keep it separate, because only now we have codes that admit metachecks for both types of checks. To provide a broader perspective on how our construction fits within the larger family of QEC codes, we refer the reader to Fig.~\ref{fig:conceptual_diagram_of_subfamilies}. \revisionblue{Scripts that construct the examples of this section with \texttt{QuantumClifford.jl} are provided in our repository~\cite{feroz_ahmed_mian_2026_24095975}.}

\begin{remark}[MM code (\(t=4, D=4\))]
\label{mm:4dtoric}
Let $G = \mathbb{Z}_\ell \times \mathbb{Z}_m \times \mathbb{Z}_p \times \mathbb{Z}_r$ with $\ell = m = p = r$, and consider the group algebra $R = \mathbb{F}_2[G] \cong \mathbb{F}_2[w,x,y,z]/\langle w^\ell-1, x^m-1, y^p-1, z^r-1 \rangle$. The MM code defined by the polynomials \revisionblue{$F_1 = 1+w$, $F_2 = 1+x$, $F_3 = 1+y$, $F_4 = 1+z$} in this quotient ring is equivalent to the 4D toric code on an $\ell \times m \times p \times r$ lattice.
\end{remark}
\begin{remark}[MM code (\(t=4, D=1\))]
\label{mm:amc}
The AMC codes of Ref.~\cite{lin2025abelianmulticyclecodessingleshot} are formulated in terms of the group algebra $\mathbb{F}_2[G]$ of a finite abelian group $G$. This approach provides significant generality, since any \revisionblue{finite} abelian group $G$ can be used. All explicit examples in~\cite{lin2025abelianmulticyclecodessingleshot} use single cyclic groups $\mathbb{Z}_\ell$ with corresponding single-variable polynomials $a_i(x) \in \mathbb{F}_2[x]/(x^\ell-1) \cong \mathbb{F}_2[\mathbb{Z}_\ell]$. \revisionblue{These are MM codes with} $G = \mathbb{Z}_\ell \times \mathbb{Z}_m \times \mathbb{Z}_p \times \mathbb{Z}_r$ with $m=p=r=1$. The group algebra $\mathbb{F}_2[G]$ is then isomorphic to the quotient ring $\mathbb{F}_2[w,x,y,z]/\langle w^\ell-1, x^m-1, y^p-1, z^r-1\rangle$, which reduces to $\mathbb{F}_2[\revisionblue{\mathbb{Z}_\ell}]$ when $m=p=r=1$. \revisionblue{More generally, every finite abelian group is isomorphic to a direct product of cyclic groups~\cite[Sec.~5.2]{dummit2004abstract}, so every AMC code is an MM code for a suitable choice of the orders $\ell_i$. These orders are not unique: two choices for the same group differ only by factors of order one and by splitting or merging factors of coprime order, since $\mathbb{Z}_m\times\mathbb{Z}_n\cong\mathbb{Z}_{mn}$ exactly when $\gcd(m,n)=1$. For example, $\mathbb{Z}_6\cong\mathbb{Z}_2\times\mathbb{Z}_3$, whereas $\mathbb{Z}_4\not\cong\mathbb{Z}_2\times\mathbb{Z}_2$. Conversely, every MM code is an AMC code, as we now show.}
\end{remark}

\begingroup\color{blue}\begin{proposition}[MM codes are AMC codes]
\label{prop:mm_amc}
Let $G = \mathbb{Z}_{\ell_1}\times\cdots\times\mathbb{Z}_{\ell_D}$, so that $S = \mathbb{F}_2[x_1,\dots,x_D]/\langle x_1^{\ell_1}-1,\dots,x_D^{\ell_D}-1\rangle \cong \mathbb{F}_2[G]$, and let $a_1,\dots,a_t \in S$. The Koszul complex $K_\bullet([a_1,\dots,a_t];S)$, with each element replaced by its regular representation, coincides with the multi-block complex $\mathrm{MBC}(\mathrm{M}(a_1),\dots,\mathrm{M}(a_t))$ of Ref.~\cite{lin2025abelianmulticyclecodessingleshot}, up to the ordering of the blocks in each degree. Consequently, the MM code and the AMC code defined by $a_1,\dots,a_t$ coincide up to the ordering of qubits and checks. For $t = 2$, the code is the scalar lifted product $\mathrm{LP}(a_1,a_2)$ of Ref.~\cite{panteleev2022almostlinear}, with $P_X = [\,a_1 \;\; a_2\,]$ and $P_Z = [\,a_2^{\top} \;\; a_1^{\top}\,]$.
\end{proposition}
The proof is given in Appendix~\ref{app:mm_homology}.
The equivalence identifies the Koszul description as a reformulation of AMC codes, with closed-form boundary maps supplied by the exterior algebra. More generally, $K_\bullet([a_1,\ldots,a_t];S)\cong\bigotimes_{i=1}^{t}K(a_i)$ is the tensor product over $S$ of $t$ scalar two-term complexes, that is, a $t$-fold lifted product of the $1\times 1$ matrices $a_1,\dots,a_t$ over the finite abelian group algebra $S$, obtained by iterating the two-factor definition of Ref.~\cite{panteleev2022almostlinear} (Sec.~\ref{sec:balanced_product_of_complexes}). An MM code is the CSS code of this complex with qubits in degree $q=\lfloor t/2\rfloor$; Definition~\ref{def:mm} selects this degree, although the same complex also defines CSS codes in other degrees. For $t = 2$ this is the lifted product $\mathrm{LP}(a_1,a_2)$, and for $t=3$ it is a three-dimensional lifted product with scalar constituent matrices~\cite{li2025transversaldimensionjump}.\endgroup

\begin{remark}
\label{mm:bb}
\revisionblue{Let $R=\mathbb F_2[G]$ with $G=\mathbb Z_{\ell_1}\times\cdots\times\mathbb Z_{\ell_D}$.} For \editcolor{$t=2, D=2$}, our framework recovers the BB codes~\cite{Bravyi_2024}; for \editcolor{$t=3, D=3$}, the TT codes~\cite{jacob2025singleshotdecodingfaulttolerantgates, menon2025magictricyclesefficientmagic}; \revisionblue{and for $t\geq4$, higher-dimensional MM/AMC instances.}
\end{remark}

\begin{remark}[MM code (\(t=2, D=2\))]
\label{mm:c2cxr}
The symmetric cyclic hypergraph product C2 code~\cite{aydin2025cyclic} with parameters 
$[[2n_c^{2}, 2k_c^2, d_c]]$, and check weight $2w$ is a 
special case of MM codes. It is obtained from two copies of a 
classical $[n_c,k_c,d_c]$ cyclic code with weight-$w$ polynomial $p(x)$ in the 
quotient ring $\mathbb{F}_2[x,y]/\langle x^{n_c}-1, y^{n_c}-1\rangle$. For example, 
the weight-6 $[[450,32,8]]$ C2 code from Table I of Ref.~\cite{aydin2025cyclic} can be obtained from univariate polynomials $p(x)=1+x+x^4$,  $p(y)=1+y+y^4$ in 
$\mathbb{F}_2[x,y]/\langle x^{15}-1,y^{15}-1\rangle$.

Similarly, the repeated cyclic hypergraph product CxR codes, introduced in Ref.~\cite{PRXQuantum.5.040328} are also special cases of MM codes. With $b = d_c$, the CxR code has parameters \revisionblue{$[[2n_c d_c, 2k_c, d_c]]$} and check weights $\omega=w+2$~\cite{aydin2025cyclic}. It is constructed  using the classical code polynomial $p(x)$ of weight $w$ and the repetition code $1+y$ in $\mathbb{F}_2[x,y]/\langle x^{n_c}-1, y^{d_c}-1\rangle$. For example, the weight-5 $[[240,8,8]]$ CxR code from Table I of Ref.~\cite{aydin2025cyclic} is obtained by using $p(x)=1+x+x^4$ and $p(y) = 1 + y$ in $\mathbb{F}_2[x,y]/\langle x^{15}-1,y^{8}-1\rangle$.
\end{remark}

\begin{remark}[MM code (\(t=2, D=2\))]
\label{mm:lacross}
The $[[2n^2, 2k^2, d]]$ La-Cross codes~\cite{pecorari2025high} with periodic boundary conditions are a subfamily of MM codes with $t=2$ and $G = \mathbb{Z}_n \times \mathbb{Z}_n$. 
For example, given a seed polynomial $h(a)=1+a+a^k$, setting $A = h(x)$ and $B = h(y)$ in the quotient ring $\mathbb{F}_2[x,y]/\langle x^n-1, y^n-1\rangle$ recovers the $[[98,18,4]]$ code with $n=7$ and $k=3$~\cite{pecorari2025high}.
\end{remark}

\begin{remark}[MM code (\(t=2, D=3\))]
\label{mm:haah-mm}
Haah's cubic code~\cite{haah2011local} on an $L \times L \times L$ lattice corresponds to the case $t=2$, $D=3$ in our MM code framework. Taking the group $G = \mathbb{Z}_L \times \mathbb{Z}_L \times \mathbb{Z}_L$ and polynomials
$A(x, y, z) = 1 + x + y + z$ , $B(x, y, z) = 1 + xy + xz + yz$ in $\mathbb{F}_2[x,y,z]/\langle x^L-1, y^L-1, z^L-1\rangle$ our MM code reproduces Haah's cubic code. For example, with $L=8$ one obtains the weight 8-limited $[[1024, 30, d]]$ code ($13 \le \editcolor {d} \le 32$) studied in Appendix B of Ref.~\cite{panteleev2022asymptoticallygoodquantumlocally}. \revisionblue{Our construction reproduces $n=1024$ and $k=30$, and an exhaustive connected-cluster search with \texttt{dist-m4ri} finds no nontrivial logical operator of weight at most $8$ in either sector, consistent with these bounds (Appendix~\ref{sec:appendix_search}).}
\end{remark}

\begin{remark}[MM code (\(t=2, D=2\))]
\label{mm:gb}
The $[[30, 8, 4]]$ GB code from Ref.~\cite{liang2025generalized} is recovered as a special case of the MM construction with shift patterns $a_{\text{shifts}} = [0, 2, 8]$ and $b_{\text{shifts}} = [0, 1, 4]$ with $\ell = 15$. Moreover, all one-dimensional generalized bicycle codes from Tables~V through~VIII of Ref.~\cite{liang2025generalized} are systematically recoverable within this framework.
\end{remark}
\begin{remark}[MM code (\(t=2, D=2\))]
\label{mm:2bga}
MM codes contain abelian 2BGA codes as a special case. When the group is a direct product of cyclic groups $G = \mathbb{Z}_m \times \mathbb{Z}_2$, the corresponding MM code construction uses the quotient ring $\mathbb{F}_2[s,x]/\langle s^2-1, x^m-1\rangle$, which is isomorphic to the group algebra $\mathbb{F}_2[\mathbb{Z}_m \times \mathbb{Z}_2]$. For instance, the $[[16, 2, 4 ]]$ code from Table~2 of Ref.~\cite{lin2023quantumtwoblockgroupalgebra} is generated by: $A(s,x) = 1 + x$, $B(s,x) = 1 + x + s + x^2 + s x + s x^3$ using the group algebra of the direct product of cyclic groups $\mathbb{Z}_4 \times \mathbb{Z}_2$.
\end{remark}

\begin{remark}[MM code (\(t=2, D=2\))]
\label{mm:mb}
\revisionblue{The MB codes of Ref.~\cite{voss2025multivariate} are defined over $\mathbb{F}_2[\mathbb{Z}_l\times\mathbb{Z}_m]$, with additional variables that are monomials in the two generators, and are therefore MM codes with $t=2$ and $D=2$.} The weight-6 $[[48, 4, 6]]$ TB code from Table 2 of~\cite{voss2025multivariate} is realized as MM code with generating polynomials $A(x,y,z) =  x^3 + y^5$ and $B(x,y,z) = x + y^2 + y^5 + z^5$ with additional variable via $z = xy$ over $S = \mathbb{F}_2[x, y] / \langle x^4 - 1, y^6 - 1 \rangle$. Given this, all TB codes in~\cite{voss2025multivariate} can be realized as MM code by choosing two integers $l$, $m$ and two generating polynomials $A$ and $B$ in $\mathbb{F}_2[x, y] / \langle x^l - 1,\; y^m - 1 \rangle$ where the additional variable $z=xy$ fully defines TB code.
\end{remark}

The MM construction is inherently commutative, built from polynomial quotient rings isomorphic to abelian group algebras. Non-abelian 2BGA codes require different techniques such as the lifted product~\cite{panteleev2022asymptoticallygoodquantumlocally} and balanced product~\cite{Breuckmann_2021}. The former construction employs left and right regular representations of the non-commutative algebra, leveraging the associativity property to ensure the required orthogonality condition for CSS codes. In the contemporary literature~\cite{salescabrera2025codesalgebrasdirectproducts, salescabrera2025dualitiesdihedralgeneralisedquaternion, willenborg2025dihedralquantumcodes}, quantum codes from non-commutative group algebras are constructed by exploiting the Wedderburn–Artin decomposition of the semisimple group algebra into matrix rings over finite fields.

\subsection{\texorpdfstring{\revisionblue{Dimension and distance in the univariate case}}{Dimension and distance in the univariate case}}
\label{sec:univariate_distance}
\begingroup\color{blue}
In the simplest setting, where $G$ is cyclic and $S\cong R_L=\mathbb F_2[x]/\langle x^L-1\rangle$, much of an MM code's parameters can be computed by hand: the number of logical qubits always, and the distance for a separable subfamily. All codes in Tables~\ref{tab:table6}--\ref{tab:table11} are of this type: they use the orders $(\ell_1,\ldots,\ell_D)=(L,1,\ldots,1)$, so that $x_i=1$ in $S$ for $i\geq2$ and $S\cong\mathbb F_2[x]/\langle x^{L}-1\rangle$. More generally, $S$ has a single cyclic factor, with $L=\prod_i\ell_i$, whenever the orders $\ell_i>1$ are pairwise coprime.

We begin with the number of logical qubits. In the univariate case the MM construction produces exactly the codes that Lin \emph{et al.} call abelian multi-cycle codes (Proposition~\ref{prop:mm_amc}), so their results apply to our codes directly.\footnote{Reference~\cite{lin2025abelianmulticyclecodessingleshot} writes $D$ for the number of defining polynomials; we write $t$.} In particular~\cite[Statement~2]{lin2025abelianmulticyclecodessingleshot}, for both odd and even $L$,
\[
 k_q=\binom tq\deg g,\qquad g=\gcd(F_1,\ldots,F_t,x^L-1),
\]
so $k_q$ is determined by the degree of their common divisor with $x^L-1$, including multiplicities when $L$ is even.

The distance is harder to obtain. In general only an upper bound is known~\cite[Statement~5]{lin2025abelianmulticyclecodessingleshot}: $d_q\leq d(C_g)$, the minimum distance of the binary cyclic code $C_g=\langle (x^L-1)/g\rangle$, for $1\leq q<t$. The algebra that gives $k_q$ exactly works by changing basis, which mixes the qubits, so it says nothing about the minimum weight of a logical operator.

There is, however, a subfamily for which the distance can be computed exactly: those where $L$ factors into pairwise coprime parts and each polynomial depends on only one of these parts. Lin et al.~\cite[Statement~4]{lin2025abelianmulticyclecodessingleshot} give the related construction with identical constituent factors; here we allow different factors. The code is then a product of $t$ small cyclic codes, and its distance is a product of their distances, by the tensor-product formula of Ref.~\cite[Theorem~17]{zeng2020minimal}.

\begin{proposition}[Separable univariate subfamily]
\label{prop:univariate_separable}
Let $t\geq2$, let $\ell_1,\ldots,\ell_t\geq2$ be pairwise coprime, and put $L=\prod_i\ell_i$. For polynomials $a_i(y)\in\mathbb F_2[y]$, define
\begin{align*}
 F_i(x)&=a_i(x^{L/\ell_i})\in R_L,\\
 g_i(y)&=\gcd\bigl(a_i(y),\,y^{\ell_i}-1\bigr),
\end{align*}
assume $\deg g_i>0$ for every $i$, and let $\delta_i$ be the minimum distance of the cyclic code
$C_i=\langle (y^{\ell_i}-1)/g_i(y)\rangle\subset\mathbb F_2[y]/\langle y^{\ell_i}-1\rangle$.
Order these distances as $\delta_{(1)}\leq\delta_{(2)}\leq\cdots\leq\delta_{(t)}$. With qubits in degree $1\leq q<t$ and checks $P_X=\partial_q$, $P_Z=\partial_{q+1}^{\top}$, the CSS code has
\begin{align*}
 n&=\binom tq L,& k&=\binom tq\prod_{i=1}^t\deg g_i,\\
 d_Z&=\prod_{i=1}^{q}\delta_{(i)},& d_X&=\prod_{i=1}^{t-q}\delta_{(i)}.
\end{align*}
That is, $d_Z$ is the product of the $q$ smallest $\delta_i$, and $d_X$ is the product of the $t-q$ smallest.
\end{proposition}
The proof is given in Appendix~\ref{app:mm_homology}.

For example, $(\ell_1,\ell_2,\ell_3,\ell_4)=(2,3,5,7)$ with $a_i(y)=1+y$ gives $L=210$, $(F_1,\ldots,F_4)=(1+x^{105},1+x^{70},1+x^{42},1+x^{30})$ and $\delta_i=\ell_i$. At $q=2$ this is the $[[1260,6,(6,6)]]$ four-dimensional toric code with side lengths $2,3,5,7$, written over a single cyclic ring. The codes found in our search do not in general have this separable form, which is why their distances are computed numerically.
\endgroup

\subsection{\texorpdfstring{\revisionblue{Homology and $X$--$Z$ duality of MM codes}}{Homology and X-Z duality of MM codes}}
\label{sec:mm_homology}
\begingroup\color{blue}
Two structural facts hold for every MM code, with no assumption on the orders $\ell_i$. The first relates the two sectors of the code, and the second describes the homology of the Koszul complex, which counts both the logical qubits and the check relations that the metachecks miss. Both follow from standard properties of Koszul complexes~\cite[Sec.~1.6]{bruns1998cohen}; we give self-contained proofs in Appendix~\ref{app:mm_homology}. For a qubit degree $q$ we write $P_X^{(q)}=\partial_q$ and $P_Z^{(q)}=\partial_{q+1}^{\top}$, so that the MM code of Definition~\ref{def:mm} uses $q=\lfloor t/2\rfloor$.

\begin{proposition}[$X$--$Z$ duality]
\label{prop:xz_duality}
Let $S=\mathbb F_2[G]$ with $G$ a finite abelian group, and let $F_1,\ldots,F_t\in S$. Let $\sigma$ be the permutation of the group-element basis of the Koszul complex that sends $g\,e_J$ to $g^{-1}e_{J^c}$, where $J^c$ is the complement of $J$ in $\{1,\ldots,t\}$; it maps $K_j$ onto $K_{t-j}$. Then, as binary matrices,
\[
 \partial_{j+1}^{\top}=\sigma\,\partial_{t-j}\,\sigma\qquad(0\leq j\leq t-1).
\]
In particular $P_Z^{(q)}=\sigma P_X^{(t-q)}\sigma$ for $1\leq q\leq t-1$. For even $t$ and $q=t/2$, the $X$- and $Z$-type codes of the MM code are equivalent up to a permutation of qubits and checks, so $d_X=d_Z$, and the metachecks satisfy $M_Z=\sigma M_X\sigma$. Moreover $\dim H_j=\dim H_{t-j}$ for every $j$, and for even $t$ the number of logical qubits $k=\dim H_{t/2}$ is even.
\end{proposition}

For the $t=4$ codes in Tables~\ref{tab:table2}--\ref{tab:table6} this has three consequences. First, a verified logical operator in one sector gives one of the same weight in the other, so we list the smaller of the two verified upper bounds for both sectors. Second, the true syndrome profiles of the two sectors coincide, which we use in Table~\ref{tab:confinement}. Third, the $X$- and $Z$-basis memory experiments of such a code are equivalent whenever the noise model, measurement schedule, and decoder respect the relabelling of qubits and checks. For odd $t$, as in Tables~\ref{tab:table7}, \ref{tab:table9} and~\ref{tab:table11}, the proposition relates the $Z$-type code in degree $q$ to the $X$-type code in degree $q+1$, so the two sectors of the MM code describe cycles of different degrees and need not agree. In the separable family of Proposition~\ref{prop:univariate_separable}, for example, $d_Z$ is a product of $q$ factor distances and $d_X$ a product of $q+1$, so that $d_X\geq d_Z$; for general codes with odd $t$, the duality imposes no such ordering.

\begin{proposition}[Koszul homology of MM codes]
\label{prop:mm_homology}
Let $S=\mathbb F_2[x_1,\ldots,x_D]/\langle x_1^{\ell_1}-1,\ldots,x_D^{\ell_D}-1\rangle$, let $F_1,\ldots,F_t\in S$, let $I=(F_1,\ldots,F_t)$ be the ideal they generate, and write $h_j=\dim_{\mathbb F_2}H_j$ for the homology of $K_\bullet([F_1,\ldots,F_t];S)$.
\begin{enumerate}
 \item[(i)] If $I=S$, the Koszul complex is exact, so $k=0$ in every degree.
 \item[(ii)] If $I\neq S$, then $h_j>0$ for every $0\leq j\leq t$.
 \item[(iii)] $I\neq S$ if and only if the $F_i$ have a common zero $(\zeta_1,\ldots,\zeta_D)$ with $\zeta_i\in\overline{\mathbb F}_2$ and $\zeta_i^{\ell'_i}=1$, where $\ell'_i$ is the odd part of $\ell_i$. Hence $k>0$ exactly when such a common zero exists.
 \item[(iv)] With qubits in an interior degree $1\leq q\leq t-1$, the $X$-type checks have $m_X-\operatorname{rank}P_X$ independent relations, of which the rows of $M_X$ span $\operatorname{rank}M_X$, and similarly for the $Z$-type checks. The numbers of relations not spanned by the metachecks are
 \begin{align}
  (m_X-\operatorname{rank}P_X)-\operatorname{rank}M_X&=h_{q-1},\\
  (m_Z-\operatorname{rank}P_Z)-\operatorname{rank}M_Z&=h_{q+1},
 \end{align}
 where $\operatorname{rank}M_X=0$ or $\operatorname{rank}M_Z=0$ if the complex supplies no such metacheck. Both numbers are positive whenever $k>0$, and then the single-shot distances $d_{ss}^X$ and $d_{ss}^Z$ are finite.
 \item[(v)] If at most one $\ell_i$ is even, then $h_j=\binom tj\dim_{\mathbb F_2}S/I$ for every $j$, so $k=\binom tq\dim_{\mathbb F_2}S/I$. If every $\ell_i$ is odd, $\dim_{\mathbb F_2}S/I$ is moreover the number of common zeros of the $F_i$ in $\mu_{\ell_1}\times\cdots\times\mu_{\ell_D}$, where $\mu_\ell$ denotes the $\ell$th roots of unity in $\overline{\mathbb F}_2$.
\end{enumerate}
\end{proposition}

Part (iv) quantifies the redundancy that the metachecks capture. For the $[[144,12,12]]$ code of Fig.~\ref{fig:single_shot}, the boundary ranks are $22,66,66,22$, so $h=(2,8,12,8,2)$. Its $X$-type checks have $96-66=30$ independent relations, of which the rows of $M_X$ span $22$; the remaining $8$ correspond to the $8$ dimensions of $H_1$, and by Proposition~\ref{prop:xz_duality} the $Z$-type checks behave in the same way. In general, whenever $k>0$ the metachecks of an MM code capture only part of the redundancy of its checks. This matters for decoders that first repair the syndrome: a repaired syndrome that satisfies all metachecks can still lie outside the image of the parity-check matrix.

Part (iii) gives a test for $k>0$ that needs no rank computation. When every $\ell_i$ is a power of two, the only admissible point is $\zeta=(1,\ldots,1)$ and $F_i(1,\ldots,1)$ is the parity of the weight of $F_i$, so $k>0$ if and only if every $F_i$ has even weight. For cyclic $G$, $\dim_{\mathbb F_2}S/I=\deg\gcd(F_1,\ldots,F_t,x^L-1)$, so part (v) recovers the formula of Statement~2 of Ref.~\cite{lin2025abelianmulticyclecodessingleshot} quoted in Sec.~\ref{sec:univariate_distance}, for odd and even $L$. Part (v) extends that formula to several cyclic factors, provided at most one order is even, and for odd orders it extends Statement~3 of Ref.~\cite{lin2025abelianmulticyclecodessingleshot} to the counting of common zeros. With two or more even orders the formula can fail: the $[[96,44]]$ code of Table~\ref{tab:table4}, with $\ell=(2,2,2,2)$, has $\dim S/I=7$ but $k=44\neq\binom42\cdot7$.
\endgroup

\section{Numerical Simulations\label{sec:numerical_simulations}}
The implementation of our MM code was carried out using the OSCAR computer algebra system~\cite{OSCAR, OSCAR-book} and integrated into the open source library QuantumClifford.jl~\cite{quantumclifford, stefan_krastanov_2025_17932783,kimlee2025quantumsavorywritesymbolicallyrun}. All explicit code instances discovered in this work are made available for direct use in the QuantumClifford.jl toolkit, which we hope will be valuable for future work in quantum error correction. Explicit matrix files are also provided in a git repository~\cite{feroz_ahmed_mian_2026_24095975}. \revisionblue{It contains the construction scripts that regenerate $P_X$, $P_Z$, $M_X$, and $M_Z$ and check the chain-complex identities, the search outputs, and the verified logical operators supporting Tables~\ref{tab:table2}--\ref{tab:table11}.} Our implementation \revisionblue{uses} OSCAR's comprehensive capabilities for computational group theory and homological algebra.

\subsection{Code-capacity Model}

\begin{figure}[tbp]
    \centering
    \includegraphics[width=8.5cm]{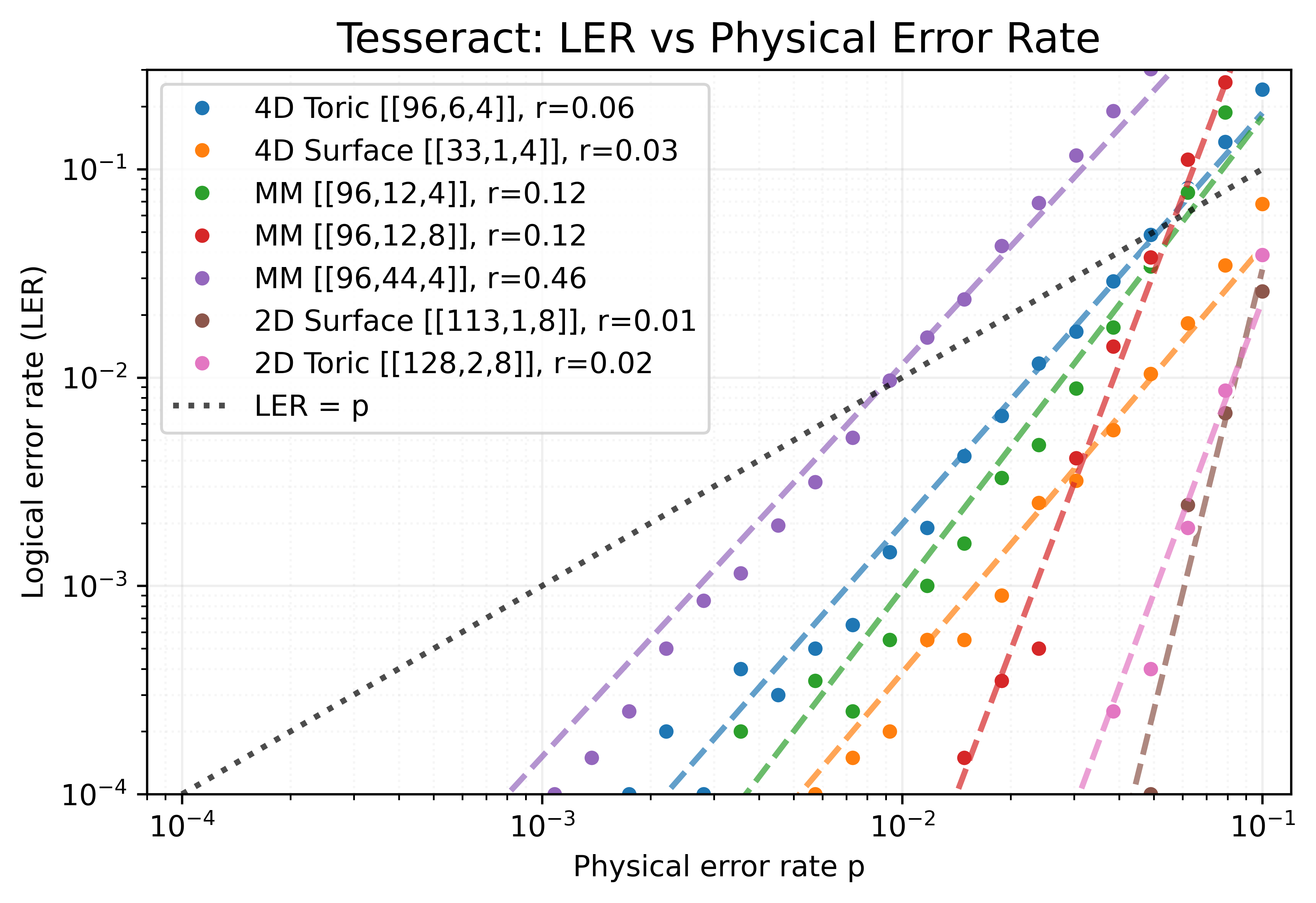}
    \caption{\textbf{Tesseract decoder performance across QECCs.} Including \revisionblue{the} MM codes \revisionblue{$[[96,12,\leq 4]]$}, \revisionblue{$[[96,12,\leq 8]]$}, $[[96,44,4]]$, single-shot decodable 4D topological codes, and 2D topological codes. The legend includes the encoding rate $r = k/n$. \revisionblue{Our codes show high rates, including an option with a rate close to one half at the same distance as the 4D toric code, at the cost of a lower pseudo-threshold.}} \label{fig:tesseract_ler_vs_p}
\end{figure}

We evaluated several MM codes under code-capacity depolarizing noise using the Tesseract search-based decoder~\cite{beni2025tesseractsearchbaseddecoderquantum}. For each code, we constructed a Stim~\cite{gidney2021stim} circuit implementing the code-capacity noise model and compiled it to a detector error model~\cite{Derks_2025} (DEM) file. In the code-capacity noise model, \revisionblue{each data qubit independently undergoes a single-qubit depolarizing channel of strength $p$ (applied with Stim's \texttt{DEPOLARIZE1}$(p)$: no error with probability $1-p$, and $X$, $Y$ or $Z$ each with probability $p/3$)} during a single error-correction step, while all stabilizer measurements are assumed to be perfect. This simplified model isolates the effect of qubit errors on the logical failure rate without introducing measurement noise. The decoder was configured with beam search parameters---a priority queue limit of 10,000 and a detector beam width of 5---and used 10 detector orderings generated from the DEM \revisionblue{by the \texttt{DetIndex} method, with repeated detector visits disabled}. \revisionblue{Further details are given in Appendix~\ref{section:single_shot_decoders}.}

Our MM codes of length $n = 96$—specifically \revisionblue{$[[96,12,\leq 4]]$}, \revisionblue{$[[96,12,\leq 8]]$}, and the higher-rate $[[96,44,4]]$—were compared against topological codes with similar parameters: a $[[96,6,4]]$ 4D toric code and a $[[33,1,4]]$ 4D surface code (both of which are rare examples of existing codes admitting metachecks \revisionblue{for both types of checks}), as well as a few 2D codes. \revisionblue{The two $[[96,12]]$ codes are labeled by the distances used to set up the simulation; the corresponding instances in Table~\ref{tab:confinement} have exact distances $4$ and $8$.} The results are shown in Fig.~\ref{fig:tesseract_ler_vs_p}. Compared to the single-shot-decodable 4D $[[96,6,4]]$ toric code with \revisionblue{the} same number of physical qubits, \revisionblue{all three MM codes have higher rates, and the $[[96,12,8]]$ code also has a higher distance}. Remarkably, the $[[96,44,4]]$ MM code provides a very high rate at the same distance (only compromising on a lower pseudo-threshold). Compared to the $[[33,1,4]]$ single-shot-decodable 4D surface code, \revisionblue{the} $[[96,12,8]]$ MM code achieves a factor of four increase in rate and \revisionblue{a} factor of two increase in distance.
\subsection{Phenomenological noise model}
\label{subsec:phenomenologicalnoisemodel}

We also test against the phenomenological noise model with $T = 2d$ rounds of syndrome extraction, where $d = \min\{d_X, d_Z\}$ is the code distance. This model is intentionally idealized to isolate the combined effect of data qubit errors and measurement errors across multiple rounds, without modeling gate-level operations explicitly; a full circuit-level analysis is presented in \revisionblue{the} next section.

\begin{figure}[tbp]
    \centering
    \includegraphics[width=1\linewidth]{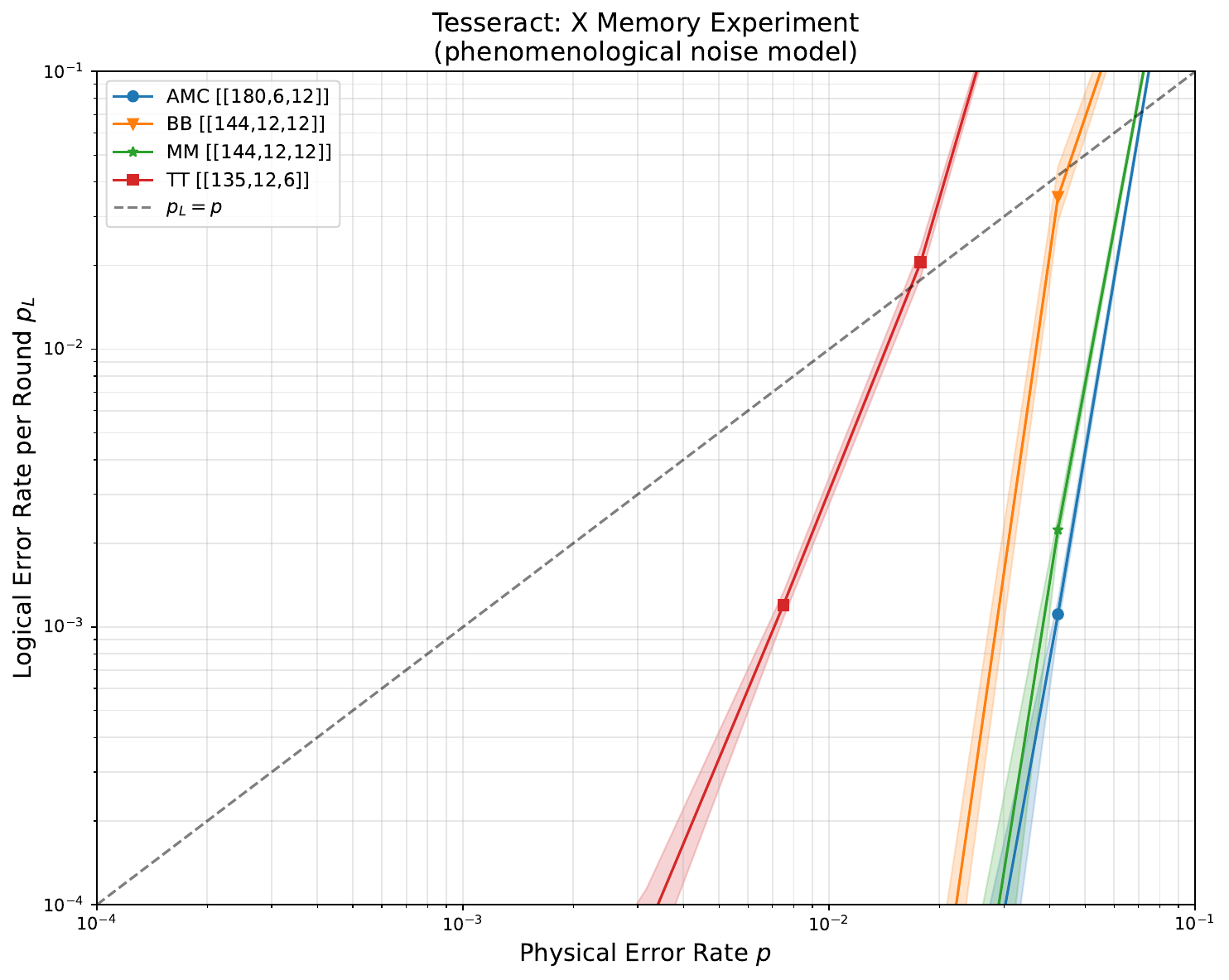}
    \caption{\textbf{Performance comparison of near-$n$ codes under phenomenological noise in the $X$-basis memory experiment with the \texttt{TesseractSinterDecoder}.} Four codes are compared: the symmetric MM $[[144,12,(12,12)]]$ code with weight-9 stabilizers checks; the BB $[[144,12,(12,12)]]$ code with weight-6 stabilizers; the AMC $[[180,6,(12,12)]]$ code with weight-6 stabilizers; and the asymmetric TT $[[135,12,(14,6)]]$ code with weight-9 $X$-type and weight-6 $Z$-type stabilizers. In the $X$-basis memory experiment, the logical $X$ observable is tracked and the circuit is vulnerable to $Z$ errors. The TT code has $d_Z = 6$ in this experiment, which is considerably smaller than the $d_Z = 12$ of the MM and BB codes, placing it at a disadvantage at this block length.}
    \label{fig:phenomenological_model_xbasis}
\end{figure}

\begin{figure}[tbp]
    \centering
    \includegraphics[width=1\linewidth]{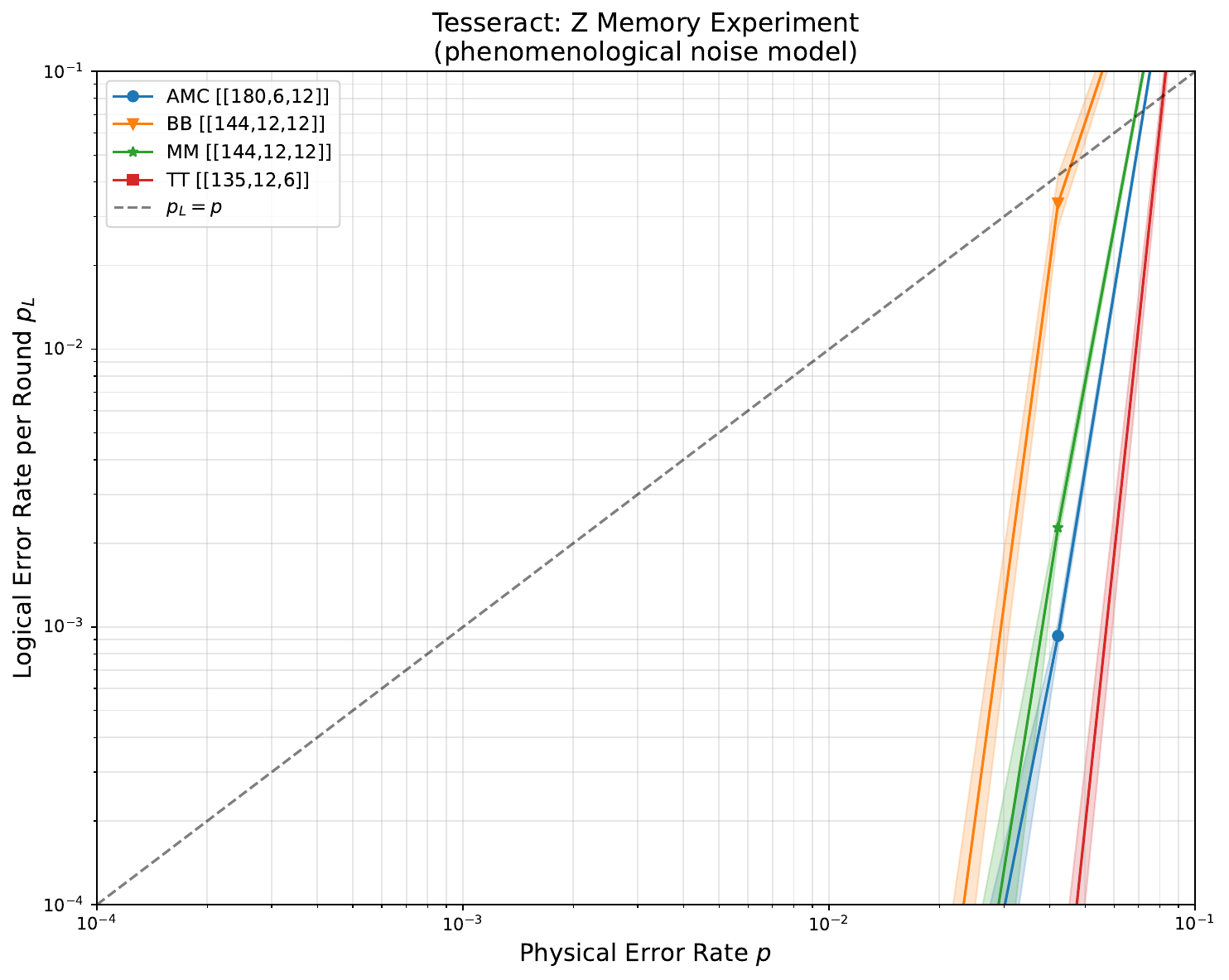}
    \caption{\textbf{Performance comparison of near-$n$ codes under phenomenological noise in the $Z$-basis memory experiment with the \texttt{TesseractSinterDecoder}.} The same codes as in Fig.~\ref{fig:phenomenological_model_xbasis}. The TT code has $d_X = 14$ in this experiment, which is larger than the $d_X = 12$ of the MM and BB codes, placing it at an advantage at this block length.}
    \label{fig:phenomenological_model_zbasis}
\end{figure}

Under the phenomenological noise model, data qubits are subjected to independent single-qubit depolarizing errors with probability $p$ at the start of each round, and each stabilizer measurement outcome is independently flipped with probability $p$. For $Z$-basis memory experiments, logical qubits are initialized in $\ket{0}$ and measured in the $Z$ basis at the end of the circuit (same, but in the $X$ basis, for $X$-basis memory experiments). Memory circuits were generated using the \texttt{make\_css\_code\_memory\_circuit} method from the LDPC package~\cite{roffe_decoding_2020} and sampled using Sinter~\cite{gidney2021stim} with the \texttt{TesseractSinterDecoder}. \revisionblue{A shot fails if the decoder mispredicts any logical observable, and we report the logical error rate per round, $p_L$, defined in Eq.~\eqref{eq:per_round} of Appendix~\ref{section:single_shot_decoders}, which also gives the remaining simulation details.} Fig.~\ref{fig:phenomenological_model_xbasis} and Fig.~\ref{fig:phenomenological_model_zbasis} present the results, showing \revisionblue{competitive behavior for our MM code instance: it has a lower logical error rate than the BB code with the same $n$ and $k$, and performs similarly to the AMC $[[180,6,12]]$ code, which encodes half as many logical qubits. The TT code is the best of the four in the $Z$-basis experiment and the worst in the $X$-basis experiment, as expected from its asymmetric distances.}

\subsection{\texorpdfstring{\revisionblue{Single-shot decoding with metachecks}}{Single-shot decoding with metachecks}}\label{subsec:singleshot}
\begin{figure*}[t]
    \centering
    \includegraphics[width=0.95\linewidth]{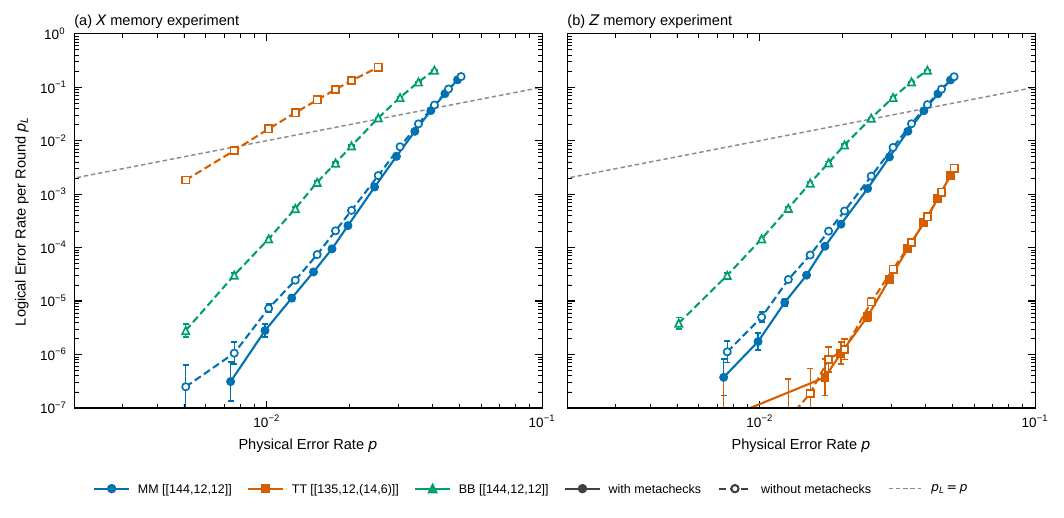}
    \caption{\revisionblue{\textbf{Single-shot decoding with metachecks under phenomenological noise.} Logical error rate per round $p_L$ after $N=8$ noisy rounds, each decoded on its own, in the $X$ (a) and $Z$ (b) memory experiments. Solid lines and filled markers: single-stage BP+OSD with the metacheck rows, Eq.~\eqref{eq:single_stage}; dashed lines and open markers: the same decoder without them. The adjacent Koszul maps supply metachecks for both types of checks of the MM $[[144,12,(12,12)]]$ code, only for the checks used in (b) for the TT $[[135,12,(14,6)]]$ code, and for neither type for the BB $[[144,12,12]]$ code. Error bars are Wilson $95\%$ intervals; points without failures and points with $P_L \geq 0.9$ over the whole run are not drawn.}}
    \label{fig:single_shot}
\end{figure*}
\begingroup\color{blue}We test single-shot decoding by correcting each round separately with the single-stage decoder of Higgott and Breuckmann~\cite{Higgott_2023}. In each of $N$ noisy rounds, every data qubit suffers an $X$ (for the $Z$ memory experiment) or $Z$ (for the $X$ memory experiment) error with probability $p$, and every bit of the measured syndrome $\mathbf{s}$ is flipped with probability $p$. After each round, the decoder receives only $\mathbf{s}$ and solves
\begin{equation}
\begin{pmatrix} H & I \\ 0 & M \end{pmatrix}\begin{pmatrix}\mathbf{e}\\ \mathbf{m}\end{pmatrix} = \begin{pmatrix}\mathbf{s}\\ M\mathbf{s}\end{pmatrix}
\label{eq:single_stage}
\end{equation}
with BP+OSD for a data-error estimate $\mathbf{e}$ and a measurement-error estimate $\mathbf{m}$, where $H$ is $P_X$ or $P_Z$ and $M$ is the corresponding metacheck matrix, so that $MH = 0$ and the metachecks act only on the measurement-error variables.

The correction $\mathbf{e}$ is applied immediately, and after $N=8$ noisy rounds a perfect round ends the run. We report the effective per-round logical error rate $p_L=1-(1-P_L)^{1/N}$, where $P_L$ is the failure probability at the end of the run; Appendix~\ref{section:single_shot_decoders} gives the remaining details.

\endgroup

\revisionblue{Figure~\ref{fig:single_shot} compares three codes of similar length whose Koszul complexes supply metachecks for both types of checks (MM $[[144,12,(12,12)]]$), for one type (TT $[[135,12,(14,6)]]$), and for neither (BB $[[144,12,12]]$). The MM code behaves similarly in the two memory bases, as Proposition~\ref{prop:xz_duality} leads one to expect, with pseudo-thresholds of about $4.1\%$, compared with about $2.5\%$ for the BB code with the same $n$ and $k$. The TT code performs well in the $Z$ memory experiment, where its measured checks have metachecks and $d_X=14$, but its pseudo-threshold drops to $0.80\%$ in the $X$ memory experiment, where $d_Z=6$ and no metacheck map is available. Because the codes also differ in check weight and redundancy, the effect of the metachecks themselves is best seen by comparing each decoder with and without the metacheck rows: most of the gap between the MM and BB codes is already present without them, and including them lowers the MM code's $p_L$ by a further factor of $1.3$ at $p=4\%$ and $2.6$--$2.9$ at $p=1\%$. Appendix~\ref{section:single_shot_decoders} gives the individual rates. How this performance scales with code size and memory duration remains to be studied.}

\subsection{Circuit-level noise model}
\label{subsec:circuitnoisemodel}

To evaluate the impact of realistic gate-level noise on code performance, and in particular, the effect of potentially larger check weights, we simulate a full circuit-level noise model. We adopt the standard single-ancilla circuit construction of Ref.~\cite{tremblay2022constant}, as implemented in the \texttt{LDPC} package~\cite{roffe_decoding_2020}. In this approach, each stabilizer generator is associated with one ancilla qubit, and CNOT gates are scheduled via a minimum edge coloring of the Tanner graph. This ensures that gates of the same color act on disjoint qubits and can therefore be executed simultaneously within a single time step.

Under the circuit-level noise model, every operation in the circuit is subject to depolarizing or flip errors at a uniform physical error rate $p$. Specifically: each CNOT gate is followed by a two-qubit depolarizing error of strength $p$; each data qubit undergoes a single-qubit depolarizing error of strength $p$ at the start of every round; each measurement outcome is flipped with probability $p$; each qubit reset is followed by a state-flip error of strength $p$ ($X$ error after $\ket{0}$ resets, $Z$ error after $\ket{+}$ resets); and each qubit idling during a CNOT layer suffers a single-qubit depolarizing error of strength $p$. Memory-experiment circuits were generated in Stim \cite{gidney2021stim} and decoded using the \texttt{TesseractSinterDecoder} via Sinter, configured with a detector beam width of $8$, beam climbing enabled, no-revisit and merge-errors flags active, a priority queue limit of $10{,}000$, and $10$ detector orderings generated via the \texttt{DetIndex} method \cite{beni2025tesseractsearchbaseddecoderquantum}\revisionblue{, with seed $2384753$ and the sampling settings of Table~\ref{tab:sim_settings}.}
\begin{figure}[tbp]
    \centering
    \includegraphics[width=1\linewidth]{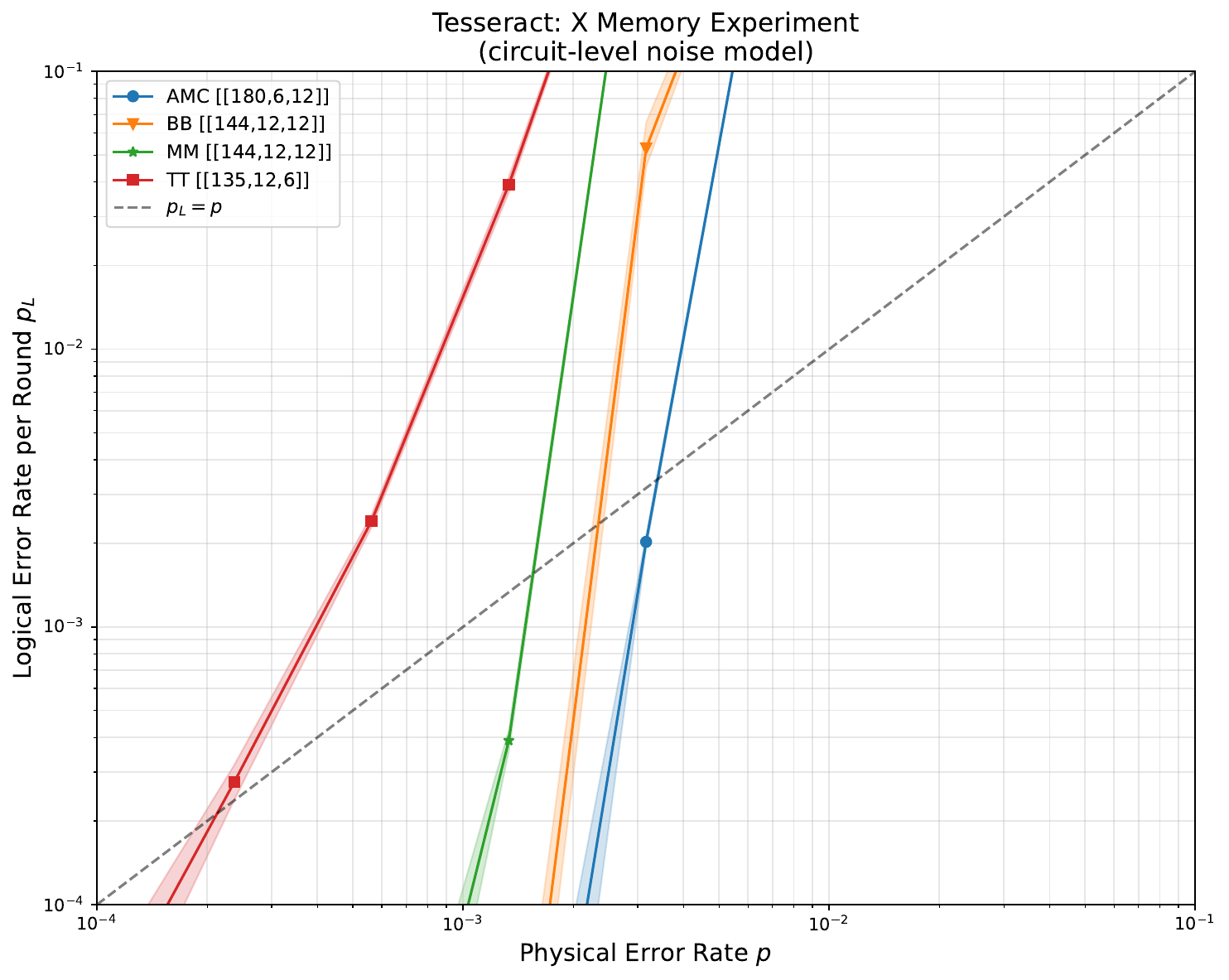}
\caption{\textbf{Performance comparison under circuit-level noise in the $X$-basis memory experiment decoded with \texttt{TesseractSinterDecoder}.} Four codes are compared: MM $[[144,12,(12,12)]]$ with weight-9 stabilizers, BB $[[144,12,(12,12)]]$ with weight-6 stabilizers, AMC $[[180,6,(12,12)]]$ with weight-6 stabilizers, and the asymmetric TT $[[135,12,(14,6)]]$ with weight-9 $X$-type and weight-6 $Z$-type stabilizers.
\label{fig:circuitlevel_model_xbasis}}
\end{figure}
\begin{figure}[tbp]
    \centering
    \includegraphics[width=1\linewidth]{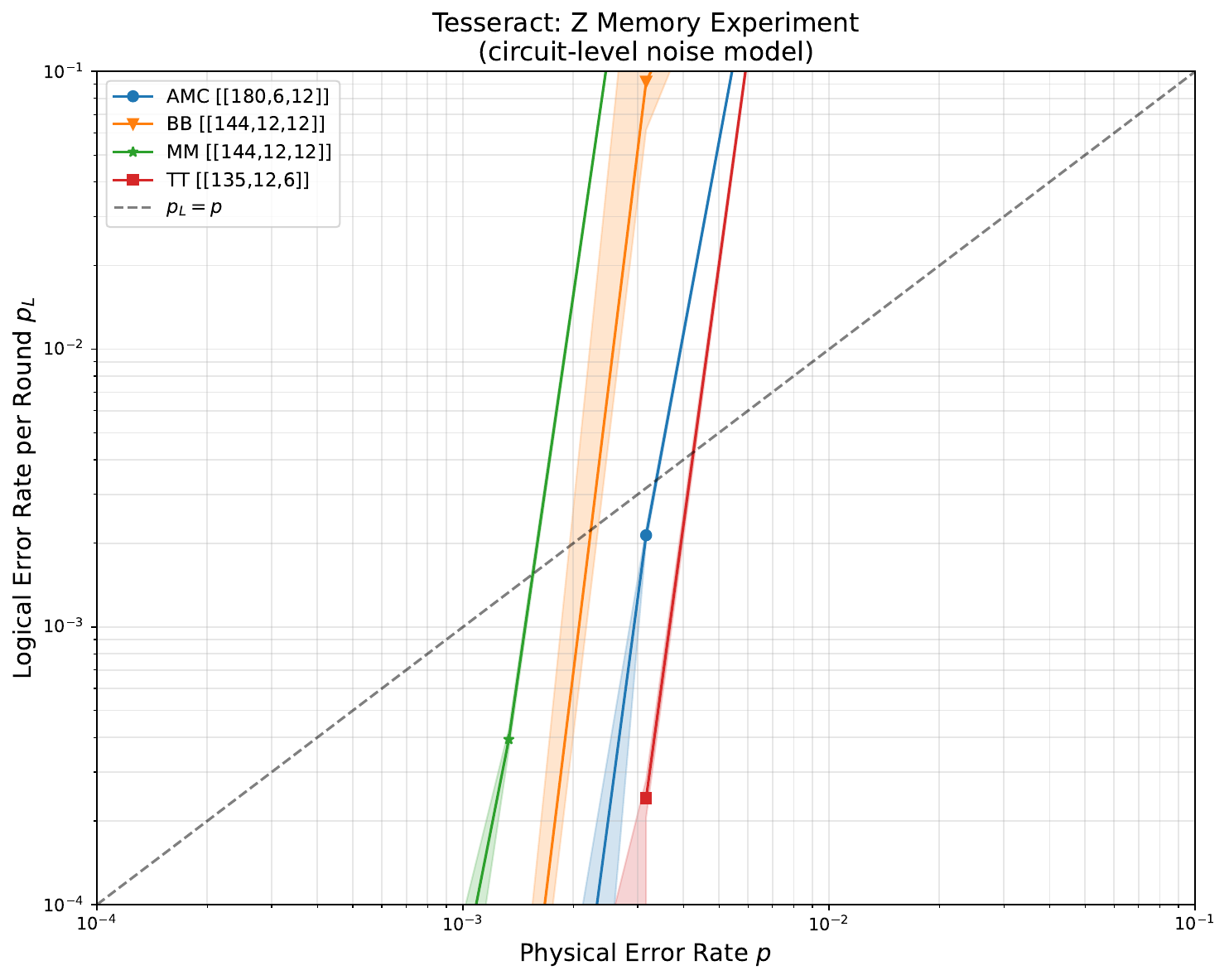}
\caption{\textbf{Performance comparison under circuit-level noise in the $Z$-basis memory experiment decoded with \texttt{TesseractSinterDecoder}.} Comparison of the same codes seen in Fig.~\ref{fig:circuitlevel_model_xbasis}. 
\label{fig:circuitlevel_model_zbasis}}
\end{figure}

Fig.~\ref{fig:circuitlevel_model_xbasis} and Fig.~\ref{fig:circuitlevel_model_zbasis} show \revisionblue{that gate-level noise changes the picture: the weight-6 BB and AMC codes now outperform the weight-9 MM code, whose pseudo-threshold of about $0.15\%$ lies below those of the BB code (about $0.23\%$) and the AMC code (about $0.34\%$). This is consistent with the cost of the heavier checks: with one ancilla per check and a minimum edge coloring, the MM circuit has $18$ CNOT layers and $2592$ idle-noise locations per round, compared with $12$ layers and $1728$ locations for the BB code (Table~\ref{tab:coloration_resources}), and the coloration schedule does not control hook errors. The TT code is limited by its $d_Z=6$ in the $X$-basis experiment and performs best in the $Z$-basis experiment. Reducing this overhead is the purpose of the weight-reduction and scheduling techniques discussed in Sec.~\ref{sec:discussion}.} We further evaluate performance using left-right syndrome extraction circuits~\cite{strikis2026high}, with results in Appendix~\ref{section:lrccircuitlevelnoisemodel}.
\subsection{Overlapping Window Decoding}
\label{subsection:owd_decoding}

\emph{Overlapping window decoding} (OWD) processes syndrome streams incrementally on bounded temporal windows, avoiding the need to store the full syndrome history~\cite{terhal2015quantum, skoric2023parallel}. Remarkably, Dennis et al.~\cite{Dennis_2002} recognized this principle as early as 2001:

\begin{quote}
\textit{In our overlapping recovery scheme, we take action to remove only these long-lived defects, leaving those of more recent vintage to be dealt with in the next recovery step.}
\end{quote}

Here, a defect refers to a syndrome change; persistent defects are more trustworthy.

In practice, OWD examines a sliding window of $w$ consecutive syndrome rounds and proposes a correction using BP+OSD (belief propagation with ordered statistics decoding). Only the first $c \leq w$ rounds\textemdash the \emph{commit} region\textemdash are finalized, while the remaining $w - c$ rounds form a \emph{buffer} that is re-decoded in the subsequent window~\cite{skoric2023parallel}. \revisionblue{In our experiments each window is decoded with min-sum BP (at most $1000$ iterations) followed by OSD of order $0$, with commit size $c = 1$.} The window advances by $c$ rounds, and when $w$ and $c$ remain constant as code distance increases, OWD is said to achieve \emph{$(w,c)$-single-shot} decoding~\cite{scruby2026high}.

\begin{figure*}
    \centering
    \includegraphics[width=1\linewidth]{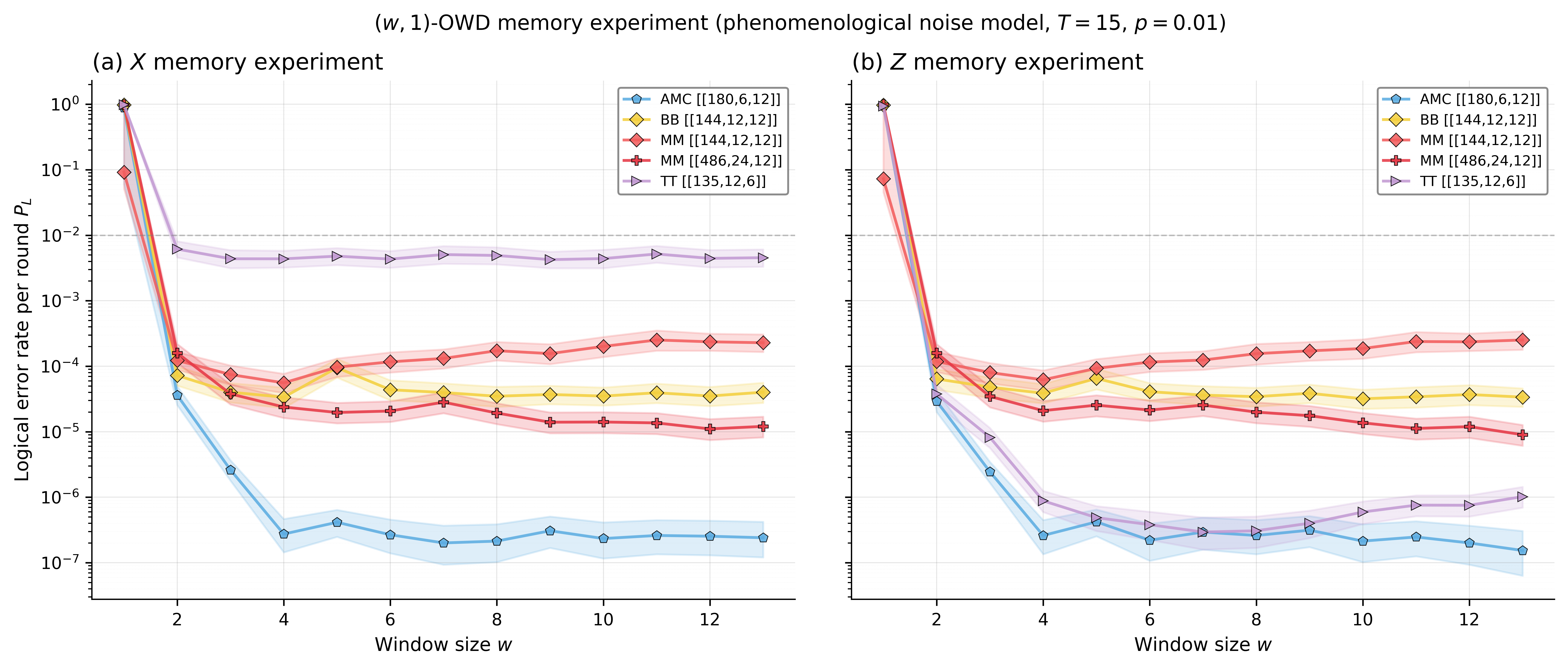}
     \caption{\textbf{Plateaus in the $X$ and $Z$ memory experiments, with phenomenological noise model.} On the vertical axis we show the per-round logical error rate $p_L$ as a function of OWD window size $w$ (on the horizontal axis) under phenomenological noise at $p = 10^{-2}$ with $T=15$ syndrome extraction rounds and commit size $c=1$. The $X$ memory experiment is shown in (a) and $Z$ in (b). We test the MM codes $[[144,12,(12,12)]]$ and $[[486,24,(12,12)]]$; the BB code [[144,12,12]]; the TT code [[135,12,(14,6)]]; and the AMC code [[180,6,(12,12)]]. The total rounds $T=15$ ensures sufficient syndrome history for the OWD decoder across window sizes up to $w=13$.}
    \label{fig:plateau_XZ_T15phenom}
\end{figure*}

\begin{figure*}
    \centering
    \includegraphics[width=1.0\linewidth]{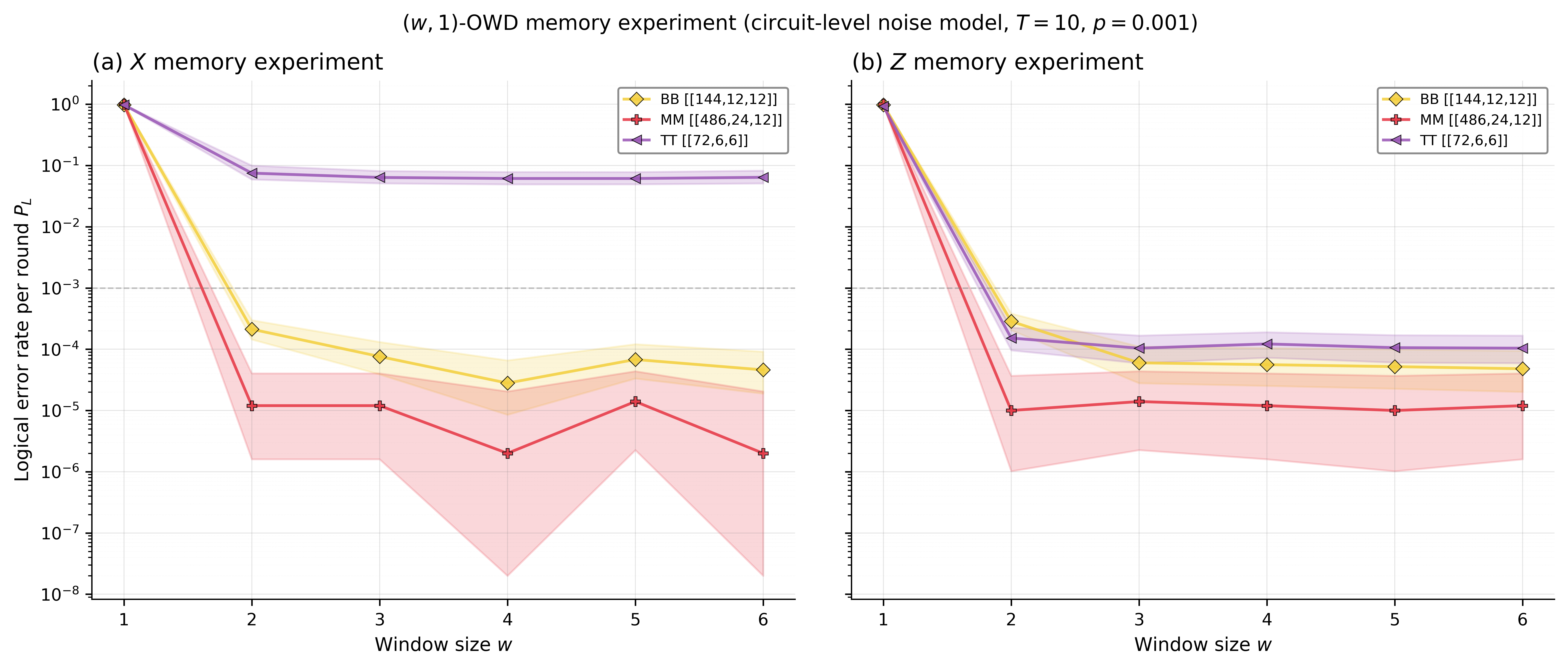}
     \caption{\editcolor{\textbf{Plateau experiment in the $X$ and $Z$ memory experiments under circuit-level noise}} ($T = 10$, $p = 10^{-3}$, commit $c = 1$). \revisionblue{The codes shown are BB $[[144,12,12]]$, MM $[[486,24,12]]$, and TT $[[72,6,6]]$.}}
    \label{fig:plateau_XZ_T10circ}
\end{figure*}

To assess the few-shot decoding performance of our codes under OWD, we perform plateau
experiments~\cite{lin2025abelianmulticyclecodessingleshot,
lin2025singleshottwoshotdecodinggeneralized,
jacob2025singleshotdecodingfaulttolerantgates}, in which the per-round logical error rate $p_L$ is measured as a function of $w$ at fixed $c$, total syndrome measurement rounds $T$, and physical error probability $p$. We conduct these experiments under both phenomenological (Fig.~\ref{fig:plateau_XZ_T15phenom}) and circuit-level (Fig.~\ref{fig:plateau_XZ_T10circ}) noise models. \revisionblue{Under phenomenological noise at $p=10^{-2}$ (Fig.~\ref{fig:plateau_XZ_T15phenom}), every code falls below the physical error rate once $w\geq2$: the MM $[[486,24,12]]$ code reaches about $10^{-5}$ per round at the largest windows, below the BB $[[144,12,12]]$ code (about $4\times10^{-5}$), the MM $[[144,12,12]]$ code stays between about $5\times10^{-5}$ and $3\times10^{-4}$, and the AMC $[[180,6,12]]$ code is lowest, near $2\times10^{-7}$. Under circuit-level noise at $p=10^{-3}$ (Fig.~\ref{fig:plateau_XZ_T10circ}), windows of two rounds already give the MM $[[486,24,12]]$ code about $10^{-5}$ per round, compared with $2$--$3\times10^{-4}$ for the BB code, although the two codes differ in length and few failures were observed for the MM code. With $w=1$, every code has a high logical failure rate (Appendix~\ref{section:single_shot_decoders}).}

\subsection{Confinement Profiles}

Confinement \revisionblue{(Definition~\ref{def:confinement})} quantifies how the weight of the measured syndrome grows with the reduced weight of a physical error \revisionblue{(Definition~\ref{def:reduced_weight})}. In a quantum CSS code this relation is formalized via boundary and coboundary confinement~\cite[see Definitions~3.44--3.45]{sriram2025diffusioncodesselfcorrectionsmallerset}, which respectively state that $Z$-type ($X$-type) errors of small reduced weight must produce syndromes under the $X$-type ($Z$-type) checks whose weight grows at least linearly with the reduced error weight. Higher confinement therefore \revisionblue{means} that low-weight physical errors necessarily flip many syndrome bits, \revisionblue{so} that their syndromes remain distinguishable even in the presence of measurement noise. \revisionblue{When this linear growth holds with constants independent of the block length across an infinite code family (good linear confinement~\cite[Definition~2.10]{tan2025singleshotuniversalityquantumldpc}), it is sufficient for single-shot decoding against both adversarial and stochastic noise~\cite{Quintavalle_2021, tan2025singleshotuniversalityquantumldpc}. For the finite instances studied here, we use confinement as a finite-size diagnostic.}

\revisionblue{Table~\ref{tab:confinement} lists finite-weight \emph{confinement profiles}~\cite{Quintavalle_2021}: for each error weight, the smallest nonzero syndrome weight observed among irreducible errors (Definition~\ref{def:irreducible}) of that weight. Computational details are given in Appendix~\ref{sec:appendix_search}.}

\renewcommand{\arraystretch}{1}
\setlength{\tabcolsep}{4pt}
\begin{table*}[tp]
\definecolor{A}{RGB}{248, 248, 255}
\definecolor{B}{RGB}{245, 255, 250}
\definecolor{C}{RGB}{255, 250, 240}
\definecolor{D}{RGB}{240, 248, 255}
\definecolor{E}{RGB}{255, 245, 238}
\definecolor{F}{RGB}{240, 255, 240}
\definecolor{G}{RGB}{255, 250, 250}
\definecolor{H}{RGB}{248, 248, 247}
\definecolor{I}{RGB}{245, 245, 245}
\definecolor{J}{RGB}{250, 250, 250}
\definecolor{K}{RGB}{245, 248, 255}
\definecolor{L}{RGB}{255, 248, 240}
\definecolor{M}{RGB}{240, 252, 245}
\definecolor{N}{RGB}{250, 245, 255}
\definecolor{O}{RGB}{245, 250, 255}
\definecolor{P}{RGB}{255, 252, 245}
\definecolor{Q}{RGB}{248, 255, 248}
\definecolor{R}{RGB}{255, 248, 248}
\definecolor{S}{RGB}{245, 250, 245}
\definecolor{T}{RGB}{250, 248, 255}
\definecolor{U}{RGB}{252, 252, 250}
\definecolor{V}{RGB}{248, 250, 255}
\definecolor{W}{RGB}{255, 252, 250}
\definecolor{X}{RGB}{250, 252, 248}
\definecolor{Y}{RGB}{252, 250, 248}
\definecolor{Z}{RGB}{248, 252, 250}
\begin{adjustbox}{max width=\textwidth, max totalheight=0.82\textheight, center}
\begin{tabular}{|c|c|c|c|c|c|c|c|c|c|}
\hline
\rowcolor{A}Code Instance & $[[n, k, (d_X, d_Z)]]$ & $\frac{k}{n}$ & $s_{\min}$ & $\tilde{w_{X}}$ & $\tilde{w_{Z}}$ & $\overline{w_{X}}$& $\overline{w_{Z}}$ & \revisionblue{Z-error profile} & \revisionblue{X-error profile}\\
\hline
\rowcolor{B}2D Planar~\cite{liang2025planar} & $[[72, 8, (4, 4)]]$ & 0.11 & 1 & 5 & 5 & 5 & 5 & 1,1 & 1,1 \\
\hline
\rowcolor{B}2D Planar~\cite{liang2025planar} & $[[98, 8, (5, 5)]]$ & 0.08 & 1 & 6 & 6 & 6 & 6 & 1,1 & 1,1 \\
\hline
\rowcolor{B}2D Planar~\cite{liang2025planar} & $[[128, 8, (6, 6)]]$ & 0.06 & 1 & 6 & 6 & 6 & 6 & 1,1 & 1,1 \\
\hline
\rowcolor{C}2D Tile~\cite{steffan2025tile, eberhardt2024pruning, liang2025planar} & $[[288, 18, (13, 13)]]$ & 0.06 & 1 & 8 & 8 & 8 & 8 & 1,1 & 1,1 \\
\hline
\rowcolor{C}2D Tile~\cite{steffan2025tile, eberhardt2024pruning, liang2025planar} & $[[512, 18, (19, 19)]]$ & 0.03 & 1 & 8 & 8 & 8 & 8 & 1,1 & 1,1\\
\hline
\rowcolor{D}4D Surface~\cite{Berthusen_2024} & $[[33, 1, (4, 4)]]$ & 0.03 & 2 & 4 & 4 & 5 & 5 & 2,2,2,3 & 2,2,2,2 \\
\hline
\rowcolor{E}4D Toric~\cite{Dennis_2002} & $[[96, 6, (4, 4)]]$ & 0.06 & 4 & 6 & 6 & 6 & 6 & 4,4,4 & 4,4,4\\
\hline
\rowcolor{F}QuantumTanner~\cite{radebold2025explicit} & $[[36, 8, (3, 3)]]$ & 0.22 & 2 & 6 & 6 & 6 & 6 & 2,2,2, &  2,2,2, \\
\hline
\rowcolor{F}QuantumTanner~\cite{radebold2025explicit} & $[[54, 11, (4, 4)]]$ & 0.20 &  2 & 6 & 6 & 6 & 6 & 2,2,2,2, & 2,2,2,2, \\
\hline
\rowcolor{F}QuantumTanner~\cite{radebold2025explicit} & $[[72, 14, (4, 4)]]$ & 0.19 & 2 & 6 & 6 & 6 & 6 & 2,2,2,2, & 2,2,2,2, \\
\hline
\rowcolor{G}QuantumTanner~\cite{leverrier2025smallquantumtannercodes} & $[[84, 2, (10, 10)]]$ &0.02 & 1 & 6 & 6 & 6 & 6 & 2,2,2,2,2,1 & 2,2,2,2,2,1\\
\hline
\rowcolor{G}QuantumTanner~\cite{leverrier2025smallquantumtannercodes} & $[[144, 8, (12, 12)]]$ & 0.05 & 2 & 9 & 9 & 9 & 9 & 2,3,3,2,2,2 & 2,3,3,2,2,2 \\
\hline
\rowcolor{G}QuantumTanner~\cite{leverrier2025smallquantumtannercodes} & $[[144, 12, (11, 11)]]$ & 0.08 & 2 & 9 & 9 & 9 & 9 & 2,3,3,2,3,2 & 2,3,3,2,3,2 \\
\hline
\rowcolor{H}QuantumTanner~\cite{wang2026check} & $[[96, 24, (4, 4)]]$ & 0.25 & 2 & 6 & 8 & 6 & 8 & 2,2,2,2 & 2,2,2,2 \\
\hline
\rowcolor{H}QuantumTanner~\cite{wang2026check} & $[[96, 30, (4, 4)]]$ & 0.31 & 2 & 8 & 8 & 8 & 8 & 2,2,2,2 & 2,2,2,2 \\
\hline
\rowcolor{H}QuantumTanner~\cite{wang2026check} & $[[128, 40, (4, 4)]]$ & 0.31 & 2 & 8 & 8 & 8 & 8 & 2,2,2,4 & 2,2,2,4 \\
\hline
\rowcolor{K}LiftedQuantumTanner~\cite{guemard2025lifts, guemard2025moderate} & $[[96, 2, (12, 12)]]$ & 0.02 & 2 & 6 & 6 & 8 & 8 & 2,2,2,2 & 2,2,2,2 \\
\hline
\rowcolor{K}LiftedQuantumTanner~\cite{guemard2025lifts, guemard2025moderate} & $[[160, 2, (16, 16)]]$ & 0.01 & 2 & 6 & 6 & 8 & 8 & 2,2,2,2, & 2,2,2,2, \\
\hline
\rowcolor{K}LiftedQuantumTanner~\cite{guemard2025lifts, guemard2025moderate} & $[[324, 4, (18, 18)]]$ & 0.01 & 2 & 5 & 5 & 5 & 5 & 2,2,2,2,2, & 2,2,2,2,3, \\
\hline
\rowcolor{L}Vanilla BivariateBicycle~\cite{Bravyi_2024} & $[[90, 8, (10, 10)]]$ & 0.08  & 2 & 6 & 6 & 6 & 6 & 3,4,3,2,3,2 & 3,4,3,2,3,2 \\
\hline
\rowcolor{L}Vanilla BivariateBicycle~\cite{Bravyi_2024} & $[[108, 8, (10, 10)]]$ & 0.07 & 2 & 6 & 6 & 6 & 6 & 3,4,3,2,3,2 & 3,4,3,2,3,2 \\
\hline
\rowcolor{L}Vanilla BivariateBicycle~\cite{Bravyi_2024} & $[[144, 12, (12, 12)]]$ & 0.08 & 3 & 6 & 6 & 6 & 6 & 3,4,3,4,3,4 & 3,4,3,4,3,4\\
\hline
\rowcolor{M}Self-Dual BivariateBicycle~\cite{liang2025selfdualbivariatebicyclecodes} & $[[64, 8, (8,8)]]$ & 0.12 & 2 & 8 & 8 & 8 & 8 & 4,2,4,2,4,2 & 4,2,4,2,4,2\\
\hline
\rowcolor{M}Self-Dual BivariateBicycle~\cite{liang2025selfdualbivariatebicyclecodes} & $[[96, 12, (8,8)]]$ & 0.12 & 2 & 8 & 8 & 8 & 8 & 4,4,4,4,4,2 & 4,4,4,4,4,2 \\
\hline
\rowcolor{M}Self-Dual BivariateBicycle~\cite{liang2025selfdualbivariatebicyclecodes} & $[[98, 14, (6, 6)]]$ & 0.14 & 2 & 8 & 8 & 8 & 8 & 4,4,2,4,2,4 & 4,4,2,4,2,4 \\
\hline
\rowcolor{O}3-Cover BivariateBicycle~\cite{liang2025selfdualbivariatebicyclecodes} & $[[96, 12, (10, 10)]]$ & 0.12 & 4 & 8 & 8 & 8 & 8 & 4,6,6,4,4,6 & 4,6,6,4,4,6\\
\hline
\rowcolor{O}7-Cover BivariateBicycle~\cite{symons2025sequences} & $[[98, 6, (12, 12)]]$ & 0.06 & 4 & 6 & 6 & 6 & 6 & 4,6,6,4,4,6 & 4,6,6,4,4,6 \\
\hline
\rowcolor{O}2-Cover BivariateBicycle~\cite{symons2025sequences} & $[[144, 14, (14,14)]]$ & 0.09 & 4 & 8 & 8 & 8 & 8 & 4,6,6,4,4,6 & 4,6,6,4,4,6 \\
\hline
\rowcolor{P}GeneralizedToric ~\cite{liang2025generalized} & $[[90, 8, (10, 10)]]$ & 0.08 & 2 & 6 & 6 & 6 & 6 & 3,4,3,2,3,2 & 3,4,3,2,3,4 \\
\hline
\rowcolor{P}GeneralizedToric ~\cite{liang2025generalized} & $[[96, 4, (12, 12)]]$ & 0.04 & 3 & 6 & 6 & 6 & 6 & 3,4,3,4,3,4 & 3,4,3,4,3,4\\
\hline
\rowcolor{P}GeneralizedToric ~\cite{liang2025generalized} & $[[98, 6, (12, 12)]]$ & 0.06 & 2 & 6 & 6 & 6 & 6 & 3,4,3,4,3,4 & 3,4,3,4,3,2 \\
\hline
\rowcolor{Q}ZSZ~\cite{Guo_2026} & $[[80,2,(8,8)]]$ & 0.02 & 1 & 6 & 6 & 6 & 6 & 3,2,3,2,1,4 & 3,2,3,2,1,2 \\
\hline
\rowcolor{Q}ZSZ~\cite{Guo_2026} & $[[144,12,(8,8)]]$ & 0.08 & 3 & 6 & 6 & 6 & 6 & 3,4,3,4,3,4 & 3,4,5,4,3,4 \\
\hline
\rowcolor{Q}ZSZ~\cite{Guo_2026} & $[[288,12,(8,8)]]$ & 0.04 & 3 & 6 & 6 & 6 & 6 & 3,4,5,4,3,4 & 3,4,5,4,3,4 \\ 
\hline
\rowcolor{R}TrivariateTricycle~\cite{jacob2025singleshotdecodingfaulttolerantgates} & $[[72, 6, (12, 6)]]$ & 0.08 & 2 & 9 & 6 & 9 & 6 & 3,2,3,2,3,2 &
6,8,10,10,12,10 \\ 
\hline
\rowcolor{R}TrivariateTricycle~\cite{jacob2025singleshotdecodingfaulttolerantgates} & $[[81, 6, (12, 6)]]$ & 0.07 & 2 & 9 & 6 & 9 & 6 & 3,2,3,2, & 6,8,6,8,8,10 \\
\hline
\rowcolor{R}TrivariateTricycle~\cite{jacob2025singleshotdecodingfaulttolerantgates} & $[[90, 3, (15, 5)]]$ & 0.03 & 2 & 6 & 4 & 6 & 4 & 2,2,2,2, & 4,6,6,6,4,6\\
\hline
\rowcolor{S}AbelianMulticycle~\cite{lin2025abelianmulticyclecodessingleshot} & $[[84, 6, (7, 7)]]$ &  0.07 & 4 & 6 & 6 & 6 & 6 & 4,6,8,8,4,6, & 4,6,8,8,4,6 \\
\hline
\rowcolor{S}AbelianMulticycle~\cite{lin2025abelianmulticyclecodessingleshot} & $[[96, 6, (8, 8)]]$ &  0.06 & 4 & 6 & 6 & 6 & 6 & 4,6,6,6,4,4, & 4,6,6,6,4,4\\
\hline
\rowcolor{S}AbelianMulticycle~\cite{lin2025abelianmulticyclecodessingleshot} & $[[108, 6, (9, 9)]]$ & 0.05 & 4 & 6 & 6 & 6 & 6 & 4,6,8,8,4,6 & 4,6,8,8,4,6\\
\hline
\rowcolor{T}MultivariateMulticycle & $[[96, 12, (4, 4)]]$ & 0.12 & 8 & \revisionblue{12} & \revisionblue{12} & \revisionblue{12} & \revisionblue{12} & 8,8,8,8 & 8,8,8,8\\ %
\hline
\rowcolor{T}MultivariateMulticycle & $[[96, 12, (8, 8)]]$ & 0.12 & 8 & 16 & 16 & 16 & 16 & 8,8,8,8,8,8 &   %
8,8,8,8,8,8\\
\hline
\rowcolor{T}MultivariateMulticycle & $[[96, 44, (4, 4)]]$ & 0.45 & 8 & 12 & 12 & 12 & 12 & 8,8,8 & 8,8,8 \\ %
\hline
\rowcolor{T}MultivariateMulticycle & $[[144, 40, (4, 4)]]$ & 0.27 & 8 & 12 & 12 & 12 & 12 & \revisionblue{8,8,8,8} & \revisionblue{8,8,8,8} \\ %
\hline
\rowcolor{T}MultivariateMulticycle & $[[216, 12, (12, 12)]]$ & 0.05 & 4 & 9 & 9 & 10 & 10 & 4,6,8,8,10,8 & \revisionblue{4,6,8,8,10,8} \\ %
\hline
\rowcolor{T}MultivariateMulticycle & $[[486, 24, (12, 12)]]$ & 0.04 & 6 & 9 & 9 & 9 & 9 & 6,10,12,14,18,18 & 6,10,12,14,18,18 \\ %
\hline
\rowcolor{T}MultivariateMulticycle & $[[486, 66, (9, 9)]]$ & 0.13 & 8 & 12 & 12 & 12 & 12 & 8,12,12,16,16,12 & 8,12,12,16,16,12 \\ %
\hline
\rowcolor{T}MultivariateMulticycle & $[[576, 64, (6, 6)]]$ & 0.11 & 8 & 12 & 12 & 12 & 12 & \revisionblue{8,8,12,8,8,8} & \revisionblue{8,8,12,8,8,8} \\ %
\hline
\end{tabular}
\end{adjustbox}
\centering
\caption{\textbf{Code parameters \editcolor{$[[n,k,(d_X,d_Z)]]$} and confinement profiles.} \revisionblue{$n$ and $k$ are the numbers of physical and logical qubits, $k/n$ is the rate, and ($\widetilde w_{X}$, $\widetilde w_{Z}$) and ($\overline{w_{X}}$, $\overline{w_{Z}}$) are the median and maximum check weights. Comparison-code distances are taken from the cited references; all eight MM distances are exact (Appendix~\ref{sec:appendix_search}). Profiles list, for each error weight $w=1,2,\dots$, the smallest nonzero syndrome weight observed among the irreducible errors visited by the search, and $s_{\min}$ is their smallest entry. For the MM codes, each profile is the entrywise minimum over the two sectors (see text).}}
\label{tab:confinement}
\end{table*}

\revisionblue{For the $t=4$ MM codes, Proposition~\ref{prop:xz_duality} relates the two sectors by a permutation of qubits and checks that maps irreducible errors to irreducible errors of the same weight and preserves syndrome weights, so the true profiles of the two sectors coincide. The time-limited search can nevertheless visit different supports in the two sectors; since every observed entry is attained by some error, we report the entrywise minimum of the two observed profiles. Because the profiles depend on the supports that the search visits, and the comparison profiles come from a separate calculation, Table~\ref{tab:confinement} characterizes the individual instances rather than ranking the confinement of the code families.} We note that \revisionblue{the} degree of confinement directly affects single-shot properties: codes with low confinement, such as certain GB codes, exhibit limited single-shot capability~\cite{lin2025singleshottwoshotdecodinggeneralized, sriram2025diffusioncodesselfcorrectionsmallerset}, because high-weight errors can produce small-weight syndromes that are indistinguishable from measurement noise.

Single-shot decoding is often colloquially attributed to one of two mechanisms: either the presence of metachecks or the expansion property of the underlying quantum expander code~\cite{Campbell_2019, Quintavalle_2021}. While this approach is widespread in the literature~\cite{Campbell_2019, Quintavalle_2021, Higgott_2023, ostrev2024classical, lin2025singleshottwoshotdecodinggeneralized, lin2025abelianmulticyclecodessingleshot}, it does not fully capture the mechanism underlying single-shot decoding. As a concrete example, the 4D toric code, whose decoder is typically analyzed in this setting, employs a parallelized local flip decoding that does not explicitly use metachecks~\cite{Dennis_2002}. \revisionblue{Its performance relies instead on a form of confinement, although not a linear one: membrane-like errors have boundaries whose weight grows sublinearly with the membrane area~\cite{lin2025abelianmulticyclecodessingleshot}.} Moreover, it is unclear whether introducing nontrivial stabilizer redundancy alone can produce single-shot capability without simultaneously inducing some form of confinement, raising the question of whether redundancy per se should be regarded as the fundamental resource enabling single-shot decoding. A similar situation arises for asymptotically good codes, which possess substantial global redundancy even though practical single-shot decoders~\cite{Gu_2024} do not explicitly exploit the use of metachecks.

\section{Discussion and Conclusions\label{sec:discussion}}
Our Multivariate Multicycle (MM) code framework provides a unification of previous QECCs, namely BB~\cite{Bravyi_2024}, MB~\cite{voss2025multivariate}, TT~\cite{jacob2025singleshotdecodingfaulttolerantgates, menon2025magictricyclesefficientmagic}, GB~\cite{PhysRevA.88.012311}, and abelian 2BGA~\cite{lin2023quantumtwoblockgroupalgebra} codes\revisionblue{, and it is a Koszul-complex formulation of the AMC codes of Ref.~\cite{lin2025abelianmulticyclecodessingleshot}, i.e.\ of iterated lifted products of $1\times1$ matrices over a finite abelian group algebra. Its main practical advantage is that the exterior algebra supplies closed-form parity-check and metacheck matrices for every $t$.} Our construction provides metachecks for \revisionblue{both types of checks when $t\geq4$, with both maps nonzero if at least one defining polynomial is nonzero} (this improves upon TT codes, which only \revisionblue{have adjacent Koszul metachecks for their $Z$-type checks}). We report a number of low-weight, high-rate, high-distance codes\revisionblue{, including collapsed 5D through 9D instances with constant check weights between 6 and 12; their tabulated distances are exact sector distances or upper bounds supported by verified logical operators. In our decoding experiments, the metacheck rows improve the single-stage decoder for the $[[144,12,12]]$ MM code, and the overlapping-window decoder reaches low logical error rates with small windows.}

\revisionblue{An important open question is whether these finite-size observations extend to an asymptotic family of MM codes. A single-shot guarantee would require control of the residual errors as the block length grows, through soundness or confinement bounds with uniform constants (Sec.~\ref{sec:single_shot_error_correction}), together with a decoder that exploits them. We expect the explicit boundary maps of the Koszul complex to be a natural starting point for such an analysis.}

Some MM codes in Appendix Tables~\ref{tab:table3} and~\ref{tab:table4} have relatively high check weights. However, they are high-rate and high-distance, therefore they \revisionblue{could} be valuable with weight-reduction techniques~\cite{hastings2016weight, hastings2021quantum, PRXQuantum.5.040302} that have been shown to preserve or even improve the effective fault-tolerant distance while lowering stabilizer weights~\cite{Tan_2025}. We expect that investigating weight-reduction techniques with MM codes\revisionblue{, including their effect on the effective distance,} will be a productive future project.

Notably, hook errors, which can reduce the effective distance of a code under noisy syndrome extraction~\cite{Gidney_2023}, \revisionblue{do not reduce the effective distance of higher-dimensional hypergraph-product codes under any single-ancilla measurement circuit, under the hypotheses of Ref.~\cite[Theorem~5]{Tan_2025}. MM codes are lifted products over a group algebra rather than hypergraph products, so extending this result to them, and computing the effective distance of the coloration and left-right circuits used in Sec.~\ref{sec:numerical_simulations} and Appendix~\ref{section:lrccircuitlevelnoisemodel}, are natural next steps.}

\revisionblue{Another} exciting future research direction is the possibility \revisionblue{of} transversal fault-tolerant non-Clifford gates for the MM codes. Recent work~\cite{barkeshli2023codimension,lin2024transversalnoncliffordgatesquantum, zhu2025transversalnoncliffordgatesqldpc, hsin2025classifyinglogicalgatesquantum, Hsin_2025, zhu2025noncliffordparallelizablefaulttolerantlogical, zhu2025topologicaltheoryqldpcnonclifford, golowich2024quantumldpccodestransversal, breuckmann2024cups, li2025poincar, zhu2026nonabelianqldpctqftformalism} has shown that the cochain complex that defines the metachecks in higher dimensional codes can be endowed with cohomological invariants, such as the cup product. For codes with suitable topological structure, cup products relate to intersection pairings through Poincaré duality~\cite{MR1867354}. The cup product descends to cohomology: changing cocycle representatives by coboundaries does not change the resulting cohomology class. Suitable scalar-valued cohomological invariants define logical phase operations. Under additional locality and algebraic conditions, these operations admit constant-depth implementations, which motivates the pursuit of higher-dimensional codes to achieve both single-shot decoding and native, fault-tolerant non-Clifford logical gates. Building on this perspective, the cups-and-gates (CG) framework~\cite{breuckmann2024cups, Hsin_2025} provides a method for constructing logical gates from cohomology invariants and identifying conditions for constant-depth implementations. Such implementations need not be transversal.
This framework naturally raises the question of whether MM codes can support well-defined cohomological invariants such as cup products, that give rise to non-Clifford fault-tolerant logical gates. Thus, in future work we plan to explore the existence of a transversal fault tolerant CCZ gate or gates higher up in the Clifford hierarchy for the MM codes.

\revisionblue{\paragraph*{Data and code availability.} The code constructions, distance-audit records, and archived decoding data are available in the accompanying repository~\cite{feroz_ahmed_mian_2026_24095975}.}

\revisionblue{\paragraph*{Use of large language models.} We used Claude (Anthropic) and ChatGPT (OpenAI) to assist with mathematical arguments, developing verification and plotting scripts, and drafting and editing text. We ran the verification scripts on the Unity computing cluster and checked the resulting outputs against the archived data. We checked all results, scripts and text, and take full responsibility for the content of this article.}

\begin{acknowledgments}
FAM conceived of this project, executed the numerical experiments, and wrote the manuscript with input from the other authors. SK suggested various benchmarks for code performance, directed manuscript outline, and manages the open source framework in which much of the development happened. OG conducted a thorough review of the presentation of the mathematics, providing insightful comments and contributing to key discussions about Koszul complexes. \revisionblue{We gratefully acknowledge Leonid Pryadko for introducing us to the LDPC-breakdown condition. We thank the anonymous referees for their careful reading of the manuscript, and one of them for supplying explicit logical operators for three of our codes.}  We thank Tommy Hofmann for contributions to our open source stack and for referring us to Matthias Zach of the OSCAR team, and we thank Matthias Zach for insightful discussions about Koszul complexes. This work was supported by NSF grants 1941583, 2346089, 2402861, 2522101.
\end{acknowledgments}
\bibliography{bibliography}
\appendix

\section{Phenomenological Noise Experiments}
\label{section:more_phenomenological_noise_experiments}
\begin{figure}[tbp]
    \centering
    \includegraphics[width=1\linewidth]{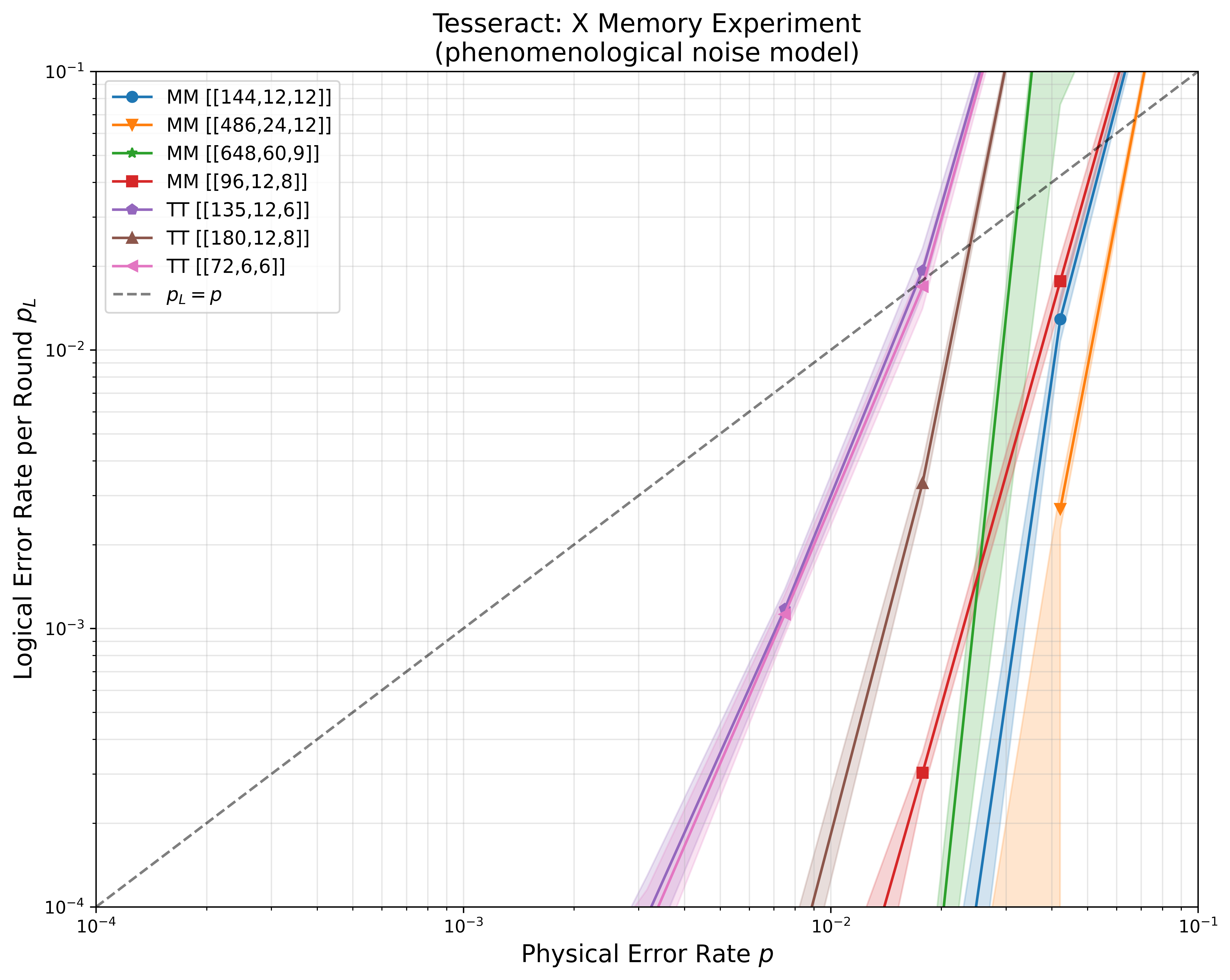}
    \caption{\revisionblue{\textbf{Phenomenological-noise performance in the $X$ memory experiment.}} Seven codes are compared: MM $[[96,12,(8,8)]]$, MM \revisionblue{$[[144,12,(\leq 12,\leq 12)]]$}, MM $[[486,24,(12,12)]]$, MM $[[648,60,(9,9)]]$, TT $[[72,6,(12,6)]]$, TT $[[135,12,(14,6)]]$ and TT $[[180,12,(20,8)]]$.}
    \label{fig:appendixfig1}
\end{figure}

\begin{figure}[tbp]
    \centering
    \includegraphics[width=1\linewidth]{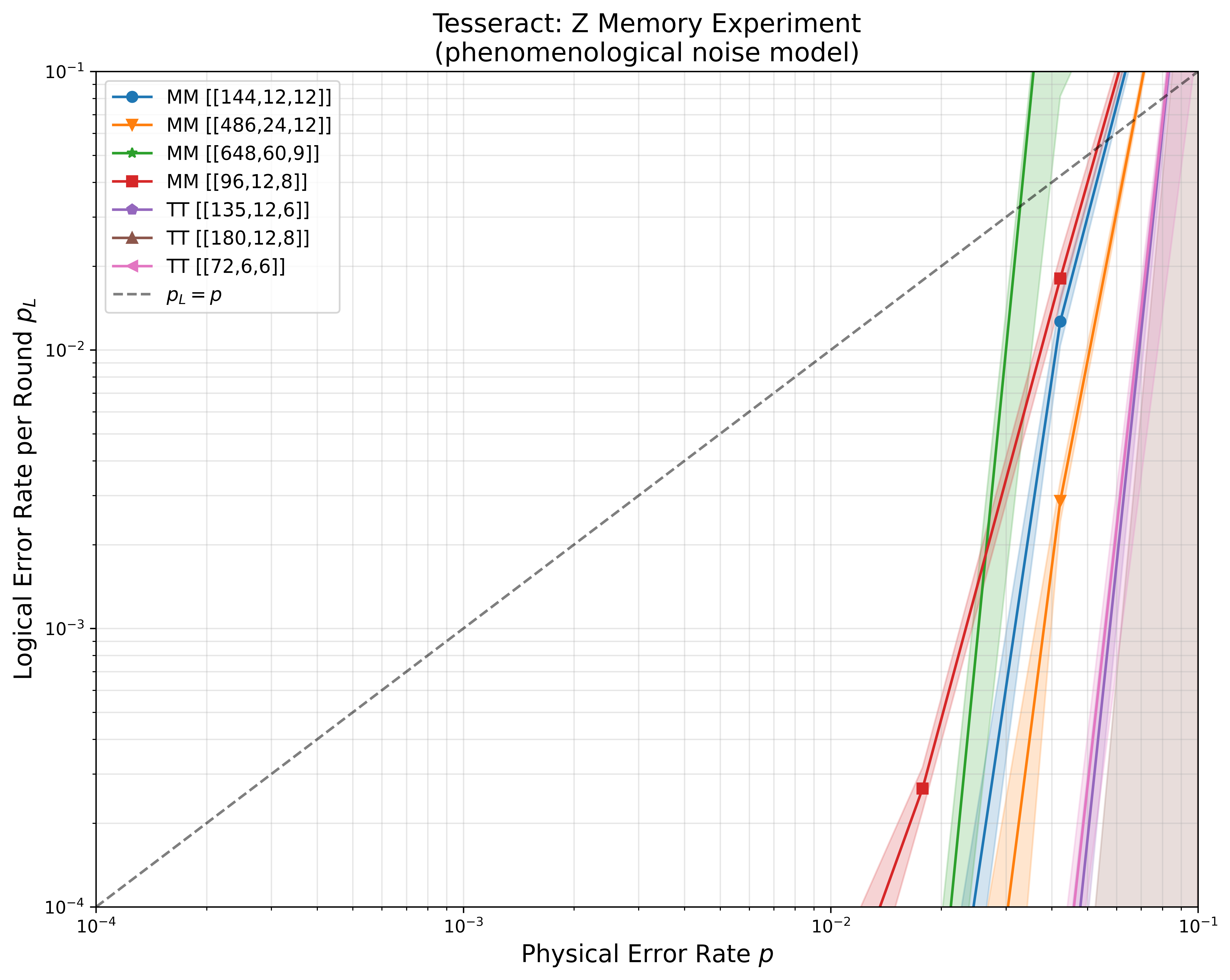}
    \caption{\textbf{MM codes consistently perform competitively against TT codes in the Z memory experiment.}  The same codes as in Fig.~\ref{fig:appendixfig1} are shown.}
    \label{fig:appendixfig2}
\end{figure}

In this appendix, we perform additional phenomenological noise experiments (see Figs.~\ref{fig:appendixfig1} and~\ref{fig:appendixfig2}) to evaluate code performance across a broader range of blocklengths. Under the phenomenological noise model, data qubits are subjected to independent single-qubit depolarizing errors with probability $p$ at the start of each round, and each stabilizer measurement outcome is independently flipped with probability $p$. Memory circuits were generated using the \texttt{make\_css\_code\_memory\_circuit} method from the \texttt{LDPC} package~\cite{roffe_decoding_2020} and sampled using \texttt{Sinter}~\cite{gidney2021stim} with the \texttt{TesseractSinterDecoder} \revisionblue{with a stopping target of $500$ logical errors or a cap of $100{,}000$} shots per physical error rate point\revisionblue{, with $T = 2d$ rounds (Table~\ref{tab:sim_settings})}.

\section{Circuit-Level Noise Experiments via Left-Right Circuits}\label{section:lrccircuitlevelnoisemodel}
In this appendix, we evaluate code performance under circuit-level noise using left-right syndrome extraction circuits (LRC)~\cite{strikis2026high}\revisionblue{, and compare them with the standard coloration circuits of the main text. A fault on an ancilla can propagate through subsequent CNOT gates into a correlated data error, known as a hook error.} If this residual error overlaps a minimum-weight logical operator, it effectively reduces the circuit distance below the code distance $d$, degrading fault-tolerance performance. Rather than mitigating hook errors through flag qubits~\cite{chamberland2018flag, chao2018quantum, lingling2020faulttolerant} or Shor-style syndrome extraction~\cite{shor1996fault}, which incur additional circuit depth and ancilla overhead, the LRC framework \revisionblue{selects CNOT orderings using estimates of the distances of these residual errors. We compare the two schedules through their logical error rates under the simulated noise model.}

\begin{figure}[tbp]
    \centering
    \includegraphics[width=1\linewidth]{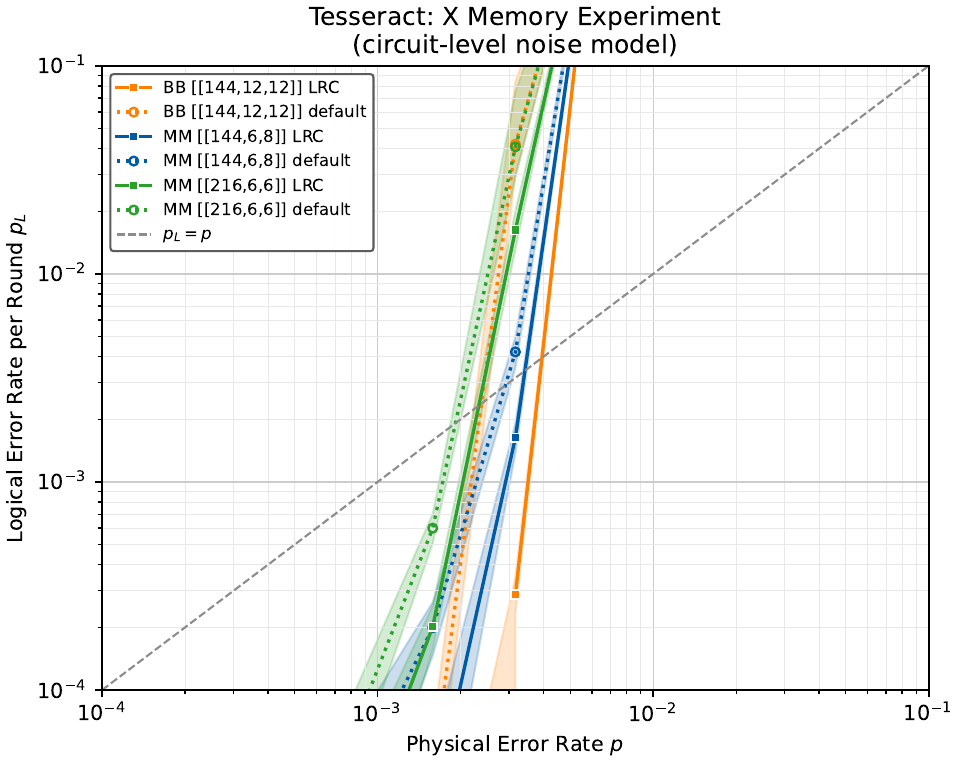}
    \caption{\textbf{Performance comparison under circuit-level noise in the $X$ memory experiment for weight-6 MM codes with 
    \texttt{TesseractSinterDecoder}.} MM~$[[144,6,(8,8)]]$ and MM~\revisionblue{$[[216,6,(6,6)]]$} are compared against BB~[[144,12,(12,12)]] under standard LDPC coloration circuits (default, dotted) and LRC circuits~\cite{strikis2026high} (LRC, solid). \revisionblue{The $[[216,6,(6,6)]]$ code was simulated with $T=24$ noisy rounds, chosen from a preliminary distance estimate of $12$; the other two codes use $T=2d$.} Both weight-6 MM codes exhibit a clear pseudo-threshold improvement under LRC scheduling, in contrast to the weight-9 MM code results shown in Figs.~\ref{fig:circuitlevel_model_xbasis} and~\ref{fig:circuitlevel_model_zbasis}.}
    \label{fig:circuitlevel_model_xbasis_w6}
\end{figure}

\begin{figure}[tbp]
    \centering
    \includegraphics[width=1\linewidth]{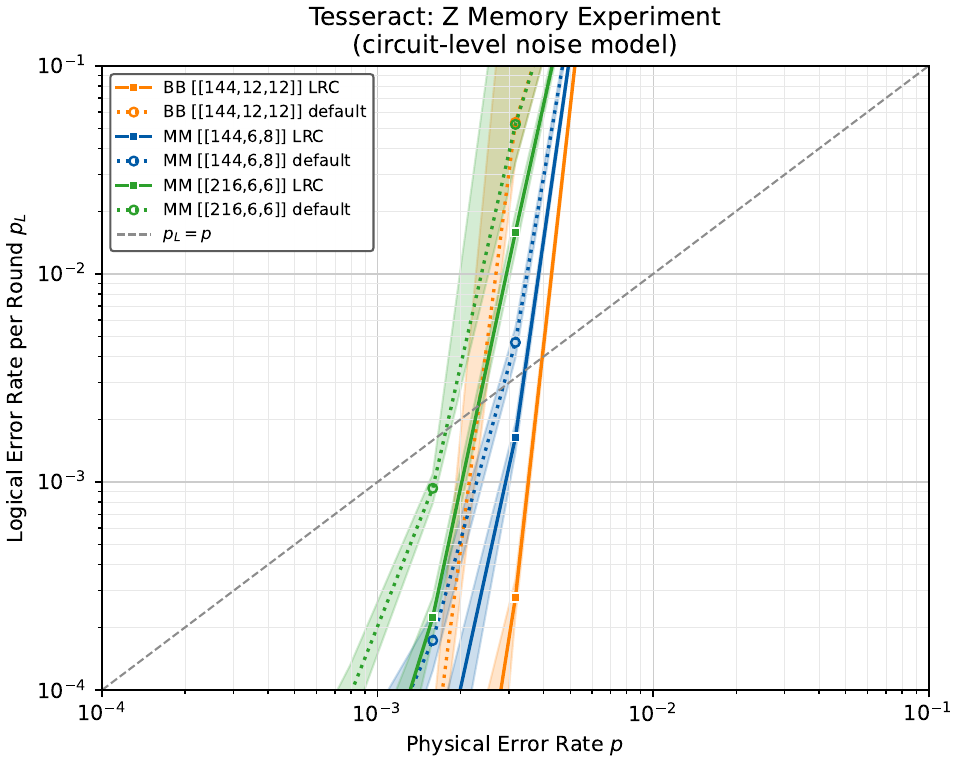}
    \caption{\textbf{Performance comparison under circuit-level noise in the $Z$ memory experiment for weight-6 MM codes with 
    \texttt{TesseractSinterDecoder}.}}
    \label{fig:circuitlevel_model_zbasis_w6}
\end{figure}

\textit{A natural question is whether the benefit of LRC scheduling depends on stabilizer check weight}. To investigate this, we evaluate our MM codes under both weight-6 and weight-9 $X$-stabilizer checks. For the weight-6 variants---specifically $[[144, 6, (8,8)]]$ and \revisionblue{$[[216, 6, (6,6)]]$}---the LRC approach yields clear improvements in both the $X$- and $Z$-basis memory experiments, as shown in Fig.~\ref{fig:circuitlevel_model_xbasis_w6} and Fig.~\ref{fig:circuitlevel_model_zbasis_w6}.

\begin{figure}[tbp]
    \centering
    \includegraphics[width=1\linewidth]{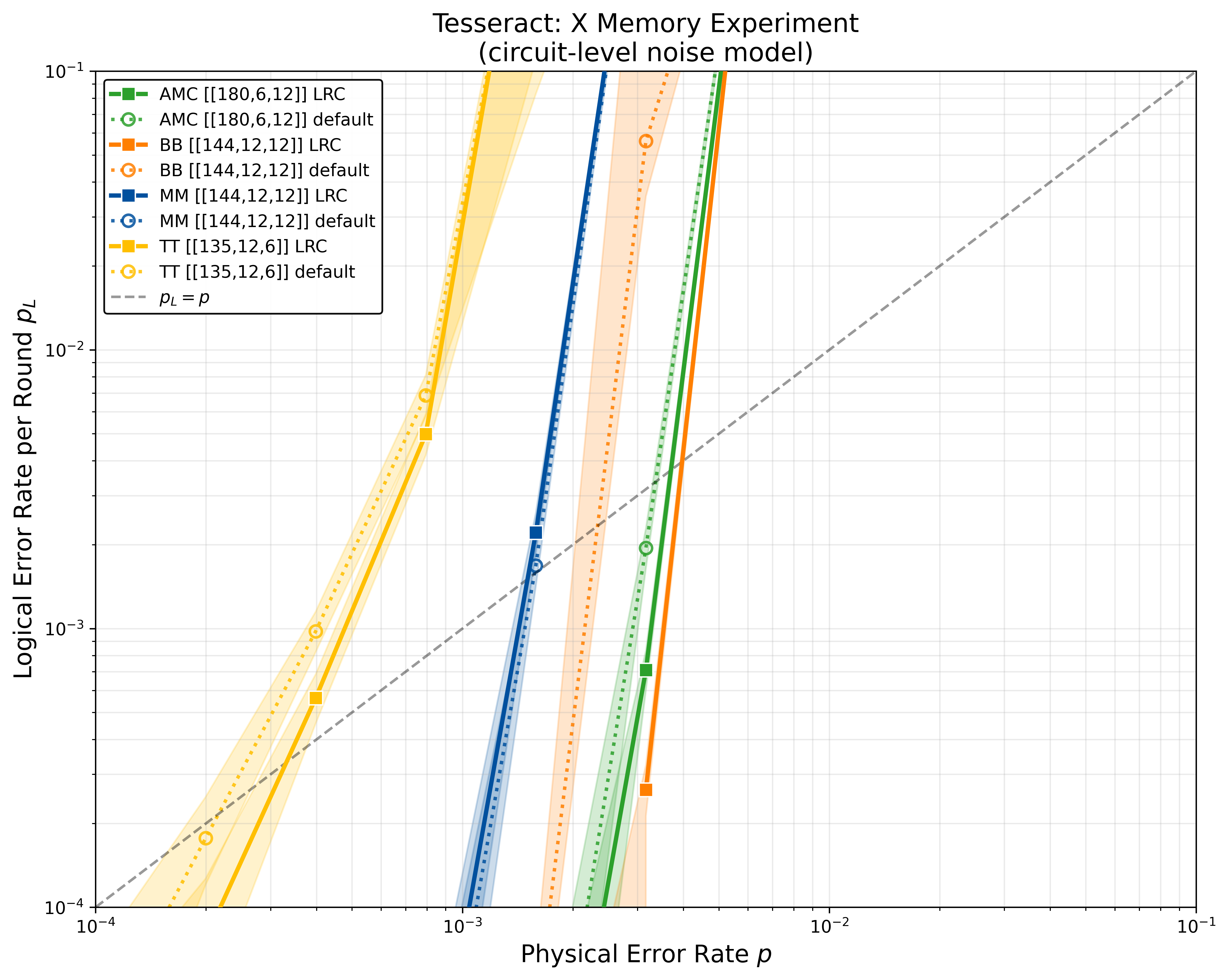}
    \caption{\textbf{Performance comparison under circuit-level noise in the $X$ memory experiment with \texttt{TesseractSinterDecoder}.} Four codes are compared under standard LDPC coloration circuits (default, dotted) and LRC circuits~\cite{strikis2026high} (LRC, solid).}
    \label{fig:circuitlevel_model_xbasis_again}
\end{figure}

\begin{figure}[tbp]
    \centering
    \includegraphics[width=1\linewidth]{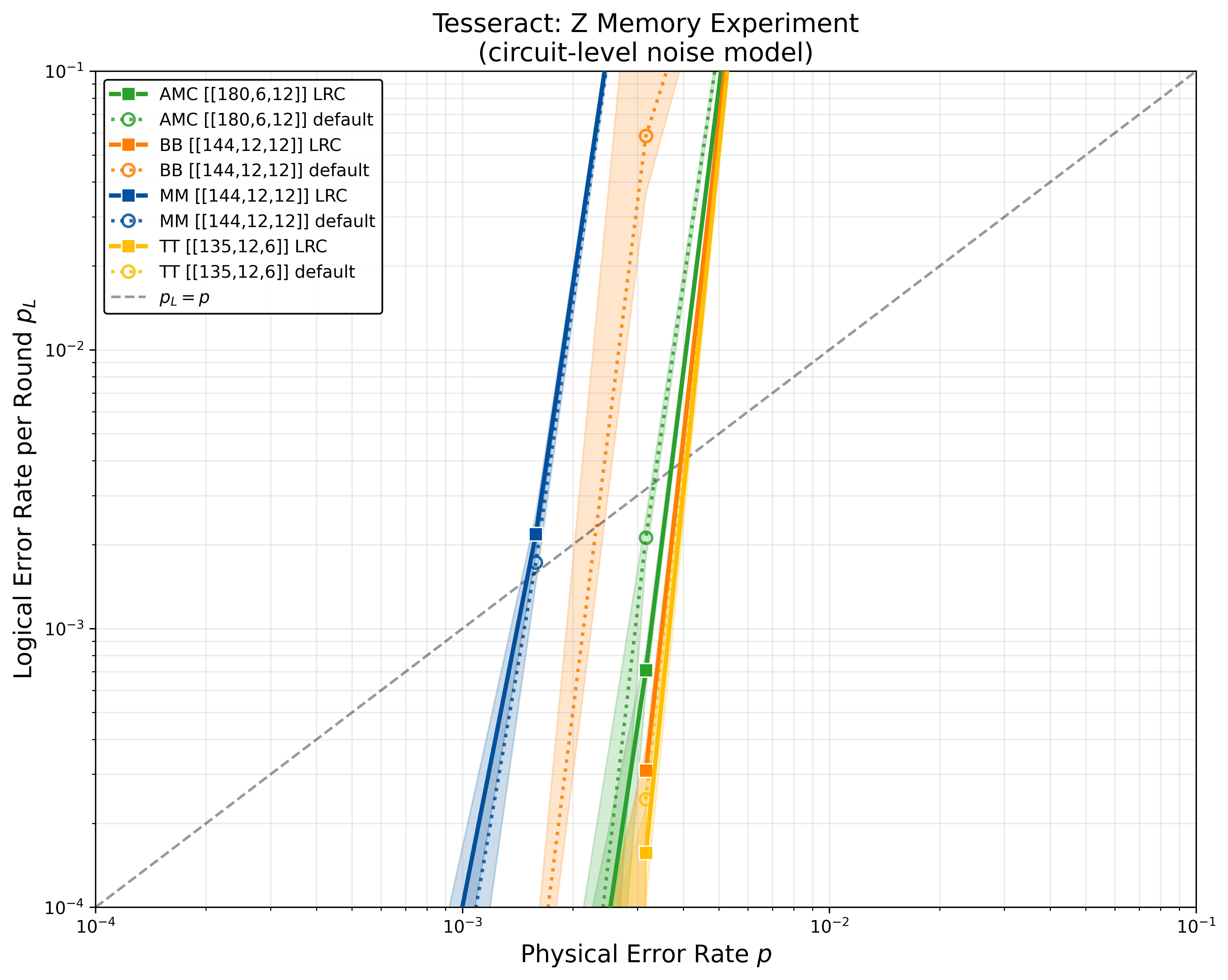}
    \caption{\textbf{Performance comparison under circuit-level noise in the $Z$ memory experiment with \texttt{TesseractSinterDecoder}.} The same four codes and two syndrome extraction approaches as in Fig.~\ref{fig:circuitlevel_model_xbasis_again} are shown.}
    \label{fig:circuitlevel_model_zbasis_again}
\end{figure}

We next compare the four codes --- BB~[[144,12,(12,12)]], MM~$[[144,12,(12,12)]]$, AMC~[[180,6,(12,12)]], and TT~[[135,12,(14,6)]] --- across both X- and Z- memory experiments in Figs.~\ref{fig:circuitlevel_model_xbasis_again} and~\ref{fig:circuitlevel_model_zbasis_again}.

For the LRC preprocessing pipeline, residual distances are estimated using min-sum BP+OSD with $200$ maximum iterations and OSD order~$3$, over $100$ attempts per residual error with a batch size of $10$, parallelised across $8$ workers. Up to $1{,}000$ colour permutations are evaluated and the top $5$ CNOT orderings are retained. The extended-code distance re-ranking step applies $1{,}000$ BP-OSD attempts with $500$ maximum iterations and OSD order~$7$. For the standard coloration circuits generated via the \texttt{LDPC} package~\cite{roffe_decoding_2020}, no CNOT ordering optimization is performed beyond the minimum edge coloring. Logical error rates are estimated via Monte Carlo sampling with \revisionblue{a cap of} $50{,}000$ shots and \revisionblue{a stopping target of} $500$ errors per physical error rate point $p$ in the range $[10^{-4}, 10^{-1}]$, with $T = 2d$ syndrome extraction rounds where $d$ is the \revisionblue{distance value used when constructing the simulation}.

\revisionblue{The resources of the archived LRC schedules are reported in Table~\ref{tab:lrc_resources}.}

\section{Search Procedure for Multivariate Multicycle Codes}
\label{sec:appendix_search}

In this appendix, we describe the computational procedure used to discover high-rate and high-distance MM codes\revisionblue{, and how their distances were verified}. The search is implemented in Julia programming language, relying on the \texttt{QECCore.jl}, \texttt{QuantumClifford.jl}, \texttt{Oscar.jl}, and \texttt{Nemo.jl} libraries for code construction. We fix a target block length $n$ and factorize it as $n=6\ell_1\ell_2\ell_3\ell_4$, enumerating all tuples $(\ell_1,\ell_2,\ell_3,\ell_4)$ with each $\ell_i\geq2$ satisfying $\ell_1\ell_2\ell_3\ell_4=n/6$. For each valid tuple, we construct the quotient ring
\begin{align}
S=\mathbb{F}_2[w,x,y,z]\Big/\bigl\langle
w^{\ell_1}-1,\,x^{\ell_2}-1,\,y^{\ell_3}-1,\,z^{\ell_4}-1
\bigr\rangle,
\end{align}
whose monomials $w^a x^b y^c z^e$ with $0\leq a<\ell_1$, $0\leq b<\ell_2$, $0\leq c<\ell_3$, $0\leq e<\ell_4$ form a standard basis of the $\mathbb{F}_2$-vector space of dimension $\ell_1\ell_2\ell_3\ell_4$. For each tuple we draw $n_{\mathrm{samples}}$ independent random instances, each consisting of four polynomials $F,G,H,I\in S$ sampled uniformly at random as sums of exactly \revisionblue{$w_{\mathrm{poly}}$} distinct monomials chosen without replacement from the full monomial basis\revisionblue{; the retained codes have $w_{\mathrm{poly}}=2$, $3$ and $4$ in Tables~\ref{tab:table2}, \ref{tab:table3} and~\ref{tab:table5}, and the inputs of Table~\ref{tab:table4} are listed explicitly}. The corresponding MM code $\mathcal{Q}=\mathrm{MM}([\ell_1,\ell_2,\ell_3,\ell_4],[F,G,H,I])$ is then constructed, and instances yielding $k=0$ logical qubits are discarded. \revisionblue{For the collapsed codes in Tables~\ref{tab:table6}--\ref{tab:table11}, we use $t=4,\ldots,9$ generators $F_i=1+\xi^{e_i}$ on the lattice $(\ell,1,\ldots,1)$, where $\xi$ is the variable of the cyclic factor and the exponents $e_i$ are distinct and drawn uniformly from $\{1,\ldots,\ell-1\}$.} To characterize sparsity, we also record the median and maximum row weights of both parity-check matrices $P_X$ and $P_Z$. \revisionblue{All arithmetic is over $\mathbb{F}_2$: exponents are reduced modulo the corresponding $\ell_i$, and repeated monomials cancel in pairs before the check matrices and their row weights are computed.}

\revisionblue{\textit{Distance verification.} Distances were first estimated with the randomized information-set decoder \textsf{sQetch}~\cite{bhardwaj2026high} and then searched with the random-window algorithm of \texttt{dist-m4ri}~\cite{Pryadko-2025-distm4ri}, saving the logical operators found. We report an upper bound only when it is the weight of a saved logical operator that we verified to have zero syndrome and to lie outside the stabilizer row space; three such operators were supplied by an anonymous referee. To establish exact distances, we ran the exhaustive connected-cluster algorithm of \texttt{dist-m4ri} on $88$ instances: all instances with $n<200$, the codes of Table~\ref{tab:confinement} and the other instances discussed in the main text, and the explicit examples of Sec.~\ref{sec:multivariate_multicycle_codes}. A weight counts as excluded only when the search completes without finding a nontrivial logical operator, and a sector distance is exact when all weights below that of a verified logical operator are excluded. For middle-degree codes with even $t$, Proposition~\ref{prop:xz_duality} transfers both upper and lower bounds between the two sectors, and we reuse a completed search for another reported instance only after checking that both check row spaces coincide. The table entries were selected before these exhaustive searches: entries whose estimated distance was at most six were omitted, except for three main-text examples whose distances are now proven exact. Seeds, budgets, completion logs and all verified logical operators are archived in the accompanying repository~\cite{feroz_ahmed_mian_2026_24095975}.}

\section{Explicit low-weight instances of MM codes}
In this appendix, we discover several instances of weight-6 MM codes for practical quantum hardware platforms in Appendix Table~\ref{tab:table2}. We also discover several novel instances of weight-6 AMC codes using our Koszul complex framework in Table~\ref{tab:table6}. Although MM codes employ geometrically non-local checks, recent work \cite{xu2024constant} suggests such codes can be implemented efficiently on reconfigurable neutral atom arrays due to the product structure inherent in many qLDPC codes, which can be exploited to implement non-local syndrome extraction circuits via parallel atom rearrangement with crossed acousto-optic deflectors (AODs), achieving effectively constant overhead. \revisionblue{The product structure of MM codes motivates studying their implementation on} this architecture.

\revisionblue{Several other constructions also produce higher-dimensional codes, including higher-dimensional hypergraph-product codes~\cite{Zeng_2019, Tan_2025}, high-dimensional cubical complexes~\cite{10756140}, and explicit and lifted quantum Tanner codes~\cite{radebold2025explicit, guemard2025moderate}. For a concrete comparison we use the quantum Tanner instances of Ref.~\cite{wang2026check} with block lengths $60 \leq n \leq 200$, which have explicit check matrices and block lengths comparable to ours; the comparison is therefore specific to these instances. }To study check weights, we compare 5D MM codes from Table~\ref{tab:table7} against quantum Tanner code instances from Ref.~\cite{wang2026check} in Fig.~\ref{fig:checkweight} for block lengths $n \in [60, 200]$. \revisionblue{Fig.~\ref{fig:checkweight} also includes 5D MM instances that are omitted from Table~\ref{tab:table7} under the estimate-based selection described in Appendix~\ref{sec:appendix_search}; the instances retained in Table~\ref{tab:table7} for this range are $[[180, 10, (12, 7)]]$, $[[190, 10, (12, 7)]]$, $[[190, 10, (13, 7)]]$ and $[[200, 10, (\leq 12, \leq 7)]]$. The distances of all audited 5D instances with $n<200$ are proven exact by the connected-cluster search of Appendix~\ref{sec:appendix_search}. All of these codes have constant check weights ($w_X = 8$, $w_Z = 6$),} while the quantum Tanner instances are $[[64, 22, (4, 4)]]$, $[[72, 19, (4, 4)]]$, $[[96, 10, (8, 4)]]$, $[[96, 24, (4, 4)]]$, $[[96, 30, (4, 4)]]$, $[[120, 23, (4, 6)]]$, $[[128, 40, (4, 4)]]$, $[[144, 6, (10, 9)]]$, $[[144, 12, (7, 7)]]$, $[[150, 62, (4, 4)]]$, $[[180, 16, (9, 7)]]$, $[[189, 24, (8, 4)]]$, and $[[192, 16, (3, 16)]]$ with variable check weights $w_X \in [6, 16]$ and $w_Z \in [6, 12]$.

\revisionblue{In Tables~\ref{tab:table2}--\ref{tab:table11}, an entry $\leq d$ denotes an upper bound supported by a verified nontrivial logical operator, and an entry without $\leq$ denotes an exact distance, established by matching upper and lower bounds. The lower bounds come from completed connected-cluster searches with \texttt{dist-m4ri}~\cite{Pryadko-2025-distm4ri}; for middle-degree codes with even $t$, bounds also transfer between the two sectors by Proposition~\ref{prop:xz_duality}. Appendix~\ref{sec:appendix_search} describes these computations. For brevity, these tables list selected instances: all instances with exact distances, the instances discussed in the text, and representative instances spanning the range of block lengths. The complete lists, with their check matrices and distance evidence, are provided in our repository~\cite{feroz_ahmed_mian_2026_24095975}.}

\onecolumngrid
\begin{center}
\refstepcounter{figure}\label{fig:checkweight}%
\includegraphics[width=0.8\textwidth]{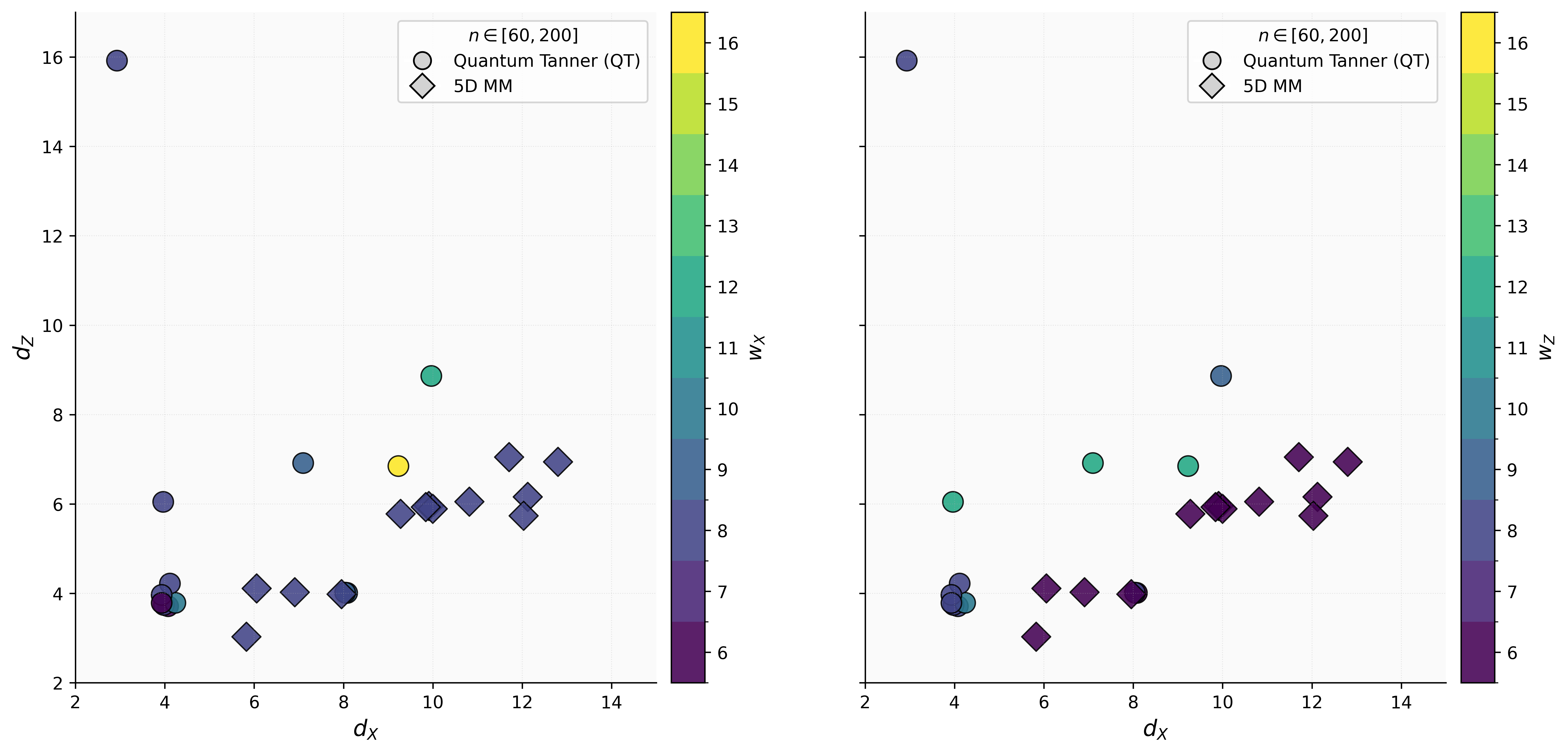}\\[4pt]
\parbox{0.86\textwidth}{\small FIG.~\thefigure.~Check weight comparison of 5D MM codes against recent small instances of quantum Tanner codes.}
\end{center}
\medskip
\definecolor{A}{RGB}{248, 248, 255}
\definecolor{B}{RGB}{245, 255, 250}
\definecolor{C}{RGB}{255, 250, 240}
\definecolor{D}{RGB}{240, 248, 255}
\definecolor{E}{RGB}{255, 245, 238}
\definecolor{F}{RGB}{240, 255, 240}
\definecolor{G}{RGB}{255, 250, 250}
\definecolor{H}{RGB}{248, 248, 247}
\definecolor{I}{RGB}{245, 245, 245}
\definecolor{J}{RGB}{250, 250, 250}
\definecolor{K}{RGB}{245, 248, 255}
\definecolor{L}{RGB}{255, 248, 240}
\definecolor{M}{RGB}{240, 252, 245}
\definecolor{N}{RGB}{250, 245, 255}
\definecolor{O}{RGB}{245, 250, 255}
\definecolor{P}{RGB}{255, 252, 245}
\definecolor{Q}{RGB}{248, 255, 248}
\definecolor{R}{RGB}{255, 248, 248}
\definecolor{S}{RGB}{245, 250, 245}
\definecolor{T}{RGB}{250, 248, 255}
\definecolor{U}{RGB}{252, 252, 250}
\definecolor{V}{RGB}{248, 250, 255}
\definecolor{W}{RGB}{255, 252, 250}
\definecolor{X}{RGB}{250, 252, 248}
\definecolor{Y}{RGB}{252, 250, 248}
\definecolor{Z}{RGB}{248, 252, 250}
\clearpage
\onecolumngrid

\begingroup
\MMTableStyle
\begin{longtable}{|*{8}{>{\columncolor{white}[\tabcolsep]}c|}}
\hline
\revisionblue{$[[n,k,(\leq d_X,\leq d_Z)]]$} & $(\ell,m,p,q)$ & $k/n$ & \revisionblue{$kd^2/n$} & $F$ & $G$ & $H$ & $I$\\ \hline
\endfirsthead
\multicolumn{8}{c}{\tablename~\thetable\ (continued)}\\
\hline
\revisionblue{$[[n,k,(\leq d_X,\leq d_Z)]]$} & $(\ell,m,p,q)$ & $k/n$ & \revisionblue{$kd^2/n$} & $F$ & $G$ & $H$ & $I$\\ \hline
\endhead
\hline
\endfoot
\caption{\textbf{Small instances of multivariate multicycle (MM) codes with stabilizer check weight $w = 6$.} The code rate is $\frac{k}{n}$, the efficiency ratio is $\frac{kd^2}{n}$\revisionblue{, with $d=\min(d_X,d_Z)$; it is an upper bound when $d$ is}.}\label{tab:table2}\\
\endlastfoot
$[[96, 12, (4,4)]]$ &  (2, 2, 2, 2)  & \revisionblue{0.125} & \revisionblue{$2.00$} &  \shortstack[c]{$wz + xyz$}  &  \shortstack[c]{$w + xyz$}  &  \shortstack[c]{$wyz + xy$}  &  \shortstack[c]{$xyz + xz$} \\ %
\hline
$[[144, 6, (8,8)]]$ &  (2, 2, 3, 2)  & \revisionblue{0.042} & \revisionblue{$2.67$} &  \shortstack[c]{$wxy + z$}  &  \shortstack[c]{$wx + wyz$}  &  \shortstack[c]{$wx + xy^2$}  &  \shortstack[c]{$xyz + y^2z$} \\ %
\hline
\revisionblue{$[[216, 6, (\leq 12,\leq 12)]]$} &  (2, 2, 3, 3)  & \revisionblue{0.028} & \revisionblue{$4.00$} &  \shortstack[c]{$wxy + xy^2$}  &  \shortstack[c]{$wy + xyz^2$}  &  \shortstack[c]{$xy^2 + y^2z$}  &  \shortstack[c]{$wxyz^2 + xz$} \\ %
\hline
$[[288, 12, (8,8)]]$ &  (2, 2, 2, 6)  & \revisionblue{0.042} & \revisionblue{$2.67$} &  \shortstack[c]{$wz^3 + xyz^2$}  &  \shortstack[c]{$wxz^2 + xyz$}  &  \shortstack[c]{$wxz^2 + w$}  &  \shortstack[c]{$wyz^5 + xyz^3$} \\ %
\hline
\revisionblue{$[[360, 6, (\leq 12,\leq 12)]]$} &  (2, 2, 3, 5)  & \revisionblue{0.017} & \revisionblue{$2.40$} &  \shortstack[c]{$xz^3 + y^2z$}  &  \shortstack[c]{$wx + wz^2$}  &  \shortstack[c]{$wxy^2 + y^2z^2$}  &  \shortstack[c]{$wxyz^3 + wxz^2$} \\ %
\hline
\revisionblue{$[[504, 6, (\leq 17,\leq 17)]]$} &  (2, 2, 3, 7)  & \revisionblue{0.012} & \revisionblue{$3.44$} &  \shortstack[c]{$wxyz^4 + wz$}  &  \shortstack[c]{$wy^2z^3 + xz^3$}  &  \shortstack[c]{$wy^2z^5 + xy^2z$}  &  \shortstack[c]{$wyz^2 + yz$} \\ %
\hline
\revisionblue{$[[600, 6, (\leq 17,\leq 17)]]$} &  (2, 2, 5, 5)  & \revisionblue{0.010} & \revisionblue{$2.89$} &  \shortstack[c]{$xy^2 + xz$}  &  \shortstack[c]{$wz^4 + yz^4$}  &  \shortstack[c]{$xz^2 + y^2z^3$}  &  \shortstack[c]{$wxz^4 + y^2z$} \\ %
\hline
$[[648, 12, (12,12)]]$ &  (2, 2, 3, 9)  & \revisionblue{0.019} & \revisionblue{$2.67$} &  \shortstack[c]{$xy^2z^4 + xy^2z^3$}  &  \shortstack[c]{$wx + xz^8$}  &  \shortstack[c]{$wz^7 + y^2z^2$}  &  \shortstack[c]{$wy^2z^2 + z^5$} \\ %
\hline
\revisionblue{$[[672, 12, (\leq 12,\leq 12)]]$} &  (2, 2, 2, 14)  & \revisionblue{0.018} & \revisionblue{$2.57$} &  \shortstack[c]{$wz^{13} + xz^9$}  &  \shortstack[c]{$wxyz^5 + yz^{13}$}  &  \shortstack[c]{$wxz^8 + wyz^{11}$}  &  \shortstack[c]{$wxz^8 + yz^{11}$} \\ %
\hline
\revisionblue{$[[672, 24, (\leq 11,\leq 11)]]$} &  (2, 2, 4, 7)  & \revisionblue{0.036} & \revisionblue{$4.32$} &  \shortstack[c]{$wy^3z + wy^3$}  &  \shortstack[c]{$wxyz^4 + wz^6$}  &  \shortstack[c]{$y^2z^2 + 1$}  &  \shortstack[c]{$wxy^3 + wz^2$} \\ %
\hline
\revisionblue{$[[792, 6, (\leq 22,\leq 22)]]$} &  (2, 2, 3, 11)  & \revisionblue{0.008} & \revisionblue{$3.67$} &  \shortstack[c]{$wxyz^2 + wxz$}  &  \shortstack[c]{$xyz^5 + y^2z^7$}  &  \shortstack[c]{$wxz^4 + z^3$}  &  \shortstack[c]{$xy^2z^3 + y^2z$} \\ %
\hline
\revisionblue{$[[792, 12, (\leq 14,\leq 14)]]$} &  (2, 2, 3, 11)  & \revisionblue{0.015} & \revisionblue{$2.97$} &  \shortstack[c]{$wxyz^3 + wxyz$}  &  \shortstack[c]{$wx + y^2z^3$}  &  \shortstack[c]{$wxyz^9 + y^2z^{10}$}  &  \shortstack[c]{$wz^5 + xy^2z^5$} \\ %
\hline
\revisionblue{$[[840, 6, (\leq 20,\leq 20)]]$} &  (2, 2, 5, 7)  & \revisionblue{0.007} & \revisionblue{$2.86$} &  \shortstack[c]{$wxy^4z + wy^4z^5$}  &  \shortstack[c]{$wy + z^5$}  &  \shortstack[c]{$wxz^2 + xy^4z^2$}  &  \shortstack[c]{$wxy^4 + wy^3z^3$} \\ %
\hline
\revisionblue{$[[864, 6, (\leq 24,\leq 24)]]$} &  (2, 2, 6, 6)  & \revisionblue{0.007} & \revisionblue{$4.00$} &  \shortstack[c]{$wy^2z^4 + wyz^3$}  &  \shortstack[c]{$xy^5z^3 + xy^4z^3$}  &  \shortstack[c]{$xy^3z + y^5z^5$}  &  \shortstack[c]{$wz^5 + 1$} \\ %
\hline
\revisionblue{$[[864, 12, (\leq 14,\leq 14)]]$} &  (2, 2, 2, 18)  & \revisionblue{0.014} & \revisionblue{$2.72$} &  \shortstack[c]{$wz^3 + xyz^7$}  &  \shortstack[c]{$wxyz^{14} + xyz^6$}  &  \shortstack[c]{$wxyz^7 + yz^{14}$}  &  \shortstack[c]{$wyz + xz^{17}$} \\ %
\hline
\revisionblue{$[[900, 12, (\leq 11,\leq 11)]]$} &  (2, 3, 5, 5)  & \revisionblue{0.013} & \revisionblue{$1.61$} &  \shortstack[c]{$wxz^4 + wyz^3$}  &  \shortstack[c]{$x^2y^4z^3 + x^2yz^2$}  &  \shortstack[c]{$wy^3z^3 + wy^3z$}  &  \shortstack[c]{$x^2y^3z^4 + xy$} \\ %
\hline
\end{longtable}
\endgroup

\bigskip

\begingroup
\MMTableStyle
\begin{longtable}{|*{7}{>{\columncolor{white}[\tabcolsep]}c|}}
\hline
\revisionblue{$[[n,k,(\leq d_X,\leq d_Z)]]$} & $(\ell,m,p,q)$ & $k/n$ & $F$ & $G$ & $H$ & $I$\\ \hline
\endfirsthead
\multicolumn{7}{c}{\tablename~\thetable\ (continued)}\\
\hline
\revisionblue{$[[n,k,(\leq d_X,\leq d_Z)]]$} & $(\ell,m,p,q)$ & $k/n$ & $F$ & $G$ & $H$ & $I$\\ \hline
\endhead
\hline
\endfoot
\caption{\textbf{Small instances of multivariate multicycle (MM) codes with stabilizer weight $w = 9$.}}\label{tab:table3}\\
\endlastfoot
$[[144, 12, (8,8)]]$ &  3, 2, 2, 2  & \revisionblue{0.083} &  \shortstack[c]{$w^2xyz + wy$\\$+ yz$}  &  \shortstack[c]{$w^2y + wxz$\\$+ x$}  &  \shortstack[c]{$w^2x + wz$\\$+ xy$}  &  \shortstack[c]{$w^2yz + w$\\$+ 1$} \\ %
\hline
$[[144, 12, (12,12)]]$ &  3, 2, 2, 2  & \revisionblue{0.083} &  \shortstack[c]{$w^2xy + w$\\$+ xyz$}  &  \shortstack[c]{$w^2x + wxy$\\$+ z$}  &  \shortstack[c]{$w^2xz + wx$\\$+ xyz$}  &  \shortstack[c]{$w^2yz + wxy$\\$+ xyz$} \\ %
\hline
$[[216, 12, (12,12)]]$ &  2, 2, 3, 3  & \revisionblue{0.056} &  \shortstack[c]{$wxy^2 + wz^2$\\$+ x$}  &  \shortstack[c]{$wxyz^2 + wy^2z$\\$+ x$}  &  \shortstack[c]{$wxy^2z^2 + xyz^2$\\$+ xz^2$}  &  \shortstack[c]{$wz + w$\\$+ xy^2z$} \\ %
\hline
\revisionblue{$[[216, 24, (\leq 9,\leq 9)]]$} &  2, 2, 3, 3  & \revisionblue{0.111} &  \shortstack[c]{$wy^2z^2 + x$\\$+ yz^2$}  &  \shortstack[c]{$wxz^2 + xy^2z$\\$+ y$}  &  \shortstack[c]{$wxy^2z + wxyz^2$\\$+ wxz^2$}  &  \shortstack[c]{$wyz + xy^2$\\$+ z$} \\ %
\hline
\revisionblue{$[[324, 12, (\leq 18,\leq 18)]]$} &  2, 3, 3, 3  & \revisionblue{0.037} &  \shortstack[c]{$wx^2y + x^2y^2$\\$+ z$}  &  \shortstack[c]{$wy + xy$\\$+ y^2z^2$}  &  \shortstack[c]{$wx^2z + wyz$\\$+ yz^2$}  &  \shortstack[c]{$wx^2y^2 + wxyz^2$\\$+ xy^2z$} \\ %
\hline
\revisionblue{$[[432, 12, (\leq 21,\leq 21)]]$} &  2, 3, 3, 4  & \revisionblue{0.028} &  \shortstack[c]{$wx^2yz^3 + wx^2z $\\$ + y^2$}  &  \shortstack[c]{$wxz^2 + x^2y $\\$ + y^2z^2$}  &  \shortstack[c]{$wxz^2 + wy $\\$ + xy^2z$}  &  \shortstack[c]{$wy + wz $\\$ + x^2y^2z^3$}\\ %
\hline
$[[486, 24, (12,12)]]$ &  3, 3, 3, 3  & \revisionblue{0.049} &  \shortstack[c]{$wyz + y^2z^2   $\\$+ y^2$}  &  \shortstack[c]{$w^2z^2 + x^2y^2z^2   $\\$+ z$}  &  \shortstack[c]{$w^2x^2z^2 + wx^2y^2z   $\\$+ wy^2z$}  &  \shortstack[c]{$w^2x^2y^2z + wz   $\\$+ x^2z^2$}\\ %
\hline
\revisionblue{$[[648, 12, (\leq 22,\leq 22)]]$} &  3, 9, 2, 2  & \revisionblue{0.019} &  \shortstack[c]{$w^2x^4yz + w^2x^2y$ \\ $+ wx^5yz$}  &  \shortstack[c]{$w^2xz + wx^7yz$ \\$+ wx^5yz$}  &  \shortstack[c]{$wx^5yz + x^8y$ $+ y$}  &  \shortstack[c]{$w^2x^8z + wx^8yz$ \\$ +x^5$}\\ %
\hline
\revisionblue{$[[648, 12, (\leq 24,\leq 24)]]$} &  3, 3, 3, 4  & \revisionblue{0.019} &  \shortstack[c]{$w^2x^2z + w^2yz^2$\\ $+z^3$}  &  \shortstack[c]{$w^2x^2z^2 + w^2xyz^2$ \\$+x^2z^2$}  &  \shortstack[c]{$w^2x^2y^2z^2 + x^2yz$ \\$+y^2z$}  &  \shortstack[c]{$w^2x^2y^2 + wx^2yz^3$ \\ $+x^2yz^3$} \\ %
\hline
\revisionblue{$[[648, 12, (\leq 31,\leq 31)]]$} &  3, 3, 3, 4  & \revisionblue{0.019} &  \shortstack[c]{$w^2y^2z + wy^2 $\\$+ yz$}  &  \shortstack[c]{$1 + w^2xy^2z^2 $\\$+ wx^2y^2z^3$}  &  \shortstack[c]{$wx^2y + wxy^2z $\\$+ x^2y^2$}  &  \shortstack[c]{$w^2yz^2 + wxz $\\$ + wz$}\\ %
\hline
\revisionblue{$[[672, 18, (\leq 20,\leq 20)]]$} &  2, 2, 2, 14  & \revisionblue{0.027} &  \shortstack[c]{$xz^5 + xz^3 $\\$+ z^{13}$}  &  \shortstack[c]{$wyz^6 + wz^5 $\\$+xyz^3$}  &  \shortstack[c]{$wxz^6 + z^{10} $\\$+ z^5$}  &  \shortstack[c]{$wxyz^{10} + wxz^4$\\$ + yz$}\\ %
\hline
\revisionblue{$[[840, 18, (\leq 46,\leq 46)]]$} &  2, 2, 5, 7  & \revisionblue{0.021} &  \shortstack[c]{$wxy^4z^3 + wxy + z^2$}  &  \shortstack[c]{$wxyz + wy^2 $\\$+ yz^5$}  &  \shortstack[c]{$wz^4 + y^3z^3 $\\$+ yz$}  &  \shortstack[c]{$wxy^3z^6 + wxy^2z^2 $\\$+ wy^4z$}\\ %
\hline
\revisionblue{$[[864, 12, (\leq 18,\leq 18)]]$} &  2, 2, 3, 12  & \revisionblue{0.014} &  \shortstack[c]{$wy^2z^5 + xz^{11}$\\$+ z^3$}  &  \shortstack[c]{$wxz^{10} + wy^2z^4$\\$+ xz^5$}  &  \shortstack[c]{$wxyz^5 + wyz^9$\\$+ yz^7$}  &  \shortstack[c]{$wxz^4 + wyz^{11}$\\$+ z^{11}$} \\ %
\hline
\revisionblue{$[[864, 12, (\leq 48,\leq 48)]]$} &  2, 2, 3, 12  & \revisionblue{0.014} &  \shortstack[c]{$wxyz^4 + xyz^9 $\\$+ y^2z^6$}  &  \shortstack[c]{$wy + xz^6 $\\$+ z^7$}  &  \shortstack[c]{$wyz^4 + yz^6 $\\$+ yz^5$}  &  \shortstack[c]{$wxyz^{11} + wxz^2 $\\$+ wx$}\\ %
\hline
\end{longtable}
\endgroup

\bigskip

\begingroup
\MMTableStyle
\begin{longtable}{|*{9}{>{\columncolor{white}[\tabcolsep]}c|}}
\hline
\revisionblue{$[[n,k,(\leq d_X,\leq d_Z)]]$} & $(\ell,m,p,q)$ & $k/n$ & $w^{\mathrm{med}}$ & $w^{\max}$ & $F$ & $G$ & $H$ & $I$\\ \hline
\endfirsthead
\multicolumn{9}{c}{\tablename~\thetable\ (continued)}\\
\hline
\revisionblue{$[[n,k,(\leq d_X,\leq d_Z)]]$} & $(\ell,m,p,q)$ & $k/n$ & $w^{\mathrm{med}}$ & $w^{\max}$ & $F$ & $G$ & $H$ & $I$\\ \hline
\endhead
\hline
\endfoot
\caption{\textbf{Parameters and explicit instances of MM codes \revisionblue{with metachecks for both types of checks}.} Here, $n$ is the number of physical qubits, $k$ the number of logical qubits, \revisionblue{the pairs give the $X$-type and $Z$-type distances or upper bounds on them, and $w^{\mathrm{med}}$ and $w^{\max}$ are the median and maximum row weights of the stabilizer checks}. The multivariate quotient ring polynomials $F(w,x,y,z)$, $G(w,x,y,z)$, $H(w,x,y,z)$, and $I(w,x,y,z)$ are defined over $\mathbb{F}_2[w, x, y, z]/\langle w^\ell-1, x^m-1, y^p-1, z^r-1 \rangle$.}\label{tab:table4}\\
\endlastfoot
$[[96, 44, (4,4)]]$ &  2, 2, 2, 2  & \revisionblue{0.458} & \revisionblue{12} & \revisionblue{12} &  $(1 + x)(1 + yz)$  &  $(1 + y)(1 + zw)$  &  $(1 + z)(1 + wx)$  &  $(1 + w)(1 + xy)$ \\ %
\hline
$[[144, 12, (8,8)]]$ &  2, 2, 2, 3  & \revisionblue{0.083} & \revisionblue{9} & \revisionblue{10} &  \shortstack[c]{$wxy + xyz$}  &  \shortstack[c]{$y + zx$\\$+ yx + zw$}  &  \shortstack[c]{$x + zyxw$}  &  \shortstack[c]{$z + y$\\$+ x + zyx$} \\ %
\hline
\revisionblue{$[[192, 12, (\leq 12,\leq 12)]]$} &  2, 2, 2, 4  & \revisionblue{0.062} & \revisionblue{12} & \revisionblue{12} &  \shortstack[c]{$wxyz^3 + xyz $\\$+ xz^3 + xz^2$}  &  \shortstack[c]{$wxz^2 + wyz^3 $\\$+ w + 1$}  &  \shortstack[c]{$wy + wz^2 $\\$+ xyz^2 + xz$}  &  \shortstack[c]{$wxyz + wy $\\$+ xz^3 + z^3$} \\ %
\hline
$[[216, 12, (12,12)]]$ &  2, 2, 3, 3  & \revisionblue{0.056} & \revisionblue{9} & \revisionblue{10} &  \shortstack[c]{$wxy + xyz$}  &  \shortstack[c]{$y + zx$\\$+ yx + zw$}  &  \shortstack[c]{$x + zyxw$}  &  \shortstack[c]{$z + y$\\$+ x + zyx$} \\ %
\hline
\revisionblue{$[[216, 12, (\leq 14,\leq 14)]]$} &  2, 2, 3, 3  & \revisionblue{0.056} & \revisionblue{12} & \revisionblue{12} &  \shortstack[c]{$wz^2 + xy^2z $\\$+ xy^2 + y $}  &  \shortstack[c]{$wxy^2z^2 + wxz^2 $\\$+ wy^2z + y^2z^2$}  &  \shortstack[c]{$wxy^2z^2 + wxyz $\\$+ xy^2z + x$}  &  \shortstack[c]{$wxz^2 + wyz $\\$+ wz^2 + xz$} \\ %
\hline
\revisionblue{$[[240, 6, (\leq 20,\leq 20)]]$} &  2, 2, 2, 5  & \revisionblue{0.025} & \revisionblue{12} & \revisionblue{12} &  \shortstack[c]{$wxz^4 + wyz^3 $\\$+ wy + xyz^4$}  &  \shortstack[c]{$wyz^4 + wz $\\$+ w + xz^3$}  &  \shortstack[c]{$wxz + wyz^4 $\\$+ wz^4 + xz^4$}  &  \shortstack[c]{$wxyz^4 + wxyz^3 $\\$+ wyz + w$} \\ %
\hline
\revisionblue{$[[336, 6, (\leq 28,\leq 28)]]$} &  2, 2, 2, 7  & \revisionblue{0.018} & \revisionblue{12} & \revisionblue{12} &  \shortstack[c]{$wxyz^4 + wz^4$ \\$+ yz^6 + yz^3$}  &  \shortstack[c]{$wyz^6 + w$ \\$+ xz^3 + yz^5$}  &  \shortstack[c]{$wz^3 + xyz^6$ \\$+ xz^3 + xz$}  &  \shortstack[c]{$wxz^4 + wz^2 $\\$+ xyz^2 + 1$}\\ %
\hline
\revisionblue{$[[432, 12, (\leq 18,\leq 18)]]$} &  2, 2, 2, 9  & \revisionblue{0.028} & \revisionblue{12} & \revisionblue{12} &  \shortstack[c]{$wxyz^6 + wxyz$ \\$+ y + z^8$}  &  \shortstack[c]{$wxyz^5 + wyz^8$ \\$+ xz^2 + yz^3$}  &  \shortstack[c]{$wxz^2 + wz^3$ \\$+ xyz^2 + yz^6$}  &  \shortstack[c]{$wyz^3 + xz^7$ \\$+ xz^2 + x$} \\ %
\hline
$[[486, 24, (12,12)]]$ &  3, 3, 3, 3  & \revisionblue{0.049} & \revisionblue{9} & \revisionblue{9} &  \shortstack[c]{$1 + wx$\\$+ x^2y$}  &  \shortstack[c]{$1 + xy$\\$+ y^2z$}  &  \shortstack[c]{$1 + yz$\\$+ wz^2$}  &  \shortstack[c]{$1 + wz$\\$+ w^2x$} \\ %
\hline
$[[486, 66, (9,9)]]$ &  3, 3, 3, 3  & \revisionblue{0.136} & \revisionblue{12} & \revisionblue{12} &  $(1 + x)(1 + yz)$  &  $(1 + y)(1 + zw)$  &  $(1 + z)(1 + wx)$  &  $(1 + w)(1 + xy)$ \\ %
\hline
$[[576, 64, (6,6)]]$ &  2, 3, 4, 4  & \revisionblue{0.111} & \revisionblue{12} & \revisionblue{12} &  $(1 + x)(1 + yz)$  &  $(1 + y)(1 + zw)$  &  $(1 + z)(1 + wx)$  &  $(1 + w)(1 + xy)$ \\ %
\hline
\revisionblue{$[[648, 60, (9,9)]]$} &  3, 3, 3, 4  & \revisionblue{0.093} & \revisionblue{12} & \revisionblue{12} &  $(1 + x)(1 + yz)$  &  $(1 + y)(1 + zw)$  &  $(1 + z)(1 + wx)$  &  $(1 + w)(1 + xy)$ \\ %
\hline
\revisionblue{$[[672, 12, (\leq 28,\leq 28)]]$} &  2, 2, 4, 7  & \revisionblue{0.018} & \revisionblue{12} & \revisionblue{12} &  \shortstack[c]{$wxz^3 + wyz^4 $\\$+ xy^3 + z^4$}  &  \shortstack[c]{$wxy^2z^4 + wxyz^2 $\\$+ wz^3 + yz$}  &  \shortstack[c]{$wxy^3z^4 + wxy^2z^6 $\\$+ wyz^6 + yz^6$}  &  \shortstack[c]{$wxyz + x $\\$+ y^2z^5 + yz^4$} \\ %
\hline
\revisionblue{$[[768, 12, (\leq 30,\leq 30)]]$} &  2, 2, 2, 16  & \revisionblue{0.016} & \revisionblue{12} & \revisionblue{12} &  \shortstack[c]{$wyz^{13} + wyz^{12} $\\$+ xyz^{11} + z^{15}$}  &  \shortstack[c]{$wxz^2 + wz^6 + $\\$xyz^{11} + yz^{12}$}  &  \shortstack[c]{$wxyz^{11} + xyz^{10} $\\$+xz^7 + z^{14}$}  &  \shortstack[c]{$wxy + wxz^9 $\\$+ wy + yz$} \\ %
\hline
\end{longtable}
\endgroup

\bigskip

\begingroup
\MMTableStyle
\begin{longtable}{|*{7}{>{\columncolor{white}[\tabcolsep]}c|}}
\hline
\revisionblue{$[[n,k,(\leq d_X,\leq d_Z)]]$} & $(\ell,m,p,q)$ & $k/n$ & $F$ & $G$ & $H$ & $I$\\ \hline
\endfirsthead
\multicolumn{7}{c}{\tablename~\thetable\ (continued)}\\
\hline
\revisionblue{$[[n,k,(\leq d_X,\leq d_Z)]]$} & $(\ell,m,p,q)$ & $k/n$ & $F$ & $G$ & $H$ & $I$\\ \hline
\endhead
\hline
\endfoot
\caption{\textbf{Multivariate multicycle (MM) codes with stabilizer weight $w = 12$.}}\label{tab:table5}\\
\endlastfoot
$[[96, 12, (8,8)]]$ &  2, 2, 2, 2  & \revisionblue{0.125} &  \shortstack[c]{$wy + wz$\\$+ w + 1$}  &  \shortstack[c]{$wxz + wy$\\$+ wz + w$}  &  \shortstack[c]{$wxyz + w$\\$+ xyz + y$}  &  \shortstack[c]{$wxyz + w$\\$+ yz + 1$} \\ %
\hline
$[[144, 6, (8,8)]]$ &  3, 2, 2, 2  & \revisionblue{0.042} &  \shortstack[c]{$wxyz + wxy$\\$+ wy + yz$}  &  \shortstack[c]{$w^2z + xyz$\\$+ xy + yz$}  &  \shortstack[c]{$w^2x + w^2yz$\\$+ w^2 + 1$}  &  \shortstack[c]{$w^2z + wy$\\$+ w + xz$} \\ %
\hline
$[[144, 6, (10,10)]]$ &  3, 2, 2, 2  & \revisionblue{0.042} &  \shortstack[c]{$w^2xz + w^2yz$\\$+ wx + z$}  &  \shortstack[c]{$w^2yz + wxz$\\$+ w + xy$}  &  \shortstack[c]{$wxy + wy$\\$+ x + y$}  &  \shortstack[c]{$wxz + xy$\\$+ x + z$} \\ %
\hline
\revisionblue{$[[144, 6, (\leq 12,\leq 12)]]$} &  3, 2, 2, 2  & \revisionblue{0.042} &  \shortstack[c]{$w^2 + wyz$\\$+ wz + w$}  &  \shortstack[c]{$w^2z + wxy$\\$+ yz + z$}  &  \shortstack[c]{$w^2y + wz$\\$+ w + 1$}  &  \shortstack[c]{$w^2z + wx$\\$+ wy + 1$} \\ %
\hline
$[[144, 12, (8,8)]]$ &  2, 3, 2, 2  & \revisionblue{0.083} &  \shortstack[c]{$x^2z + x^2$\\$+ yz + z$}  &  \shortstack[c]{$wx^2y + w$\\$+ x^2 + yz$}  &  \shortstack[c]{$wxy + w$\\$+ x^2y + z$}  &  $wxyz + wxy + x^2y + xz$\\ %
\hline
$[[144, 12, (10,10)]]$ &  3, 2, 2, 2  & \revisionblue{0.083} &  \shortstack[c]{$w^2x + xyz$\\$+ x + y$}  &  \shortstack[c]{$w^2y + wxy$\\$+ wx + z$}  &  \shortstack[c]{$w^2yz + w^2$\\$+ yz + z$}  &  $w^2x + wxyz + wyz + 1$\\ %
\hline
\revisionblue{$[[144, 12, (\leq 12,\leq 12)]]$} &  3, 2, 2, 2  & \revisionblue{0.083} &  \shortstack[c]{$w^2xz + w^2yz$\\$+ wy + xy$}  &  \shortstack[c]{$wxy + wyz$\\$+ wz + xy$}  &  \shortstack[c]{$w^2xy + w^2$\\$+ wyz + 1$}  &  $w^2xz + w^2z + wxz + xy$\\ %
\hline
\end{longtable}
\endgroup

\bigskip

\begingroup
\MMTableStyle
\begin{longtable}{|*{7}{>{\columncolor{white}[\tabcolsep]}c|}}
\hline
\revisionblue{$[[n,k,(\leq d_X,\leq d_Z)]]$} & $\ell$ & $k/n$ & $F$ & $G$ & $H$ & $I$\\ \hline
\endfirsthead
\multicolumn{7}{c}{\tablename~\thetable\ (continued)}\\
\hline
\revisionblue{$[[n,k,(\leq d_X,\leq d_Z)]]$} & $\ell$ & $k/n$ & $F$ & $G$ & $H$ & $I$\\ \hline
\endhead
\hline
\endfoot
\caption{\textbf{Collapsed 4-dimensional AMC codes via the Multivariate Multicycle (MM) framework.} \revisionblue{These codes include instances related to Ref.~\cite{lin2025abelianmulticyclecodessingleshot}; the distances given here are from our own search.} The multivariate quotient ring polynomials $F(w,x,y,z)$, $G(w,x,y,z)$, $H(w,x,y,z)$, and $I(w,x,y,z)$ are defined over $\mathbb{F}_2[w, x, y, z]/\langle w^\ell-1, x-1, y-1, z-1 \rangle$. \revisionblue{All these codes have weight-6 $X$ and $Z$ stabilizer checks.}}\label{tab:table6}\\
\endlastfoot
$[[162, 6, (9,9)]]$ &  27  & \revisionblue{0.037} &  \shortstack[c]{$1 + w^{7}$}  &  \shortstack[c]{$1 + w^{13}$}  &  \shortstack[c]{$1 + w^{15}$}  &  \shortstack[c]{$1 + w^{24}$} \\ %
\hline
$[[168, 6, (9,9)]]$ &  28  & \revisionblue{0.036} &  \shortstack[c]{$1 + w^{4}$}  &  \shortstack[c]{$1 + w^{7}$}  &  \shortstack[c]{$1 + w^{13}$}  &  \shortstack[c]{$1 + w^{19}$} \\ %
\hline
$[[168, 12, (7,7)]]$ &  28  & \revisionblue{0.071} &  \shortstack[c]{$1 + w^{4}$}  &  \shortstack[c]{$1 + w^{6}$}  &  \shortstack[c]{$1 + w^{8}$}  &  \shortstack[c]{$1 + w^{18}$} \\ %
\hline
$[[174, 6, (11,11)]]$ &  29  & \revisionblue{0.034} &  \shortstack[c]{$1 + w^{2}$}  &  \shortstack[c]{$1 + w^{8}$}  &  \shortstack[c]{$1 + w^{20}$}  &  \shortstack[c]{$1 + w^{24}$} \\ %
\hline
$[[180, 6, (10,10)]]$ &  30  & \revisionblue{0.033} &  \shortstack[c]{$1 + w^{3}$}  &  \shortstack[c]{$1 + w^{13}$}  &  \shortstack[c]{$1 + w^{21}$}  &  \shortstack[c]{$1 + w^{28}$} \\ %
\hline
$[[186, 6, (11,11)]]$ &  31  & \revisionblue{0.032} &  \shortstack[c]{$1 + w$}  &  \shortstack[c]{$1 + w^{14}$}  &  \shortstack[c]{$1 + w^{19}$}  &  \shortstack[c]{$1 + w^{23}$} \\ %
\hline
\revisionblue{$[[216, 6, (\leq 12,\leq 12)]]$} &  36  & \revisionblue{0.028} &  \shortstack[c]{$1 + w^{3}$}  &  \shortstack[c]{$1 + w^{21}$}  &  \shortstack[c]{$1 + w^{26}$}  &  \shortstack[c]{$1 + w^{35}$} \\ %
\hline
\revisionblue{$[[246, 6, (\leq 12,\leq 12)]]$} &  41  & \revisionblue{0.024} &  \shortstack[c]{$1 + w$}  &  \shortstack[c]{$1 + w^{12}$}  &  \shortstack[c]{$1 + w^{23}$}  &  \shortstack[c]{$1 + w^{37}$} \\ %
\hline
\revisionblue{$[[300, 6, (\leq 13,\leq 13)]]$} &  50  & \revisionblue{0.020} &  \shortstack[c]{$1 + w^{12}$}  &  \shortstack[c]{$1 + w^{16}$}  &  \shortstack[c]{$1 + w^{21}$}  &  \shortstack[c]{$1 + w^{23}$} \\ %
\hline
\revisionblue{$[[360, 6, (\leq 15,\leq 15)]]$} &  60  & \revisionblue{0.017} &  \shortstack[c]{$1 + w^{4}$}  &  \shortstack[c]{$1 + w^{18}$}  &  \shortstack[c]{$1 + w^{25}$}  &  \shortstack[c]{$1 + w^{44}$} \\ %
\hline
\revisionblue{$[[420, 6, (\leq 15,\leq 15)]]$} &  70  & \revisionblue{0.014} &  \shortstack[c]{$1 + w^{2}$}  &  \shortstack[c]{$1 + w^{13}$}  &  \shortstack[c]{$1 + w^{32}$}  &  \shortstack[c]{$1 + w^{56}$} \\ %
\hline
\revisionblue{$[[474, 6, (\leq 17,\leq 17)]]$} &  79  & \revisionblue{0.013} &  \shortstack[c]{$1 + w^{7}$}  &  \shortstack[c]{$1 + w^{10}$}  &  \shortstack[c]{$1 + w^{12}$}  &  \shortstack[c]{$1 + w^{61}$} \\ %
\hline
\end{longtable}
\endgroup

\bigskip

\begingroup
\MMTableStyle
\begin{longtable}{|*{8}{>{\columncolor{white}[\tabcolsep]}c|}}
\hline
\revisionblue{$[[n,k,(\leq d_X,\leq d_Z)]]$} & $\ell$ & $k/n$ & $F$ & $G$ & $H$ & $I$ & $J$\\ \hline
\endfirsthead
\multicolumn{8}{c}{\tablename~\thetable\ (continued)}\\
\hline
\revisionblue{$[[n,k,(\leq d_X,\leq d_Z)]]$} & $\ell$ & $k/n$ & $F$ & $G$ & $H$ & $I$ & $J$\\ \hline
\endhead
\hline
\endfoot
\caption{\textbf{Collapsed 5-dimensional (5D) multivariate multicycle (MM) codes.} The multivariate quotient ring polynomials $F(v,w,x,y,z)$, $G(v,w,x,y,z)$, $H(v,w,x,y,z)$, $I(v,w,x,y,z)$ and $J(v,w,x,y,z)$ are defined over $\mathbb{F}_2[v,w, x, y, z]/\langle v^\ell-1, w-1, x-1, y-1, z-1 \rangle$. Notably, all these 5D MM codes have weight-8 X stabilizer checks and weight-6 Z stabilizer checks.}\label{tab:table7}\\
\endlastfoot
$[[180, 10, (12,7)]]$ &  18  & \revisionblue{0.056} &  \shortstack[c]{$1 + v$}  &  \shortstack[c]{$1 + v^5$}  &  \shortstack[c]{$1 + v^{10}$}  &  \shortstack[c]{$1 + v^{11}$}  &  \shortstack[c]{$1 + v^{14}$} \\ %
\hline
$[[190, 10, (12,7)]]$ &  19  & \revisionblue{0.053} &  \shortstack[c]{$1 + v$}  &  \shortstack[c]{$1 + v^4$}  &  \shortstack[c]{$1 + v^{5}$}  &  \shortstack[c]{$1 + v^{11}$}  &  \shortstack[c]{$1 + v^{17}$} \\ %
\hline
$[[190, 10, (13,7)]]$ &  19  & \revisionblue{0.053} &  \shortstack[c]{$1 + v^4$}  &  \shortstack[c]{$1 + v^{10}$}  &  \shortstack[c]{$1 + v^{12}$}  &  \shortstack[c]{$1 + v^{16}$}  &  \shortstack[c]{$1 + v^{18}$} \\ %
\hline
\revisionblue{$[[200, 10, (\leq 12,\leq 7)]]$} &  20  & \revisionblue{0.050} &  \shortstack[c]{$1 + v$}  &  \shortstack[c]{$1 + v^2$}  &  \shortstack[c]{$1 + v^{9}$}  &  \shortstack[c]{$1 + v^{14}$}  &  \shortstack[c]{$1 + v^{15}$} \\ %
\hline
\revisionblue{$[[210, 10, (\leq 14,\leq 7)]]$} &  21  & \revisionblue{0.048} &  \shortstack[c]{$1 + v^3$}  &  \shortstack[c]{$1 + v^4$}  &  \shortstack[c]{$1 + v^{8}$}  &  \shortstack[c]{$1 + v^{10}$}  &  \shortstack[c]{$1 + v^{15}$} \\ %
\hline
\revisionblue{$[[250, 10, (\leq 15,\leq 8)]]$} &  25  & \revisionblue{0.040} &  \shortstack[c]{$1 + v$}  &  \shortstack[c]{$1 + v^4$}  &  \shortstack[c]{$1 + v^{9}$}  &  \shortstack[c]{$1 + v^{15}$}  &  \shortstack[c]{$1 + v^{22}$} \\ %
\hline
\revisionblue{$[[290, 10, (\leq 15,\leq 8)]]$} &  29  & \revisionblue{0.034} &  \shortstack[c]{$1 + v$}  &  \shortstack[c]{$1 + v^{3}$}  &  \shortstack[c]{$1 + v^{5}$}  &  \shortstack[c]{$1 + v^{6}$}  &  \shortstack[c]{$1 + v^{15}$} \\ %
\hline
\revisionblue{$[[380, 10, (\leq 20,\leq 8)]]$} &  38  & \revisionblue{0.026} &  \shortstack[c]{$1 + v^{2}$}  &  \shortstack[c]{$1 + v^{9}$}  &  \shortstack[c]{$1 + v^{16}$}  &  \shortstack[c]{$1 + v^{17}$}  &  \shortstack[c]{$1 + v^{35}$} \\ %
\hline
\revisionblue{$[[460, 10, (\leq 22,\leq 10)]]$} &  46  & \revisionblue{0.022} &  \shortstack[c]{$1 + v^{9}$}  &  \shortstack[c]{$1 + v^{16}$}  &  \shortstack[c]{$1 + v^{20}$}  &  \shortstack[c]{$1 + v^{42}$}  &  \shortstack[c]{$1 + v^{43}$} \\ %
\hline
\revisionblue{$[[540, 10, (\leq 25,\leq 12)]]$} &  54  & \revisionblue{0.019} &  \shortstack[c]{$1 + v$}  &  \shortstack[c]{$1 + v^{8}$}  &  \shortstack[c]{$1 + v^{37}$}  &  \shortstack[c]{$1 + v^{41}$}  &  \shortstack[c]{$1 + v^{51}$} \\ %
\hline
\revisionblue{$[[610, 10, (\leq 26,\leq 11)]]$} &  61  & \revisionblue{0.016} &  \shortstack[c]{$1 + v^{23}$}  &  \shortstack[c]{$1 + v^{32}$}  &  \shortstack[c]{$1 + v^{39}$}  &  \shortstack[c]{$1 + v^{43}$}  &  \shortstack[c]{$1 + v^{49}$} \\ %
\hline
\revisionblue{$[[700, 10, (\leq 27,\leq 11)]]$} &  70  & \revisionblue{0.014} &  \shortstack[c]{$1 + v^{26}$}  &  \shortstack[c]{$1 + v^{36}$}  &  \shortstack[c]{$1 + v^{41}$}  &  \shortstack[c]{$1 + v^{46}$}  &  \shortstack[c]{$1 + v^{59}$} \\ %
\hline
\revisionblue{$[[780, 10, (\leq 26,\leq 11)]]$} &  78  & \revisionblue{0.013} &  \shortstack[c]{$1 + v^{43}$}  &  \shortstack[c]{$1 + v^{48}$}  &  \shortstack[c]{$1 + v^{50}$}  &  \shortstack[c]{$1 + v^{59}$}  &  \shortstack[c]{$1 + v^{72}$} \\ %
\hline
\revisionblue{$[[860, 10, (\leq 28,\leq 12)]]$} &  86  & \revisionblue{0.012} &  \shortstack[c]{$1 + v^{20}$}  &  \shortstack[c]{$1 + v^{23}$}  &  \shortstack[c]{$1 + v^{49}$}  &  \shortstack[c]{$1 + v^{55}$}  &  \shortstack[c]{$1 + v^{64}$} \\ %
\hline
\end{longtable}
\endgroup

\bigskip

\begingroup
\MMTableStyle
\begin{longtable}{|*{9}{>{\columncolor{white}[\tabcolsep]}c|}}
\hline
\revisionblue{$[[n,k,(\leq d_X,\leq d_Z)]]$} & $\ell$ & $k/n$ & $F$ & $G$ & $H$ & $I$ & $J$ & $K$\\ \hline
\endfirsthead
\multicolumn{9}{c}{\tablename~\thetable\ (continued)}\\
\hline
\revisionblue{$[[n,k,(\leq d_X,\leq d_Z)]]$} & $\ell$ & $k/n$ & $F$ & $G$ & $H$ & $I$ & $J$ & $K$\\ \hline
\endhead
\hline
\endfoot
\caption{\textbf{Collapsed 6-dimensional (6D) multivariate multicycle (MM) codes.} The multivariate quotient ring polynomials $F(u,v,w,x,y,z)$, $G(u,v,w,x,y,z)$, $H(u,v,w,x,y,z)$, $I(u,v,w,x,y,z)$, $J(u,v,w,x,y,z)$ and $K(u,v,w,x,y,z)$ are defined over $\mathbb{F}_2[u,v,w, x, y, z]/\langle u^\ell-1,v-1, w-1, x-1, y-1, z-1 \rangle$. All these 6D MM codes have weight-8 X stabilizer checks and weight-8 Z stabilizer checks.}\label{tab:table8}\\
\endlastfoot
\revisionblue{$[[200, 20, (\leq 7,\leq 7)]]$} &  10  & \revisionblue{0.100} &  \shortstack[c]{$1 + u$}  &  \shortstack[c]{$1 + u^3$}  &  \shortstack[c]{$1 + u^6$}  &  \shortstack[c]{$1 + u^7$}  &  \shortstack[c]{$1 + u^8$}  &  \shortstack[c]{$1 + u^9$} \\ %
\hline
\revisionblue{$[[340, 20, (\leq 10,\leq 10)]]$} &  17  & \revisionblue{0.059} &  \shortstack[c]{$1 + u^5$}  &  \shortstack[c]{$1 + u^7$}  &  \shortstack[c]{$1 + u^9$}  &  \shortstack[c]{$1 + u^{11}$}  &  \shortstack[c]{$1 + u^{14}$}  &  \shortstack[c]{$1 + u^{15}$} \\ %
\hline
\revisionblue{$[[420, 20, (\leq 12,\leq 12)]]$} &  21  & \revisionblue{0.048} &  \shortstack[c]{$1 + u^{7}$}  &  \shortstack[c]{$1 + u^{11}$}  &  \shortstack[c]{$1 + u^{13}$}  &  \shortstack[c]{$1 + u^{15}$}  &  \shortstack[c]{$1 + u^{18}$}  &  \shortstack[c]{$1 + u^{19}$} \\ %
\hline
\revisionblue{$[[600, 20, (\leq 14,\leq 14)]]$} &  30  & \revisionblue{0.033} &  \shortstack[c]{$1 + u^{11}$}  &  \shortstack[c]{$1 + u^{12}$}  &  \shortstack[c]{$1 + u^{16}$}  &  \shortstack[c]{$1 + u^{17}$}  &  \shortstack[c]{$1 + u^{21}$}  &  \shortstack[c]{$1 + u^{29}$} \\ %
\hline
\revisionblue{$[[740, 20, (\leq 18,\leq 18)]]$} &  37  & \revisionblue{0.027} &  \shortstack[c]{$1 + u^{5}$}  &  \shortstack[c]{$1 + u^{11}$}  &  \shortstack[c]{$1 + u^{20}$}  &  \shortstack[c]{$1 + u^{23}$}  &  \shortstack[c]{$1 + u^{33}$}  &  \shortstack[c]{$1 + u^{35}$} \\ %
\hline
\revisionblue{$[[880, 20, (\leq 15,\leq 15)]]$} &  44  & \revisionblue{0.023} &  \shortstack[c]{$1 + u^{13}$}  &  \shortstack[c]{$1 + u^{20}$}  &  \shortstack[c]{$1 + u^{21}$}  &  \shortstack[c]{$1 + u^{25}$}  &  \shortstack[c]{$1 + u^{26}$}  &  \shortstack[c]{$1 + u^{41}$} \\ %
\hline
\revisionblue{$[[1040, 20, (\leq 18,\leq 18)]]$} &  52  & \revisionblue{0.019} &  \shortstack[c]{$1 + u^{4}$}  &  \shortstack[c]{$1 + u^{21}$}  &  \shortstack[c]{$1 + u^{35}$}  &  \shortstack[c]{$1 + u^{42}$}  &  \shortstack[c]{$1 + u^{47}$}  &  \shortstack[c]{$1 + u^{50}$} \\ %
\hline
\revisionblue{$[[1180, 20, (\leq 22,\leq 22)]]$} &  59  & \revisionblue{0.017} &  \shortstack[c]{$1 + u^{14}$}  &  \shortstack[c]{$1 + u^{29}$}  &  \shortstack[c]{$1 + u^{36}$}  &  \shortstack[c]{$1 + u^{40}$}  &  \shortstack[c]{$1 + u^{47}$}  &  \shortstack[c]{$1 + u^{51}$} \\ %
\hline
\end{longtable}
\endgroup

\bigskip

\begingroup
\MMTableStyle
\begin{longtable}{|*{10}{>{\columncolor{white}[\tabcolsep]}c|}}
\hline
\revisionblue{$[[n,k,(\leq d_X,\leq d_Z)]]$} & $\ell$ & $k/n$ & $F$ & $G$ & $H$ & $I$ & $J$ & $K$ & $L$\\ \hline
\endfirsthead
\multicolumn{10}{c}{\tablename~\thetable\ (continued)}\\
\hline
\revisionblue{$[[n,k,(\leq d_X,\leq d_Z)]]$} & $\ell$ & $k/n$ & $F$ & $G$ & $H$ & $I$ & $J$ & $K$ & $L$\\ \hline
\endhead
\hline
\endfoot
\caption{\textbf{Collapsed 7-dimensional (7D) multivariate multicycle (MM) codes.} \revisionblue{The multivariate quotient ring polynomials $F(\tau,u,v,w,x,y,z)$, $G(\tau,u,v,w,x,y,z)$, $H(\tau,u,v,w,x,y,z)$, $I(\tau,u,v,w,x,y,z)$, $J(\tau,u,v,w,x,y,z)$, $K(\tau,u,v,w,x,y,z)$ and $L(\tau,u,v,w,x,y,z)$ are defined over $\mathbb{F}_2[\tau,u,v,w, x, y, z]/\langle \tau^\ell-1,u-1,v-1, w-1, x-1, y-1, z-1 \rangle$.} All these 7D MM codes have weight-10 X stabilizer checks and weight-8 Z stabilizer checks.}\label{tab:table9}\\
\endlastfoot
\revisionblue{$[[385, 35, (\leq 10,\leq 7)]]$} &  11  & \revisionblue{0.091} &  \revisionblue{\shortstack[c]{$1+ \tau$}}  &  \revisionblue{\shortstack[c]{$1+ \tau^2$}}  &  \revisionblue{\shortstack[c]{$1+ \tau^3$}}  &  \revisionblue{\shortstack[c]{$1+ \tau^4$}}  &  \revisionblue{\shortstack[c]{$1+ \tau^6$}}  &  \revisionblue{\shortstack[c]{$1+ \tau^8$}}  &  \revisionblue{\shortstack[c]{$1+ \tau^{10}$}} \\ %
\hline
\revisionblue{$[[490, 35, (\leq 11,\leq 7)]]$} &  14  & \revisionblue{0.071} &  \revisionblue{\shortstack[c]{$1+ \tau$}}  &  \revisionblue{\shortstack[c]{$1+ \tau^3$}}  &  \revisionblue{\shortstack[c]{$1+ \tau^7$}}  &  \revisionblue{\shortstack[c]{$1+ \tau^8$}}  &  \revisionblue{\shortstack[c]{$1+ \tau^9$}}  &  \revisionblue{\shortstack[c]{$1+ \tau^{11}$}}  &  \revisionblue{\shortstack[c]{$1+ \tau^{12}$}} \\ %
\hline
\revisionblue{$[[560, 35, (\leq 15,\leq 8)]]$} &  16  & \revisionblue{0.062} &  \revisionblue{\shortstack[c]{$1+ \tau^{2}$}}  &  \revisionblue{\shortstack[c]{$1+ \tau^{4}$}}  &  \revisionblue{\shortstack[c]{$1+ \tau^{6}$}}  &  \revisionblue{\shortstack[c]{$1+ \tau^{7}$}}  &  \revisionblue{\shortstack[c]{$1+ \tau^{11}$}}  &  \revisionblue{\shortstack[c]{$1+ \tau^{13}$}}  &  \revisionblue{\shortstack[c]{$1+ \tau^{15}$}} \\ %
\hline
\revisionblue{$[[700, 35, (\leq 16,\leq 10)]]$} &  20  & \revisionblue{0.050} &  \revisionblue{\shortstack[c]{$1+ \tau^{6}$}}  &  \revisionblue{\shortstack[c]{$1+ \tau^{9}$}}  &  \revisionblue{\shortstack[c]{$1+ \tau^{12}$}}  &  \revisionblue{\shortstack[c]{$1+ \tau^{13}$}}  &  \revisionblue{\shortstack[c]{$1+ \tau^{17}$}}  &  \revisionblue{\shortstack[c]{$1+ \tau^{18}$}}  &  \revisionblue{\shortstack[c]{$1+ \tau^{19}$}} \\ %
\hline
\revisionblue{$[[805, 35, (\leq 18,\leq 11)]]$} &  23  & \revisionblue{0.043} &  \revisionblue{\shortstack[c]{$1+ \tau^{4}$}}  &  \revisionblue{\shortstack[c]{$1+ \tau^{7}$}}  &  \revisionblue{\shortstack[c]{$1+ \tau^{10}$}}  &  \revisionblue{\shortstack[c]{$1+ \tau^{12}$}}  &  \revisionblue{\shortstack[c]{$1+ \tau^{18}$}}  &  \revisionblue{\shortstack[c]{$1+ \tau^{20}$}}  &  \revisionblue{\shortstack[c]{$1+ \tau^{22}$}} \\ %
\hline
\revisionblue{$[[945, 35, (\leq 18,\leq 9)]]$} &  27  & \revisionblue{0.037} &  \revisionblue{\shortstack[c]{$1+ \tau^{6}$}}  &  \revisionblue{\shortstack[c]{$1+ \tau^{7}$}}  &  \revisionblue{\shortstack[c]{$1+ \tau^{9}$}}  &  \revisionblue{\shortstack[c]{$1+ \tau^{10}$}}  &  \revisionblue{\shortstack[c]{$1+ \tau^{12}$}}  &  \revisionblue{\shortstack[c]{$1+ \tau^{16}$}}  &  \revisionblue{\shortstack[c]{$1+ \tau^{23}$}} \\ %
\hline
\end{longtable}
\endgroup

\bigskip

\begingroup
\MMTableStyle
\begin{longtable}{|*{11}{>{\columncolor{white}[\tabcolsep]}c|}}
\hline
\revisionblue{$[[n,k,(\leq d_X,\leq d_Z)]]$} & $\ell$ & $k/n$ & $F$ & $G$ & $H$ & $I$ & $J$ & $K$ & $L$ & $M$\\ \hline
\endfirsthead
\multicolumn{11}{c}{\tablename~\thetable\ (continued)}\\
\hline
\revisionblue{$[[n,k,(\leq d_X,\leq d_Z)]]$} & $\ell$ & $k/n$ & $F$ & $G$ & $H$ & $I$ & $J$ & $K$ & $L$ & $M$\\ \hline
\endhead
\hline
\endfoot
\caption{\textbf{Collapsed 8-dimensional (8D) multivariate multicycle (MM) codes.} \revisionblue{The multivariate quotient ring polynomials $F(s,\tau,u,v,w,x,y,z)$, $G(s,\tau,u,v,w,x,y,z)$, $H(s,\tau,u,v,w,x,y,z)$, $I(s,\tau,u,v,w,x,y,z)$, $J(s,\tau,u,v,w,x,y,z)$, $K(s,\tau,u,v,w,x,y,z)$, $L(s,\tau,u,v,w,x,y,z)$ and $M(s,\tau,u,v,w,x,y,z)$ are defined over $\mathbb{F}_2[s,\tau,u,v,w, x, y, z]/\langle s^\ell-1, \tau-1,u-1,v-1, w-1, x-1, y-1, z-1 \rangle$.} All these 8D MM codes have weight-10 X stabilizer checks and weight-10 Z stabilizer checks.}\label{tab:table10}\\
\endlastfoot
$[[630, 70, (9,9)]]$ &  9  & \revisionblue{0.111} &  \revisionblue{\shortstack[c]{$1+ s$}}  &  \revisionblue{\shortstack[c]{$1+ s^2$}}  &  \revisionblue{\shortstack[c]{$1+ s^3$}}  &  \revisionblue{\shortstack[c]{$1+ s^4$}}  &  \revisionblue{\shortstack[c]{$1+ s^5$}}  &  \revisionblue{\shortstack[c]{$1+ s^6$}}  &  \revisionblue{\shortstack[c]{$1+ s^7$}}  &  \revisionblue{\shortstack[c]{$1+ s^8$}} \\ %
\hline
\revisionblue{$[[700, 70, (\leq 8,\leq 8)]]$} &  10  & \revisionblue{0.100} &  \revisionblue{\shortstack[c]{$1+ s$}}  &  \revisionblue{\shortstack[c]{$1+ s^2$}}  &  \revisionblue{\shortstack[c]{$1+ s^3$}}  &  \revisionblue{\shortstack[c]{$1+ s^4$}}  &  \revisionblue{\shortstack[c]{$1+ s^5$}}  &  \revisionblue{\shortstack[c]{$1+ s^6$}}  &  \revisionblue{\shortstack[c]{$1+ s^7$}}  &  \revisionblue{\shortstack[c]{$1+ s^9$}} \\ %
\hline
\revisionblue{$[[980, 70, (\leq 10,\leq 10)]]$} &  14  & \revisionblue{0.071} &  \revisionblue{\shortstack[c]{$1+ s^{2}$}}  &  \revisionblue{\shortstack[c]{$1+ s^{3}$}}  &  \revisionblue{\shortstack[c]{$1+ s^{7}$}}  &  \revisionblue{\shortstack[c]{$1+ s^{9}$}}  &  \revisionblue{\shortstack[c]{$1+ s^{10}$}}  &  \revisionblue{\shortstack[c]{$1+ s^{11}$}}  &  \revisionblue{\shortstack[c]{$1+ s^{12}$}}  &  \revisionblue{\shortstack[c]{$1+ s^{13}$}} \\ %
\hline
\revisionblue{$[[1260, 70, (\leq 12,\leq 12)]]$} &  18  & \revisionblue{0.056} &  \revisionblue{\shortstack[c]{$1+ s^{4}$}}  &  \revisionblue{\shortstack[c]{$1+ s^{5}$}}  &  \revisionblue{\shortstack[c]{$1+ s^{6}$}}  &  \revisionblue{\shortstack[c]{$1+ s^{7}$}}  &  \revisionblue{\shortstack[c]{$1+ s^{11}$}}  &  \revisionblue{\shortstack[c]{$1+ s^{13}$}}  &  \revisionblue{\shortstack[c]{$1+ s^{15}$}}  &  \revisionblue{\shortstack[c]{$1+ s^{16}$}} \\ %
\hline
\revisionblue{$[[1540, 70, (\leq 12,\leq 12)]]$} &  22  & \revisionblue{0.045} &  \revisionblue{\shortstack[c]{$1+ s^{3}$}}  &  \revisionblue{\shortstack[c]{$1+ s^{5}$}}  &  \revisionblue{\shortstack[c]{$1+ s^{7}$}}  &  \revisionblue{\shortstack[c]{$1+ s^{8}$}}  &  \revisionblue{\shortstack[c]{$1+ s^{14}$}}  &  \revisionblue{\shortstack[c]{$1+ s^{15}$}}  &  \revisionblue{\shortstack[c]{$1+ s^{17}$}}  &  \revisionblue{\shortstack[c]{$1+ s^{20}$}} \\ %
\hline
\revisionblue{$[[1820, 70, (\leq 13,\leq 13)]]$} &  26  & \revisionblue{0.038} &  \revisionblue{\shortstack[c]{$1+ s^{3}$}}  &  \revisionblue{\shortstack[c]{$1+ s^{9}$}}  &  \revisionblue{\shortstack[c]{$1+ s^{10}$}}  &  \revisionblue{\shortstack[c]{$1+ s^{12}$}}  &  \revisionblue{\shortstack[c]{$1+ s^{15}$}}  &  \revisionblue{\shortstack[c]{$1+ s^{18}$}}  &  \revisionblue{\shortstack[c]{$1+ s^{19}$}}  &  \revisionblue{\shortstack[c]{$1+ s^{22}$}} \\ %
\hline
\end{longtable}
\endgroup

\bigskip

\begingroup
\MMTableStyle
\begin{longtable}{|*{12}{>{\columncolor{white}[\tabcolsep]}c|}}
\hline
\revisionblue{$[[n,k,(\leq d_X,\leq d_Z)]]$} & $\ell$ & $k/n$ & $F$ & $G$ & $H$ & $I$ & $J$ & $K$ & $L$ & $M$ & $N$\\ \hline
\endfirsthead
\multicolumn{12}{c}{\tablename~\thetable\ (continued)}\\
\hline
\revisionblue{$[[n,k,(\leq d_X,\leq d_Z)]]$} & $\ell$ & $k/n$ & $F$ & $G$ & $H$ & $I$ & $J$ & $K$ & $L$ & $M$ & $N$\\ \hline
\endhead
\hline
\endfoot
\caption{\textbf{Collapsed 9-dimensional (9D) multivariate multicycle (MM) codes.} All these 9D MM codes have weight-12 X stabilizer checks and weight-10 Z stabilizer checks. \revisionblue{The multivariate quotient ring polynomials $F(r,s,\tau,u,v,w,x,y,z)$, $G(r,s,\tau,u,v,w,x,y,z)$, $H(r,s,\tau,u,v,w,x,y,z)$, $I(r,s,\tau,u,v,w,x,y,z)$, $J(r,s,\tau,u,v,w,x,y,z)$, $K(r,s,\tau,u,v,w,x,y,z)$, $L(r,s,\tau,u,v,w,x,y,z)$, $M(r,s,\tau,u,v,w,x,y,z)$ and $N(r,s,\tau,u,v,w,x,y,z)$ are defined over $\mathbb{F}_2[r,s,\tau,u,v,w, x, y, z]/\langle r^\ell-1, s-1, \tau-1,u-1,v-1, w-1, x-1, y-1, z-1 \rangle$.}}\label{tab:table11}\\
\endlastfoot
\revisionblue{$[[1386, 126, (\leq 11,\leq 9)]]$} &  11  & \revisionblue{0.091} &  \revisionblue{\shortstack[c]{$1+ r^2$}}  &  \revisionblue{\shortstack[c]{$1+ r^3$}}  &  \revisionblue{\shortstack[c]{$1+ r^4$}}  &  \revisionblue{\shortstack[c]{$1+ r^5$}}  &  \revisionblue{\shortstack[c]{$1+ r^6$}}  &  \revisionblue{\shortstack[c]{$1+ r^7$}}  &  \revisionblue{\shortstack[c]{$1+ r^8$}}  &  \revisionblue{\shortstack[c]{$1+ r^9$}}  &  \revisionblue{\shortstack[c]{$1+ r^{10}$}} \\ %
\hline
\revisionblue{$[[1512, 126, (\leq 12,\leq 9)]]$} &  12  & \revisionblue{0.083} &  \revisionblue{\shortstack[c]{$1+ r$}}  &  \revisionblue{\shortstack[c]{$1+ r^2$}}  &  \revisionblue{\shortstack[c]{$1+ r^3$}}  &  \revisionblue{\shortstack[c]{$1+ r^5$}}  &  \revisionblue{\shortstack[c]{$1+ r^7$}}  &  \revisionblue{\shortstack[c]{$1+ r^8$}}  &  \revisionblue{\shortstack[c]{$1+ r^9$}}  &  \revisionblue{\shortstack[c]{$1+ r^{10}$}}  &  \revisionblue{\shortstack[c]{$1+ r^{11}$}} \\ %
\hline
\revisionblue{$[[1638, 126, (\leq 13,\leq 10)]]$} &  13  & \revisionblue{0.077} &  \revisionblue{\shortstack[c]{$1+ r$}}  &  \revisionblue{\shortstack[c]{$1+ r^2$}}  &  \revisionblue{\shortstack[c]{$1+ r^4$}}  &  \revisionblue{\shortstack[c]{$1+ r^5$}}  &  \revisionblue{\shortstack[c]{$1+ r^6$}}  &  \revisionblue{\shortstack[c]{$1+ r^7$}}  &  \revisionblue{\shortstack[c]{$1+ r^8$}}  &  \revisionblue{\shortstack[c]{$1+ r^{10}$}}  &  \revisionblue{\shortstack[c]{$1+ r^{11}$}} \\ %
\hline
\revisionblue{$[[1764, 126, (\leq 14,\leq 10)]]$} &  14  & \revisionblue{0.071} &  \revisionblue{\shortstack[c]{$1+ r$}}  &  \revisionblue{\shortstack[c]{$1+ r^2$}}  &  \revisionblue{\shortstack[c]{$1+ r^3$}}  &  \revisionblue{\shortstack[c]{$1+ r^4$}}  &  \revisionblue{\shortstack[c]{$1+ r^5$}}  &  \revisionblue{\shortstack[c]{$1+ r^6$}}  &  \revisionblue{\shortstack[c]{$1+ r^7$}}  &  \revisionblue{\shortstack[c]{$1+ r^{9}$}}  &  \revisionblue{\shortstack[c]{$1+ r^{13}$}} \\ %
\hline
\end{longtable}

\endgroup

\twocolumngrid
\section{\texorpdfstring{\revisionblue{Simulation details}}{Simulation details}}\label{section:single_shot_decoders}
\revisionblue{Table~\ref{tab:sim_settings} summarizes the experiments and recorded settings. Figure~\ref{fig:single_shot} uses \texttt{ldpc} 2.4.1. The distance audit uses Julia 1.12.6, with Julia package versions recorded in the accompanying repository, \texttt{dist-m4ri} 0.9.0 (commit \texttt{1f46277}), and \textsf{sQetch} 0.1.0. The runs for Figs.~\ref{fig:phenomenological_model_xbasis}, \ref{fig:phenomenological_model_zbasis}, \ref{fig:circuitlevel_model_xbasis} and~\ref{fig:circuitlevel_model_zbasis} use Python 3.13.3, Stim and Sinter 1.15.0, \texttt{tesseract-decoder} 0.1.1.dev20260414023431, \texttt{ldpc} 2.4.1, NumPy 2.2.6 and SciPy 1.17.1. The accompanying repository contains the input matrices, circuits and detector error models used for sampling, raw statistics, plotting script, settings and file hashes for these runs~\cite{feroz_ahmed_mian_2026_24095975}. For the remaining experiments, the saved notebook records Stim and Sinter 1.16.0 and \texttt{tesseract-decoder} 0.1.1.dev20260709013459; these notebook versions do not establish the versions used in earlier runs. Available run logs and results for these experiments are also provided in the repository.}

\subsection{\texorpdfstring{\revisionblue{Protocol of the memory experiments}}{Protocol of the memory experiments}}
In the code-capacity experiment of Fig.~\ref{fig:tesseract_ler_vs_p}, each circuit measures all stabilizers noiselessly, applies the depolarizing channel once, and measures all stabilizers noiselessly again; detectors compare the two rounds. A shot is counted as a logical failure if the decoder mispredicts any of the $2k$ logical $X$ and $Z$ observables. The sampled values of $p$ and numbers of shots are listed in Table~\ref{tab:sim_settings}, and the dashed lines in Fig.~\ref{fig:tesseract_ler_vs_p} are least-squares power-law fits to the points with a nonzero logical error rate.

In the circuit-level experiments of Figs.~\ref{fig:circuitlevel_model_xbasis} and~\ref{fig:circuitlevel_model_zbasis}, the pseudo-thresholds are interpolated between sampled points a factor of $2.4$ apart, so they are coarse. At $p\approx0.13\%$ the per-round logical error rate of the MM code is about $4\times10^{-4}$, compared with below $10^{-5}$ for the BB and AMC codes.

\revisionblue{The following protocol applies to the multi-round memory experiments of Sec.~\ref{sec:numerical_simulations} and
Appendices~\ref{section:more_phenomenological_noise_experiments} and~\ref{section:lrccircuitlevelnoisemodel};
the single-shot and overlapping-window experiments are specified separately. Here $d$ is the distance
value used in the simulation and shown in the figure legends. For the TT code this is the smaller sector
distance, $d=6$, so in Figs.~\ref{fig:phenomenological_model_xbasis}, \ref{fig:phenomenological_model_zbasis},
\ref{fig:circuitlevel_model_xbasis} and~\ref{fig:circuitlevel_model_zbasis} we use $T=24$ for the MM, BB
and AMC codes and $T=12$ for the TT code. Memory circuits were generated with the
\texttt{make\_css\_code\_memory\_circuit} method of the LDPC package~\cite{roffe_decoding_2020} and sampled
with Sinter~\cite{gidney2021stim} using the \texttt{TesseractSinterDecoder}, with the decoder settings given
for the circuit-level model.}

\revisionblue{From the failure probability $P_L$ of a whole $T$-round run, the per-round logical error rate is
\begin{equation}
p_L=\tfrac12\bigl[1-(1-2P_L)^{1/T}\bigr]\qquad (P_L\leq\tfrac12),
\label{eq:per_round}
\end{equation}
and $p_L = 1 - \tfrac12\bigl[1-(2P_L-1)^{1/T}\bigr]$ for $P_L>\tfrac12$, as computed by
\texttt{sinter.plot\_error\_rate}. For a single binary logical outcome, the first branch inverts the independent-flip model. Because a shot fails if any logical observable is mispredicted, we use this conversion as an effective per-round normalization rather than as a measured stationary failure probability. It reduces to $p_L\approx P_L/T$ when $P_L\ll1$. We call the physical error rate at which
$p_L = p$ the pseudo-threshold; unlike a threshold, it is a finite-size crossing that refers to a
single code and decoder. The shaded region around each curve contains the hypotheses whose likelihood is within a
factor of $10^{3}$ of the maximum-likelihood estimate (the Sinter default, about $\pm 3.7$ standard
deviations for Gaussian statistics); points without an observed logical error are not drawn. The noise
model, decoder settings, number of rounds, sampled values of $p$ and stopping rule of every experiment
are listed in Table~\ref{tab:sim_settings}, and the circuit resources in
Table~\ref{tab:coloration_resources}.}

\begin{table*}[tp]
\color{blue}
\centering
\footnotesize
\def\simcell#1#2{\parbox[t]{#1}{\raggedright\strut #2\strut}}%
\begin{tabular}{|l|l|l|l|l|l|}
\hline
Fig. & \simcell{2.1cm}{Experiment} & \simcell{3.4cm}{Noise model} & Decoder & Rounds & \simcell{5.0cm}{Sampled $p$; stopping rule per point} \\
\hline
\ref{fig:tesseract_ler_vs_p} & \simcell{2.1cm}{code capacity} & \simcell{3.4cm}{depolarizing on data, perfect measurements} & Tesseract-B & 1 & \simcell{5.0cm}{30 values in $[10^{-4},10^{-1}]$; $2\times10^{4}$ shots} \\
\ref{fig:phenomenological_model_xbasis}, \ref{fig:phenomenological_model_zbasis} & \simcell{2.1cm}{memory} & \simcell{3.4cm}{phenomenological} & Tesseract-A & 24, TT 12 & \simcell{5.0cm}{9 values in $[10^{-4},10^{-1}]$; $10^3$ errors or $10^5$ shots} \\
\ref{fig:single_shot} & \simcell{2.1cm}{single-shot} & \simcell{3.4cm}{phenomenological, $X$ or $Z$ flips} & BP+OSD-S & $8+1$ & \simcell{5.0cm}{13 values in $[0.5\%,5\%]$; $10^{3}$ failures or $2\times10^{6}$ shots} \\
\ref{fig:circuitlevel_model_xbasis}, \ref{fig:circuitlevel_model_zbasis} & \simcell{2.1cm}{memory} & \simcell{3.4cm}{circuit-level} & Tesseract-A & 24, TT 12 & \simcell{5.0cm}{9 values in $[10^{-4},10^{-1}]$; $5\times10^3$ errors or $2\times10^5$ shots} \\
\ref{fig:plateau_XZ_T15phenom} & \simcell{2.1cm}{OWD plateau} & \simcell{3.4cm}{phenomenological} & \shortstack[l]{BP+OSD-W\\$c=1$} & 15 & \simcell{5.0cm}{$p=10^{-2}$, $w=1,\dots,13$; stopping rule not recorded} \\
\ref{fig:plateau_XZ_T10circ} & \simcell{2.1cm}{OWD plateau} & \simcell{3.4cm}{circuit-level} & \shortstack[l]{BP+OSD-W\\$c=1$} & 10 & \simcell{5.0cm}{$p=10^{-3}$, $w=1,\dots,6$; stopping rule not recorded} \\
\ref{fig:appendixfig1}, \ref{fig:appendixfig2} & \simcell{2.1cm}{memory} & \simcell{3.4cm}{phenomenological} & Tesseract-A & $2d$ & \simcell{5.0cm}{9 values in $[10^{-4},10^{-1}]$; $500$ errors or $10^5$ shots} \\
\ref{fig:circuitlevel_model_xbasis_w6}--\ref{fig:circuitlevel_model_zbasis_again} & \simcell{2.1cm}{memory, coloration vs.\ LRC} & \simcell{3.4cm}{circuit-level} & Tesseract-A & \simcell{1.2cm}{$2d$; $24$ for $[[216,6]]$} & \simcell{5.0cm}{values in $[10^{-4},10^{-1}]$; $500$ errors or $5\times10^{4}$ shots} \\
\hline
\end{tabular}
\caption{\textbf{Settings of the numerical experiments.} $d$ denotes the distance value used in the simulation and figure legend; it may be an upper bound. Memory-experiment $p$ values are logarithmically spaced; single-shot sampling points are listed in the archived results. Sampling stops when either the shot or error limit is reached, with possible batch overshoot. Tesseract-A (saved run settings for Figs.~\ref{fig:phenomenological_model_xbasis}--\ref{fig:phenomenological_model_zbasis} and~\ref{fig:circuitlevel_model_xbasis}--\ref{fig:circuitlevel_model_zbasis}; saved notebook configuration for the appendix experiments): detector beam $8$, beam climbing, no detector revisiting, merged errors, priority-queue limit $10^{4}$, $10$ \texttt{DetIndex} orderings, seed $2384753$. Tesseract-B: detector beam $5$, priority-queue limit $10^{4}$, $10$ \texttt{DetIndex} orderings, no detector revisiting. BP+OSD-S: min-sum BP with scaling factor $0.625$, at most $30$ iterations, and exhaustive OSD of order $10$ (Sec.~\ref{subsec:singleshot}). BP+OSD-W: min-sum BP with at most $1000$ iterations and OSD of order $0$ in each overlapping window. ``Not recorded'' means the setting was not saved with the results.}
\label{tab:sim_settings}

\end{table*}

\revisionblue{Table~\ref{tab:coloration_resources} lists the resources of the coloration circuits used in Figs.~\ref{fig:phenomenological_model_xbasis}, \ref{fig:phenomenological_model_zbasis}, \ref{fig:circuitlevel_model_xbasis} and~\ref{fig:circuitlevel_model_zbasis}, extracted from the archived Stim circuits. Each round measures all $X$-type checks and then all $Z$-type checks, with one ancilla per check, and schedules the CNOT gates of each type by a minimum edge coloring of its Tanner graph; the number of CNOT layers per round is therefore $w_X+w_Z$. In every layer the CNOT targets are disjoint. We count as idle-noise locations the single-qubit depolarizing channels applied, during CNOT layers, to qubits that do not take part in a CNOT; the data-noise locations are the channels applied to every data qubit at the start of a round. Both counts refer to the circuit-level model; the phenomenological circuits have the same gates but carry noise only on the data qubits at the start of each round and on the measurements. The X- and Z-basis circuits of a code have identical resources.}

\begin{table*}[tp]
\color{blue}
\centering
\small
\renewcommand{\arraystretch}{1.15}
\begin{tabular}{|l|r|r|r|c|r|r|r|r|r|}
\hline
Code & $n$ & $a_X$ & $a_Z$ & $(w_X,w_Z)$ & CNOT layers & CNOTs & Idle-noise loc. & Data-noise loc. & Rounds $T$ \\
\hline
MM $[[144,12,12]]$ & 144 & 96 & 96 & $(9,9)$ & 18 & 1728 & 2592 & 144 & 24 \\
BB $[[144,12,12]]$ & 144 & 72 & 72 & $(6,6)$ & 12 & 864 & 1728 & 144 & 24 \\
AMC $[[180,6,12]]$ & 180 & 120 & 120 & $(6,6)$ & 12 & 1440 & 2160 & 180 & 24 \\
TT $[[135,12,(14,6)]]$ & 135 & 45 & 135 & $(9,6)$ & 15 & 1215 & 2295 & 135 & 12 \\
\hline
\end{tabular}
\caption{\textbf{Resources of the coloration circuits of Figs.~\ref{fig:phenomenological_model_xbasis}, \ref{fig:phenomenological_model_zbasis}, \ref{fig:circuitlevel_model_xbasis} and~\ref{fig:circuitlevel_model_zbasis}.} $a_X$ and $a_Z$ count the ancillas measuring $X$- and $Z$-type checks, $w_X,w_Z$ are the check weights, and the CNOT layers, CNOT gates, idle-noise locations and data-noise locations are counted per syndrome-extraction round in the repeated bulk of the circuit. An idle-noise location is one qubit target of a single-qubit depolarizing channel during a CNOT layer.}
\label{tab:coloration_resources}
\end{table*}

\revisionblue{Table~\ref{tab:lrc_resources} reports the resources of the archived left-right circuits. One ancilla is assigned to each measured check. We count CNOT layers as intervals separated by \texttt{TICK} instructions that contain CNOT gates, within one iteration of the repeated bulk circuit. This excludes initialization and termination. Idle-noise locations are specified by the qubit targets of the \texttt{DEPOLARIZE1} instructions in the archived circuits; in the bulk circuit these targets are disjoint from the CNOT participants in the same interval. The table describes the saved LRC schedules; the coloration circuits of the main-text experiments are described in Table~\ref{tab:coloration_resources}.}

\begin{table*}[tp]
\color{blue}
\centering
\small
\renewcommand{\arraystretch}{1.15}
\begin{tabular}{|l|r|r|r|r|c|r|}
\hline
Code & $n$ & $a_X$ & $a_Z$ & CNOT layers & $(w_X,w_Z)$ & Idle-noise locations \\
\hline
BB & 144 & 72 & 72 & 8 & $(6,6)$ & 288 \\
MM (weight 6) & 144 & 96 & 96 & 12 & $(6,6)$ & 868 \\
MM (weight 6) & 216 & 144 & 144 & 12 & $(6,6)$ & 1353 \\
AMC & 180 & 120 & 120 & 12 & $(6,6)$ & 1018 \\
MM (weight 9) & 144 & 96 & 96 & 17 & $(9,9)$ & 1244 \\
TT & 135 & 45 & 135 & 14 & $(9,6)$ & 1194 \\
\hline
\end{tabular}
\caption{\textbf{Resources of the archived LRC schedules.} $a_X$ and $a_Z$ count ancillas measuring $X$ and $Z$ checks; $w_X,w_Z$ are their check weights. CNOT layers and idle-noise locations are counted per repeated bulk iteration. An idle-noise location is one qubit target of a \texttt{DEPOLARIZE1} instruction, so repeated occurrences on the same qubit count separately.}
\label{tab:lrc_resources}
\end{table*}

\begingroup\color{blue}
\subsection{Details of the single-shot experiment}
Figure~\ref{fig:single_shot} uses three codes whose distances $(d_X,d_Z)=(12,12)$, $(14,6)$ and $(12,12)$ are established by an exhaustive connected-cluster search with \texttt{dist-m4ri} (Appendix~\ref{sec:appendix_search}). The $X$-type checks of the MM code satisfy $30$ independent relations, of which the metachecks span $22$ (Proposition~\ref{prop:mm_homology}(iv)), while those of the BB code satisfy $6$. Pseudo-thresholds, the crossings of $p_L=p$, are obtained by log--log interpolation between the sampled points. The pseudo-thresholds of the MM code round to $4.07\%$ in both memory bases, and at $p=1\%$ its effective per-round logical error rates are $2.8\times10^{-6}$ ($X$, $45$ failures) and $1.8\times10^{-6}$ ($Z$, $28$ failures). The BB code has pseudo-thresholds of $2.46\%$ ($X$) and $2.47\%$ ($Z$), and at the sampled points with $p=1$--$2\%$ its $p_L$ is roughly $30$--$84$ times that of the MM code. Without the metacheck rows, the MM code reaches $p_L=7.3\times10^{-6}$ ($X$) and $5.0\times10^{-6}$ ($Z$) at $p=1\%$, compared with $1.5\times10^{-4}$ for the BB code. For the TT code in the $Z$ memory experiment, the metacheck rows change $p_L$ only slightly. Data, scripts, and checks of the metacheck matrices against QuantumClifford.jl are provided in our repository~\cite{feroz_ahmed_mian_2026_24095975}.

In the single-shot experiment of Sec.~\ref{subsec:singleshot}, the residual error after each correction is carried into the next round. The final perfect round represents a transversal readout of the data qubits; it is decoded with $H$ alone, and the shot fails if the remaining error is a nontrivial logical operator. Like Eq.~\eqref{eq:per_round}, the per-round rate $p_L$ normalizes a finite-duration failure probability rather than measuring a stationary rate. For a type of check without an adjacent Koszul metacheck map, the decoder uses $(H\;\,I)$ only. The metacheck equations are algebraically redundant, since multiplying $H\mathbf{e}+\mathbf{m}=\mathbf{s}$ by $M$ gives $M\mathbf{m}=M\mathbf{s}$. They add no independent syndrome information, but change the graph on which BP+OSD operates. BP uses the min-sum rule with scaling factor $0.625$ and at most $30$ iterations, followed by exhaustive OSD of order $10$. Error bars are Wilson $95\%$ intervals, and the sampling settings are listed in Table~\ref{tab:sim_settings}.

\subsection{Window size one and the LDPC-breakdown condition}
With $w=1$, the windowed decoder gives high logical failure rates for all three codes under circuit-level noise, and under phenomenological noise it reaches only about $0.07$--$0.09$ per round for the MM $[[144,12,12]]$ code. These $w=1$ points differ from the single-shot experiment of Sec.~\ref{subsec:singleshot} in three ways: the latter uses independent $X$ or $Z$ flips instead of depolarizing noise, the single-stage decoder of Eq.~\eqref{eq:single_stage} with the metacheck rows and OSD of order $10$ instead of windowed BP+OSD of order $0$, and a perfect final round.

\revisionblue{The LDPC-breakdown condition of Refs.~\cite{lin2025singleshottwoshotdecodinggeneralized, lin2025abelianmulticyclecodessingleshot} offers a simple comparison between data-error and measurement-error explanations of a syndrome. For an irreducible error $e$ (Definition~\ref{def:irreducible}) whose syndrome has weight $s$ in a single round, explaining that syndrome by measurement errors over a window of $w$ rounds has lower weight than $e$ when
\begin{equation}
s\,w<|e|.
\label{eq:ldpc_breakdown}
\end{equation}
For the observed profile $6,10,12,14,18,18$ of the MM $[[486,24,12]]$ code in Table~\ref{tab:confinement}, the ratios $|e|/s$ are at most $1/3$ over the reported prefix, so the complementary inequality $s\,w\geq|e|$ already holds at $w=1$; for the reported BB profile $3,4,3,4,3,4$, the weight-six entry requires $w\geq2$. In the experiments, however, every code needs $w\geq2$, and under phenomenological noise the MM $[[486,24,12]]$ code keeps improving at larger windows. These features lie beyond a minimum-weight criterion evaluated on observed profiles, so we take the window-size dependence of BP+OSD from the decoding experiments themselves.}
\endgroup

\section{Lifted-Product Structure of MM Codes}
\label{sec:appendix_lp_example}
\begingroup\color{blue}

This appendix makes the lifted-product description of MM codes explicit. Each defining polynomial $F_i$ is a $1\times1$ boundary matrix of the two-term complex $K(F_i)=(S\xleftarrow{F_i}S)$, and the Koszul complex is their tensor product over $S$. Boundary matrices act on column vectors, and with qubits in degree $q$ the check matrices are $P_X=\partial_q$ and $P_Z=\partial_{q+1}^{\top}$. The \texttt{QuantumClifford.jl} commands that construct every example below are provided in the accompanying repository~\cite{feroz_ahmed_mian_2026_24095975}.

\paragraph*{A four-fold example.} The $[[96,12,(4,4)]]$ MM code has $t=4$, $S=\mathbb F_2[w,x,y,z]/\langle w^2-1,x^2-1,y^2-1,z^2-1\rangle\cong\mathbb F_2[\mathbb Z_2^4]$ and
\begin{gather*}
F_1=wz+xyz,\quad F_2=w+xyz,\\
F_3=wyz+xy,\quad F_4=xyz+xz.
\end{gather*}
Grouping the constituents in pairs gives the two factor complexes $\mathcal C=K_\bullet([F_1,F_2];S)$ and $\mathcal D=K_\bullet([F_3,F_4];S)$, which are scalar lifted-product codes $[[32,8,2]]$, and
\begin{equation}
K_\bullet([F_1,F_2,F_3,F_4];S)\cong\mathcal C\otimes_S\mathcal D .
\label{eq:app_mm_grouped_product}
\end{equation}
The MM code places the qubits in degree two, $(\mathcal C\otimes_S\mathcal D)_2\cong S^6$, giving $6\cdot16=96$ qubits. The middle boundary maps give $P_X$ and $P_Z$, while the outer ones are built from the factor checks,
\[
M_X=\partial_1=[\,P_X^{\mathcal C}\;\;P_X^{\mathcal D}\,],\qquad M_Z=\partial_4^{\top}=[\,P_Z^{\mathcal D}\;\;P_Z^{\mathcal C}\,].
\]
A three-term presentation $\mathrm{LP}(A,B)$ has no such outer maps, although its checks may still be redundant. Connected-cluster searches with \texttt{dist-m4ri} give $d_X=d_Z=4$ for the MM code and $d_X=d_Z=2$ for each factor.

\paragraph*{Three generators.} For $t=3$, the Koszul differential $\partial(re_J)=\sum_{i\in J}F_ire_{J\setminus\{i\}}$ with $(F_1,F_2,F_3)=(a,b,c)$ and qubits in degree one gives
\[
P_X=[\,\rho(a)\;\;\rho(b)\;\;\rho(c)\,],\qquad
P_Z=\begin{bmatrix}\rho(b^*)&\rho(a^*)&0\\ \rho(c^*)&0&\rho(a^*)\\ 0&\rho(c^*)&\rho(b^*)\end{bmatrix},
\]
where $\rho$ is the regular representation and $a^*$ is the image of $a$ under $g\mapsto g^{-1}$. These are the scalar three-fold lifted-product check matrices of Li, Preskill and Xu~\cite{li2025transversaldimensionjump}; the remaining map supplies $M_Z=\partial_3^{\top}$. For example, $a=x^2y+x^2y^2$, $b=1+xy^2$ and $c=x+x^2y$ over $\mathbb F_2[x,y]/\langle x^3-1,y^3-1\rangle$ give $(n,k)=(27,3)$, and the two-generator component $(a,b)$ gives $(n,k)=(18,2)$.

\paragraph*{Full versus two-matrix check presentation.} Truncating a factor complex before taking the tensor product changes the code: for the three-dimensional toric code with $F_i=1+x_i$ on $\mathbb Z_3^3$, replacing $K_\bullet([F_1,F_2];S)$ by $S\xleftarrow{[F_1\;F_2]}S^2$ keeps $n=81$ but gives $k=11$ instead of $3$. More generally, in the convention $n=\ell(n_An_B+r_Ar_B)$ with $\ell=\dim_{\mathbb F_2}R$, a two-matrix lifted product of $A\in R^{r_A\times n_A}$ and $B\in R^{r_B\times n_B}$ has $m_X=\ell r_An_B$ and $m_Z=\ell n_Ar_B$ check rows, counting redundant ones.

\begin{proposition}[Obstruction to a two-matrix check presentation]
\label{prop:not_two_fold}
The standard two-matrix lifted-product presentation satisfies $n^2\geq4m_Xm_Z$, whereas the full Koszul check matrices satisfy $n^2<4m_Xm_Z$ whenever $t\geq3$ and $1\leq q\leq t-1$. These check matrices therefore cannot coincide, even up to row and column permutations, with those of a two-matrix lifted product.
\end{proposition}

\begin{proof}
For the two-matrix lifted product, $n^2-4m_Xm_Z=\ell^2(n_An_B-r_Ar_B)^2\geq0$. For the Koszul construction, $n=\binom{t}{q}s$, $m_X=\binom{t}{q-1}s$ and $m_Z=\binom{t}{q+1}s$ with $s=\dim_{\mathbb F_2}S$, so
\[
\frac{n^2}{m_Xm_Z}=\left(1+\frac1q\right)\left(1+\frac1{t-q}\right),
\]
and both factors are at most $2$, with equality in both only for $t=2$, $q=1$.
\end{proof}

A different generating set of the same stabilizer group might still admit a two-matrix presentation. Finally, the qubit degree matters: placing the qubits in degree $q=1,2,3$ of the four-fold example gives $[[64,8]]$, $[[96,12]]$ and $[[64,8]]$ codes, and only $q=2$ has metacheck maps on both sides. In general both maps are present when $2\leq q\leq t-2$, which $q=\lfloor t/2\rfloor$ satisfies for every $t\geq4$; the degree-one and degree-three codes are exchanged by the duality of Proposition~\ref{prop:xz_duality}.

\endgroup

\section{\texorpdfstring{\revisionblue{Univariate MM codes as quasi-cyclic codes}}{Univariate MM codes as quasi-cyclic codes}}
\label{app:quasicyclic}
\begingroup\color{blue}For $D = 1$, write $R_\ell = \mathbb{F}_2[x]/\langle x^\ell - 1\rangle$, so that $K_q \cong R_\ell^{\binom{t}{q}}$. The binary symplectic representation of the stabilizer group is then an $R_\ell$-submodule of $R_\ell^{2\binom{t}{q}}$, that is, an even-index quasi-cyclic code of index $2\binom{t}{q}$, generated by the images of the $\binom{t}{q-1}$ block rows of $P_X$ and the $\binom{t}{q+1}$ block rows of $P_Z$. For $t = 4$ and $q = 2$, this is an index-12 module with at most eight generators. The symplectic self-orthogonality of such multi-generator quasi-cyclic codes is characterized in general by Guan et al.~\cite{guan2025symplectic}; their distance bounds concern index-two codes with a single generator and therefore do not carry over directly to the high-index MM codes. The ring $R_\ell$ is a principal ideal ring, and for odd $\ell$ it is semisimple and decomposes into finite fields, the cyclic constituents, according to the factorization of $x^\ell - 1$; Sec.~\ref{sec:univariate_distance} discusses the resulting dimension and distance formulas, including an exactly solvable subfamily. The multivariate ring $S$ is also semisimple when every $\ell_i$ is odd; as a finite commutative semisimple algebra, it decomposes into finite fields. If some $\ell_i$ is even, $S$ is not semisimple, so this field decomposition does not apply. \revisionblue{With exactly one even $\ell_i$ every ideal of $S$ is still principal; with two or more even $\ell_i$ ideals need not be principal~\cite{fisher1976principal}.}
In this picture, the adjacent Koszul maps supply relations among the natural stabilizer generators, and further relations exist precisely when the adjacent homology is nonzero (Proposition~\ref{prop:mm_homology}(iv)).
\endgroup

\begingroup\color{blue}
\begin{example}[An index-12 quasi-cyclic stabilizer module]
Let
\[
R_5=\mathbb{F}_2[x]/\langle x^5-1\rangle
\]
and choose four Koszul generators
\[
a=1+x,\quad b=1+x^2,\quad
c=1+x^3,\quad d=1+x^4.
\]
Place the qubits in $K_2\cong R_5^6$, with block order
$(12,13,14,23,24,34)$. Thus, the binary code has
$n=6\cdot5=30$ physical qubits.

Write $\overline{f}(x)=f(x^{-1})$ in $R_5$.
Here,
$\overline{a}=d$, $\overline{b}=c$,
$\overline{c}=b$, and $\overline{d}=a$.
In the column convention, the polynomial check matrices
are
\[
P_X=
\begin{pmatrix}
b&c&d&0&0&0\\
a&0&0&c&d&0\\
0&a&0&b&0&d\\
0&0&a&0&b&c
\end{pmatrix},
\qquad
P_Z=
\begin{pmatrix}
b&c&0&d&0&0\\
a&0&c&0&d&0\\
0&a&b&0&0&d\\
0&0&0&a&b&c
\end{pmatrix}.
\]
Each polynomial entry denotes its $5\times5$ binary
regular-representation matrix. In particular, the
involution $f\mapsto\overline{f}$ accounts for the
binary transpose in $P_Z=\partial_3^\top$.

The binary symplectic stabilizer module is
\[
\mathcal{S}
=
\operatorname{rowspan}_{R_5}
\begin{pmatrix}
P_X&0\\
0&P_Z
\end{pmatrix}
\subseteq R_5^{12}.
\]
It therefore has at most eight module generators.
For example, its first generator is
\[
g=(b,c,d,0,0,0\mid0,0,0,0,0,0).
\]
The vectors $g,xg,\ldots,x^4g$ represent five cyclically
shifted binary stabilizers. Applying these shifts to
all eight module generators spans the binary stabilizer
space.

The index $12$ counts the six $X$-component blocks and
six $Z$-component blocks, each of length five. Thus the
symplectic representation has $60$ binary coordinates,
although the code has only $30$ physical qubits.
\end{example}
\endgroup

\section{\texorpdfstring{\revisionblue{Proofs of the propositions of Section~\ref{sec:multivariate_multicycle_codes}}}{Proofs of the propositions of Section VI}}
\label{app:mm_homology}
\begingroup\color{blue}
This appendix proves Propositions~\ref{prop:mm_amc}, \ref{prop:univariate_separable}, \ref{prop:xz_duality} and~\ref{prop:mm_homology}. The last two use only linear algebra and the explicit Koszul formula, Eq.~\eqref{eq:explicit_formula}. Throughout, $S=\mathbb F_2[G]$ with $G=\mathbb Z_{\ell_1}\times\cdots\times\mathbb Z_{\ell_D}$, $F_1,\ldots,F_t\in S$, $I=(F_1,\ldots,F_t)$, and $K_j=\bigoplus_{|J|=j}S\,e_J$ with
\[
 \partial(r\,e_J)=\sum_{i\in J}F_i\,r\,e_{J\setminus\{i\}},
\]
where no signs appear because the characteristic is two. Every $\partial_j$ commutes with multiplication by elements of $S$, so each homology group $H_j$ is an $S$-module as well as an $\mathbb F_2$-vector space.

\subsection{Proof of Proposition~\ref{prop:mm_amc}}
Write $K(a) = (S \xleftarrow{a} S)$ for the two-term complex in degrees $0$ and $1$. The Koszul complex satisfies $K_\bullet([a_1,\dots,a_t];S) \cong K_\bullet([a_1,\dots,a_{t-1}];S)\otimes_S K(a_t)$~\cite{eisenbud2013commutative, weibel1994introduction}. For any complex $(Q_\bullet,\partial^Q)$ of $S$-modules, the tensor product $Q_\bullet \otimes_S K(a)$ has degree-$i$ term $Q_{i-1} \oplus Q_i$, where the first summand is $Q_{i-1}\otimes K(a)_1$ and the second is $Q_i \otimes K(a)_0$, and, in characteristic $2$, differential
\[
\partial_i = \begin{bmatrix} \partial^Q_{i-1} & 0 \\ a\,I & \partial^Q_{i} \end{bmatrix},
\qquad
\partial_1 = \begin{bmatrix} a & \partial^Q_1 \end{bmatrix},
\qquad
\partial_{t} = \begin{bmatrix} \partial^Q_{t-1} \\ a \end{bmatrix},
\]
which is the recursion of Eqs.~(23)--(25) of Ref.~\cite{lin2025abelianmulticyclecodessingleshot} with $N = \mathrm{M}(a)$; the signs in that recursion vanish modulo $2$. Both complexes start from the same two-term complex $K(a_1)$, so by induction on $t$ they coincide up to the ordering of blocks in each degree. Applying the regular representation, which is an injective ring homomorphism from $S$ to binary matrices (multilevel circulant matrices in the product-group basis), preserves this identification. For $t = 2$, the column-convention matrices are $\partial_1 = [\,a_1 \;\; a_2\,]$ and $\partial_2 = [\,a_2 \;\; a_1\,]^{\top}$, so $P_X = \partial_1$ and $P_Z = \partial_2^{\top} = [\,a_2^{\top} \;\; a_1^{\top}\,]$. These are the check matrices of $\mathrm{LP}(a_1,a_2)$, in which the transpose of a circulant corresponds to the conjugation $g \mapsto g^{-1}$ of the group algebra.

\subsection{Proof of Proposition~\ref{prop:univariate_separable}}
\revisionblue{Write $A_i=\mathbb F_2[y_i]/\langle y_i^{\ell_i}-1\rangle$. Because the $\ell_i$ are pairwise coprime, $y_i\mapsto x^{L/\ell_i}$ is an algebra isomorphism $\bigotimes_i A_i\to R_L$ that only relabels qubits, so it preserves Hamming weight. Under it, $F_i$ acts only on $A_i$, and the Koszul complex is the tensor product over $\mathbb F_2$ of the two-term complexes $A_i\xrightarrow{a_i}A_i$. For the $i$th factor, $H_1=\ker(a_i)=C_i$ and $H_0=A_i/a_iA_i$ both have dimension $\deg g_i$; the degree-one distance is $\delta_i$, and the degree-zero distance is one, since some coordinate vector has nonzero image in the nonzero quotient. The K\"unneth formula gives $k$. Applying Theorem~17 of Ref.~\cite{zeng2020minimal} one factor at a time gives $d_Z=\min_J\prod_{i\in J}\delta_i$, where $J$ runs over the $q$-element subsets of $\{1,\ldots,t\}$: a lowest-weight $Z$ logical is a minimum-weight codeword of $C_i$ in each factor $i\in J$ and a single nonzero coordinate in each other factor. The minimum is attained by the $q$ smallest $\delta_i$. For $d_X$, transpose each factor and reverse the degrees: transposing a circulant replaces $y_i$ by $y_i^{-1}$, which only relabels coordinates, so the degree-zero and degree-one distances become $\delta_i$ and one, and the same theorem gives $d_X$ as the product of the $t-q$ smallest $\delta_i$.}

This proof uses coprimality only to write the code over a single cyclic ring. If instead $G=A_1\times\cdots\times A_t$ for arbitrary finite abelian groups $A_i$, and each $F_i\in\mathbb F_2[A_i]$ is a non-unit, the same argument gives $d_Z$ and $d_X$ as the products of the $q$ and $t-q$ smallest distances of the abelian codes $\ker F_i\subset\mathbb F_2[A_i]$. For instance, $F_i=1+x_i$ over $\mathbb Z_L^4$ gives the four-dimensional toric code with $d_X=d_Z=L^2$.

\subsection{Proof of Proposition~\ref{prop:xz_duality}}
Write $\rho(F)$ for the matrix of multiplication by $F$ in the group basis. For $g\in G$, $\rho(g)$ is the permutation matrix of $h\mapsto gh$, so $\rho(g)^{\top}=\rho(g)^{-1}=\rho(g^{-1})$, and by linearity $\rho(F)^{\top}=\rho(\overline F)$ with $\overline F=\sum_gF_g\,g^{-1}$. Let $\alpha$ be the permutation matrix of $g\mapsto g^{-1}$. Then $\alpha\rho(g)\alpha$ sends $h$ to $(gh^{-1})^{-1}=g^{-1}h$, so $\alpha\rho(F)\alpha=\rho(\overline F)=\rho(F)^{\top}$.

Now compare block patterns. The block of $\partial_{j+1}^{\top}$ in row $K$ ($|K|=j+1$) and column $J$ ($|J|=j$) is $\rho(F_i)^{\top}$ if $J\subset K$ with $K\setminus J=\{i\}$, and zero otherwise. The block of $\partial_{t-j}$ in row $K^c$ ($|K^c|=t-j-1$) and column $J^c$ ($|J^c|=t-j$) is $\rho(F_i)$ if $K^c\subset J^c$ with $J^c\setminus K^c=\{i\}$, and zero otherwise. Since $J\subset K$ if and only if $K^c\subset J^c$, and then $K\setminus J=J^c\setminus K^c$, the two patterns agree under $J\mapsto J^c$, $K\mapsto K^c$. Conjugating each block by $\alpha$ turns $\rho(F_i)$ into $\rho(F_i)^{\top}$. Together these give $\partial_{j+1}^{\top}=\sigma\,\partial_{t-j}\,\sigma$, where $\sigma$ sends $g\,e_J$ to $g^{-1}e_{J^c}$. Taking $j=q$ gives $P_Z^{(q)}=\sigma P_X^{(t-q)}\sigma$, and taking $j=q+1$ with $t=2q$ gives $M_Z=\partial_{q+2}^{\top}=\sigma\,\partial_{q-1}\,\sigma=\sigma M_X\sigma$. Permutation matrices preserve rank, kernels and Hamming weight, so all weight-based quantities agree in the two sectors. Taking ranks in $\partial_{j+1}^{\top}=\sigma\partial_{t-j}\sigma$ gives $\operatorname{rank}\partial_{j+1}=\operatorname{rank}\partial_{t-j}$ for all $j$, and since $\dim K_j=\dim K_{t-j}$, it follows that $h_j=\dim K_j-\operatorname{rank}\partial_j-\operatorname{rank}\partial_{j+1}=h_{t-j}$. For even $t$, the alternating sum $\sum_j(-1)^jh_j$ equals $\sum_j(-1)^j\dim K_j=s\sum_j(-1)^j\binom tj=0$, because the ranks cancel in pairs; pairing degree $j$ with degree $t-j$ then gives $(-1)^{t/2}h_{t/2}=-2\sum_{j<t/2}(-1)^jh_j$, so $k=h_{t/2}$ is even.

This is the self-duality of the Koszul complex~\cite[Prop.~1.6.10]{bruns1998cohen}, $e_J\mapsto e_{J^c}$, combined with the fact that the binary transpose of $\rho(F)$ is $\rho(\overline F)$.

\subsection{Proof of Proposition~\ref{prop:mm_homology}}
\textit{(i) The ideal $I$ acts as zero on homology.} Fix $i$ and define $h_i\colon K_j\to K_{j+1}$ by $h_i(r\,e_J)=r\,e_{J\cup\{i\}}$ if $i\notin J$ and $h_i(r\,e_J)=0$ if $i\in J$. For $i\notin J$,
\begin{align}
 \partial h_i(r e_J)&=F_i\,r\,e_J+\sum_{m\in J}F_m\,r\,e_{(J\setminus\{m\})\cup\{i\}},\\
 h_i\partial(r e_J)&=\sum_{m\in J}F_m\,r\,e_{(J\setminus\{m\})\cup\{i\}},
\end{align}
so $\partial h_i+h_i\partial=F_i\cdot\mathrm{id}$. The case $i\in J$ is similar. Multiplication by $F_i$ is therefore null-homotopic and acts as zero on every $H_j$, and so does every element of $I$~\cite[Prop.~1.6.5]{bruns1998cohen}. If $I=S$, then $1\in I$ acts as zero, so every $H_j$ vanishes.

\textit{(ii) Every $H_j$ is nonzero when $I\neq S$.} A finite-dimensional commutative algebra is a finite product of local algebras, $S\cong S_1\times\cdots\times S_r$~\cite[Thm.~8.7]{atiyah2018introduction}, and since the differentials are $S$-linear the Koszul complex and its homology split in the same way. On a factor where some $F_i$ is a unit, the homology vanishes by (i). If $I\neq S$, some factor $S_m$ has all $F_i$ in its maximal ideal, and in a finite-dimensional local algebra these elements are nilpotent. On such a factor we add one generator at a time. Let $K'$ be the Koszul complex of $F_1,\ldots,F_{t-1}$, so that $K_j=K'_{j-1}\oplus K'_j$ as in the proof of Proposition~\ref{prop:mm_amc}. This gives a short exact sequence of complexes $0\to K'\to K\to K'[-1]\to0$, whose long exact homology sequence
\begin{align*}
 &\cdots\to H_j(K')\xrightarrow{\;F_t\;}H_j(K')\to H_j(K)\\
 &\qquad\to H_{j-1}(K')\xrightarrow{\;F_t\;}H_{j-1}(K')\to\cdots
\end{align*}
has connecting map multiplication by $F_t$. Exactness gives an injection $H_j(K')/F_tH_j(K')\hookrightarrow H_j(K)$ and a surjection of $H_j(K)$ onto the kernel of $F_t$ on $H_{j-1}(K')$. A nilpotent map on a nonzero finite-dimensional space has a nonzero kernel and a nonzero cokernel. Starting from $H_0=S_m\neq0$ for no generators, induction on the number of generators shows that every $H_j$ with $0\leq j\leq t$ is nonzero. This is the rigidity of the Koszul complex~\cite[Exercise~1.6.31]{bruns1998cohen} in the case of grade zero.

\textit{(iii) Maximal ideals as common zeros.} A maximal ideal $\mathfrak m$ of $S$ has a finite field as quotient, so it is the kernel of a ring homomorphism $\chi\colon S\to\overline{\mathbb F}_2$. Such a $\chi$ is determined by $\zeta_i=\chi(x_i)$, which must satisfy $\zeta_i^{\ell_i}=1$. Writing $\ell_i=2^{a_i}\ell'_i$ with $\ell'_i$ odd, $x^{\ell_i}-1=(x^{\ell'_i}-1)^{2^{a_i}}$ in characteristic two, so $\zeta_i^{\ell'_i}=1$. Conversely, every such tuple defines a homomorphism. Hence $I\neq S$, that is $I$ lies in some maximal ideal, exactly when the $F_i$ vanish at a common tuple of this form. Combined with (i) and (ii), $k=h_q>0$ exactly when such a common zero exists.

\textit{(iv) Counting relations.} Relations among the $X$-type checks are row vectors $r$ with $rP_X=0$. They form the left kernel of $\partial_q$, of dimension $m_X-\operatorname{rank}\partial_q$ with $m_X=\dim K_{q-1}$. The rows of $M_X=\partial_{q-1}$ are such relations and span a space of dimension $\operatorname{rank}\partial_{q-1}$. The difference is $\dim K_{q-1}-\operatorname{rank}\partial_q-\operatorname{rank}\partial_{q-1}$. By rank--nullity and the definition of $H_{q-1}$,
\begin{align}
 \dim K_{q-1}-\operatorname{rank}\partial_{q-1}&=\dim\ker\partial_{q-1},\\
 \dim\ker\partial_{q-1}-\operatorname{rank}\partial_q&=h_{q-1},
\end{align}
so the difference equals $h_{q-1}$.
For the $Z$-type checks, $rP_Z=0$ means $\partial_{q+1}r^{\top}=0$, and the rows of $M_Z=\partial_{q+2}^{\top}$ span $\operatorname{im}\partial_{q+2}$, so the difference is $\dim\ker\partial_{q+1}-\operatorname{rank}\partial_{q+2}=h_{q+1}$. Both are positive when $k>0$ by (ii). Finally, $\ker M_X/\operatorname{im}P_X=H_{q-1}\neq0$, so some syndrome $s\notin\operatorname{im}P_X$ satisfies $M_Xs=0$ and $d_{ss}^X$ is finite. On the $Z$ side, $\ker M_Z/\operatorname{im}P_Z$ is the cohomology group $H^{q+1}$, which has the same dimension $h_{q+1}$.

\textit{(v) At most one even order.} Write $G=P\times Q$, where $P$ is the Sylow $2$-subgroup (the elements of $2$-power order) and $|Q|$ is odd; $P$ is cyclic exactly when at most one $\ell_i$ is even. For odd $m$ the polynomial $x^m-1$ has no repeated factors, so $\mathbb F_2[Q]$ is a product of finite fields $\kappa_m$, and $S\cong\prod_m\kappa_m[P]$. Suppose $P$ is cyclic of order $N=2^a$ with generator $p$. Since $(p-1)^N=p^N-1=0$, each factor is $S_m\cong\kappa_m[\varepsilon]/(\varepsilon^N)$ with $\varepsilon=p-1$. In this ring every nonzero element has the form $u\,\varepsilon^e$ with $u$ a unit, obtained by factoring out the lowest power of $\varepsilon$. If all $F_i$ vanish on the factor, the differentials vanish there and the formula below is immediate. Otherwise let $g=F_{i_0}$ be the image with the smallest exponent $e$; then every image is $F_i=c_ig$ for some $c_i\in S_m$. The $S_m$-linear automorphism $\varphi$ of $S_m^t$ with $\varphi(e_i)=e_i+c_ie_{i_0}$ for $i\neq i_0$ satisfies $f=f'\circ\varphi$, where $f(e_i)=F_i$, $f'(e_{i_0})=g$ and $f'(e_i)=0$ otherwise, so applying $\varphi$ to each factor of $e_{j_1}\wedge\cdots\wedge e_{j_k}$ commutes with the differentials of the exterior-algebra description in Sec.~\ref{sec:koszul_complexes} and gives an isomorphism of Koszul complexes $K(F_1,\ldots,F_t)\cong K(g,0,\ldots,0)$ on that factor (see also~\cite[Prop.~1.6.8]{bruns1998cohen}). This isomorphism changes Hamming weights, so it says nothing about distances. In $K(g,0,\ldots,0)$ only the generator $g$ acts, and
\[
 H_j\cong(S_m/gS_m)^{\binom{t-1}{j}}\oplus\operatorname{Ann}(g)^{\binom{t-1}{j-1}}.
\]
For $g=u\,\varepsilon^e$, the annihilator is $(\varepsilon^{N-e})$, and multiplication by $\varepsilon^{N-e}$ maps $S_m$ onto it with kernel $(\varepsilon^e)=gS_m$, so $\operatorname{Ann}(g)\cong S_m/gS_m$. Since $\binom{t-1}{j}+\binom{t-1}{j-1}=\binom tj$ and $gS_m$ is the image of $I$ in $S_m$, the factor contributes $\binom tj\dim_{\mathbb F_2}(S_m/gS_m)$ to $h_j$, and summing over the factors gives $h_j=\binom tj\dim_{\mathbb F_2}S/I$. If every $\ell_i$ is odd, $P$ is trivial and each factor is a field $\mathbb F_{2^{r_m}}$, which contributes to $S/I$ exactly when all $F_i$ vanish there. Such a factor corresponds to an orbit of $r_m$ points of $\mu_{\ell_1}\times\cdots\times\mu_{\ell_D}$ under $\zeta\mapsto\zeta^2$, so $\dim S/I$ is the number of common zeros. With two or more even orders, the image of $I$ in a local factor need not be principal, and this argument does not apply.
\endgroup

\end{document}